\documentclass[final,3p,times]{elsarticle}

\usepackage{amssymb}
\usepackage{amsmath}

\usepackage{algorithm}%
\usepackage{algorithmic}%

\usepackage{tikz}
\usetikzlibrary{arrows.meta}
\usepackage{subfig}
\usepackage{bm}
\usepackage{xfrac}

\usepackage{xcolor}

\newcommand{\vecr}[1]{\bm{#1}}
\newcommand{\matx}[1]{\bm{#1}}
\newcommand{\x}{\vecr{x}}
\newcommand{\xiv}{\vecr{\xi}}
\newcommand{\g}{\vecr{\gamma}}
\renewcommand{\L}{\vecr{\Lambda}}
\newcommand{\s}{\vecr{s}}

\newcommand{\fof}[1]{\hspace{-0.3ex}\left( #1 \right)}
\newcommand{\bof}[1]{\left( #1 \right)}
\newcommand{\fofx}{\fof{\x}}

\newcommand{\dpart}[2]{\frac{\partial #1}{\partial #2}}

\journal{Computer Methods in Applied Mechanics and Engineering}

\begin{document}

\begin{frontmatter}



\title{Large-Scale Topology Optimisation of Time-dependent Thermal Conduction Using Space-Time Finite Elements and a Parallel Space-Time Multigrid Preconditioner}

\author[sdu]{Joe Alexandersen\corref{cor1}} 
\ead{joal@sdu.dk}
\cortext[cor1]{Corresponding author}
            
\author[sdu]{Magnus Appel} 

\affiliation[sdu]{organization={Institute of Mechanical and Electrical Engineering, University of Southern Denmark},
            addressline={Campusvej 55}, 
            city={Odense M},
            postcode={DK-5230}, 
            country={Denmark}}

\begin{abstract}
This paper presents a novel space–time topology optimisation framework for time-dependent thermal conduction problems, aiming to significantly reduce the time-to-solution. By treating time as an additional spatial dimension, we discretise the governing equations using a stabilised continuous Galerkin space–time finite element method. The resulting large all-at-once system is solved using an iterative Krylov solver preconditioned with a parallel space–time multigrid method employing a semi-coarsening strategy. Implemented in a fully parallel computing framework, the method yields a parallel-in-time method that demonstrates excellent scalability on a distributed-memory supercomputer, solving problems up to 4.2 billion degrees of freedom. Comparative studies show up to $52\times$ speed-up over traditional time-stepping approaches, with only moderate increases in total computational cost in terms of core-hours. The framework is validated on benchmark problems with both time-constant and time-varying designs, and its flexibility is demonstrated through variations in material properties. These results establish the proposed space-time method as a promising approach for large-scale time-dependent topology optimisation in thermal applications.
\end{abstract}

\begin{graphicalabstract}
\begin{figure*}[h!]
    \centering
    \includegraphics[width=\linewidth]{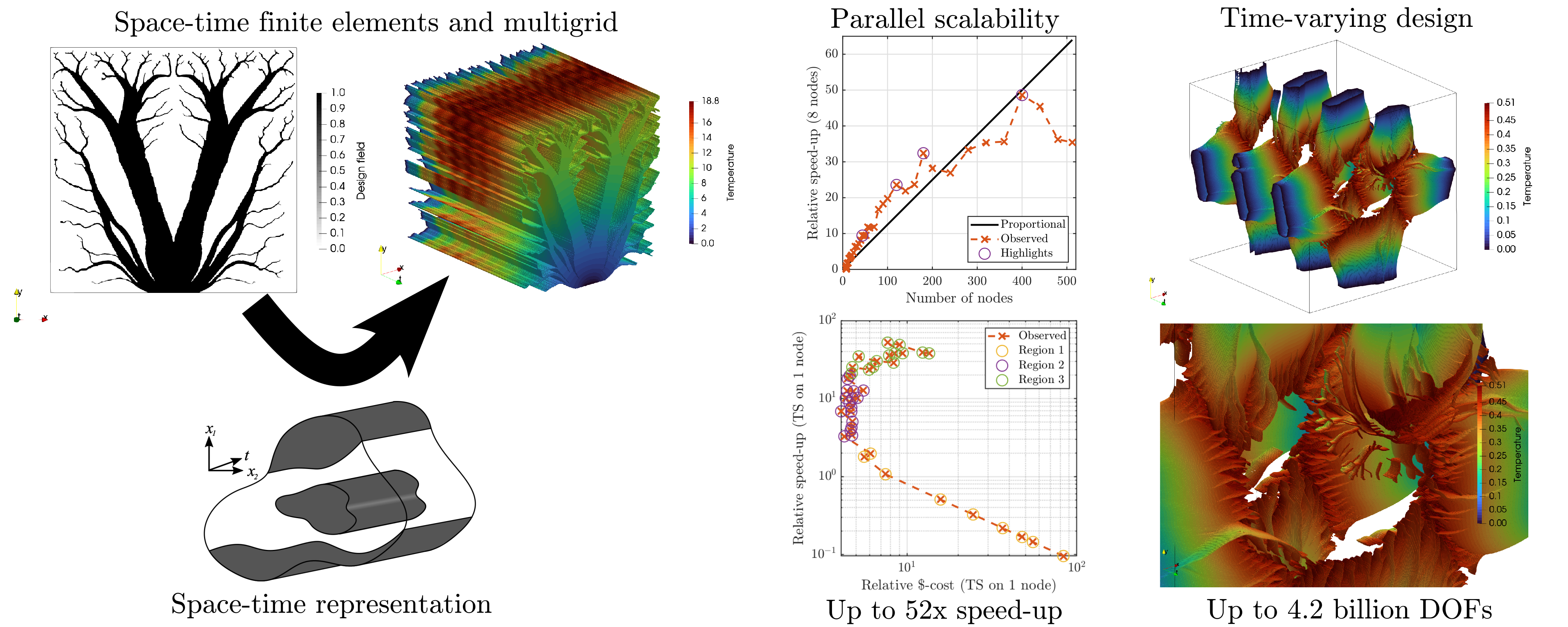}
\end{figure*}
\end{graphicalabstract}

\begin{highlights}

\item Novel space-time topology optimisation framework for transient diffusion
\item Applies a multigrid preconditioner to solve large all-at-once transient systems
\item Achieves up to $52\times$ speed-up over traditional time-stepping methods
\item Solves systems with up to 4.2 billion DOFs using full space-time parallelism
\item Demonstrated on time-constant and time-varying design fields for various material properties
\end{highlights}


\end{frontmatter}



\section{Introduction}

\subsection{Motivation}

Topology optimisation is a powerful computational design method, originally developed for stiffness design in structural mechanics \citep{Bendsoee1988,Bendsoee2004}. It has since been applied to a variety of physical problems \citep{Deaton2014}, such as electromagnetics, photonics, and fluid dynamics.
It is an iterative process, where the performance of the system is analysed using one or several computer simulations. The design is updated until the performance is satisfactorily optimised, a demanding process usually requiring around 100-1000 simulations. This means that each simulation must be as fast as possible in order to reduce the overall time-to-solution for the user.
The current practical limit for topology optimisation is the treatment of static problems, where time-stepping is avoided and computational cost is kept low.
For optimisation of practical problems, the computational time of time-stepping methods is impractical in an academic timescale and infeasible in an industrial timescale, because even simple time-dependent problems can take days to optimise.

Parallel computing can reduce the computational time of each time step, splitting the work over multiple computers simultaneously performing computations. 
However, Amdahl's law of strong scaling states that efficiency is limited by how much of the algorithm is parallelised \citep{Amdahl1967}.
Since the time-stepping approach is inherently serial, it will limit the attainable performance to that gained from parallelising each time step.
To overcome these limitations, we will consider time as an additional dimension, taking a space-time view of the physical problem, and solve the full time-dependent system using an ``all-at-once'' approach.
This increases the size of the equation system significantly and raises memory requirements, but parallel computing can now be applied to the entire computational process.
Large distributed-memory supercomputers handle the memory requirements and by parallelising the solution process in both space and time simultaneously, Amdahl's law dictates that a higher efficiency can be achieved \citep{Amdahl1967}.

However, increasingly larger supercomputers will not solve the problem alone. Thus, a space-time multigrid method with semi-coarsening is introduced as a preconditioner for a Krylov solver.
This is implemented in a fully parallel computational framework and demonstrated on a distributed-memory supercomputer for problems up to 4.2 billion degrees-of-freedom. Scaling and comparison studies show up to $52\times$ speed-up relative to a reference time-stepping solver using a moderate increase in core-hour cost.

\subsection{Literature}

Due to the rather broad scope of the paper, we focus on showing the overall development over the years in the different relevant areas: topology optimisation using iterative solvers; topology optimisation for time-dependent problems; computational effort reduction; and parallel-in-time methods. But due to the broad scope, a lot of references are covered in total to give a good overview for the reader.

\subsubsection{Topology optimisation using iterative solvers}

For topology optimisation of steady problems, a lot of work has been published on accelerating the solution process using high-performance computing (HPC) and iterative linear solvers \citep{Mukherjee2021}. Some early works applied a range of different preconditioners, such as pointwise Jacobi \citep{Borrvall2001}, Finite Element Tearing and Interconnection (FETI) \citep{Evgrafov2008}, and  Factorized Sparse Approximate Inverse (FSAI) \citep{Aage2013}. \citet{Amir2014} showed the first application of geometric multigrid (GMG) preconditioning for topology optimisation in a serial programming setting. This was subsequently extended by \citet{Aage2015} to a parallel programming setting using the Message Passing Interface (MPI) and the Portable, Extensible Toolkit for Scientific Computation (PETSc) \citep{petsc-user-ref}. \citet{Aage2015} open-sourced their modular and extendable code, which forms the basis of the presented work. Subsequently, the use of GMG has been applied for super-high resolution \citep{Aage2017,Baandrup2020}, other physics \citep{Alexandersen2016,Hoeghoej2020,Rogie2022,Zhou2025}, adaptive meshing \citep{Herrero-Perez2023}, and isogeometric discretisations \citep{Luo2024}. Algebraic multigrid (AMG) has also been applied for topology optimisation \citep{Mukherjee2021}, but is outside the scope of the present work.

\subsubsection{Topology optimisation for time-dependent problems}

The earliest example of application to time-dependent problems of what can be considered ``topology optimisation'', is arguably the application to linear dynamic loading by \citet{Min1999}. Subsequently, \citet{Li2001} applied the Evolutionary Structural Optimization (ESO) method to time-dependent thermal problems coupled to a mechanical stress model.
For time-dependent thermal problems: Zhuang and co-workers \citep{Zhuang2013,Zhuang2015} treated pure conduction problems; \citet{Wu2019} showed clearly that transient effects have importance for conductive designs; and \citet{Zeng2020} showed that transient effects have importance for conjugate heat transfer designs. For thermal energy storage and phase change materials: \citet{Srinivas2006} treated heat sinks with phase change materials; \citet{Pizzolato2017} treated heat transfer enhancement in latent heat thermal energy storage; \citet{Lundgaard2019} treated porous thermal energy storage; \citet{Yao2021} combined forced convection cooling with latent heat storage; and \citet{Christensen2023} used phase-change materials to reduce temperature fluctuations. For thermomechanical problems: \citet{Li2004} treated thermomechanical actuators; and \citet{Guo2024} demonstrated coupled heat transfer and structural dynamics.

Topology optimisation uses the adjoint sensitivity method, where the extension to time-dependent problems is well documented \citep{Michaleris1994,Dahl2008}. However, the computational cost
of traditional time-stepping procedures has severely limited applications to practical problems. Not only is the solution generated step-by-step through time, but the full forward solution must be saved and used during an also time-dependent backward problems to generate the adjoint solution. Thus, the computational time of topology optimisation of time-dependent systems can easily increase to multiple days. Despite this being an inherent problem, surprisingly few references actually discuss the computational time when treating time-dependent problems. 
A few examples of computational time have been identified: \citet{Christensen2023} lists their time-to-solution as up to 40 hours for two-dimensional phase change heat transfer;
\citet{Elesin2014} list their time-to-solution as ``several days'' for three-dimensional photonics; 
\citet{Coffin2016} list their time-to-solution as 12 days for two-dimensional natural convection \citep{Coffin2016}; \citet{Kristiansen2022} mentions 18 hours using model reduction methods, but estimates 22 days using a full order model for linear dynamics; \citet{Makhija2020} list their time-to-solution as 42 days for two-dimensional laminar flow.
All of the above treat rather simple cases and, thus, it is obvious that something needs to be done to make topology optimisation of time-dependent problems feasible in reasonable time.

\subsubsection{Reducing computational effort for time-dependent problems}

Due to the large computational effort of treating time-dependent problems, several methods have been proposed to reduce it. 
The adjoint method requires to store the full forward time history, which can require a lot of memory. To reduce memory requirements, checkpointing \citep{Griewank1992,Griewank2000} is common for optimisation of time-dependent problems. Although it reduces the memory consumption since the full forward solution is no longer stored, the computational time further increases accordingly, since the forward solution now has to be regenerated during the backward adjoint solve. \citet{Margetis2023} suggested to use data compression and reduced basis methods to speed up the re-computation, although still being slower than full storage methods.
Another method to reduce the memory consumption is the local-in-time adjoint developed by \citet{Yamaleev2010}, which \citet{Chen2017} and \citet{Yaji2018} applied to topology optimisation of fluid and conjugate heat transfer, respectively. Recently,
\citet{Theulings2024} combined checkpointing with the local-in-time adjoint method for thermal and fluid topology optimisation.

To reduce the computational time, mostly model reduction methods and reduced basis methods have been applied. Hyun et al. \citep{Hyun2019,Hyun2021} proposed to use an eigenvalue problem to capture the transient thermal behaviour and \citet{Onodera2025} coupled the approach to dynamics. \citet{Isiklar2024} used a harmonic assumption instead to reduce the computational cost. \citet{Yan2024} used the Lyapunov equation and Proper Orthogonal Decomposition for thermal problems.
Modal reduction was used by \citet{Hooijkamp2018} and \citet{Kolk2018} to reduce the problem to a finite low number of independent ordinary differential equations. Reduced basis methods and Krylov reduction methods have been applied for dynamics by \citet{Li2024,Li2025} and \citet{Kristiansen2022}, respectively. Lastly, \citet{Theulings2024} recently proposed the parallel local-in-time adjoint for thermal and fluid problems, demonstrating a decent speed-up at the cost of potential errors.

\subsubsection{Parallel-in-time methods}

Although seemingly counter-intuitive, it is possible to parallelise in time, just as in space.
In fact, there is a long history of parallel-in-time methods as covered by the review papers by \citet{Gander2015} and \citet{Ong2020}. Some examples include Parareal \citep{Lions2001}, ``multigrid reduction in time'' (MGRIT) \citep{Friedhoff2012}, the parallel time-stepping method \citep{Womble1990}, and the ``Parallel Full Approximation Scheme in Space and Time'' (PFASST) \citep{Emmett2012}. A few of these methods have been applied to optimisation problems, like size optimisation \citep{Hahne2023} and optimal control problems \citep{Du2013, Minion2018, Gunther2019}.

However, to our knowledge, only two parallel-in-time methods have been applied to topology optimisation in the literature. As previously mentioned, \citet{Theulings2024} introduced the parallel local-in-time method, which although speeding up the solution process has the severe disadvantage that it becomes unreliable when using too many processors. Recently, we introduced a one-shot Parareal method for topology optimisation of time-dependent thermal conduction problems \citep{Appel2024}, which has significantly better accuracy but is limited by relatively low scalability.

To address these problems, this work applies Space-Time MultiGrid (STMG) methods \citep{Hackbusch1983} as a preconditioner for an iterative solver, because these are among the fastest parallel-in-time methods \citep{Falgout2017}. STMG works similarly to spatial multigrid methods, in that a hierarchy of coarser grids is defined, after which smoothers and coarse grid corrections are applied to solve the fine grid problem.
The important difference is that STMG methods are applied to space-time domains, with time treated as being another dimension of the domain, as opposed to purely spatial domains. 
Since time is qualitatively different from the dimensions of space, space-time problems are inherently anisotropic. Thus, to ensure high convergence rates, semi-coarsening in the direction of strongest coupling is an option \citep{Horton1995}. Recently, we introduced a semi-coarsening strategy for STMG of high-contrast time-dependent thermal conduction problems \citep{Appel2025} in one spatial dimension plus time, denoted (1+1)D. This semi-coarsening strategy will herein be applied in the GMG preconditioner for large-scale topology optimisation problems of time-dependent thermal conduction in (2+1)D space-time.

\subsection{Contributions}

This work attempts to reduce the time-to-solution for topology optimisation of time-dependent thermal conduction problems by at least one order of magnitude. The paper builds on the initial work previously presented at conferences by \citet{Alexandersen2023} and \citet{Alexandersen2025}.
We use a space-time formulation of the governing equation, considering time simply as an additional dimension akin to the spatial ones, and apply a stabilised continuous Galerkin (CG) space-time finite element (STFE) method to form an all-at-once system. This system is then solved using an iterative Krylov solver with a space-time multigrid (STMG) preconditioner based on a semi-coarsening strategy. This is implemented in a fully parallel computing framework, which allows for very good scalability and solutions of systems of equations up to 4.2 billion degrees-of-freedom (DOFs) on supercomputers in short time.

The paper presents the proof-of-concept and initial results based on a relatively simple implementation with simple smoothers and coarse solver. We study the performance of the method for two different problems and various settings, present scaling results and projected speed-ups, as well as time-dependent design fields. However, there are several ingredients and parameters of the method, which are kept constant and not fully explored, which will be the subject of future research.

\subsection{Paper layout}

The paper is laid out as follows: Section \ref{sec:physics} introduces the governing equations, design representation and example problems; Section \ref{sec:method} presents the space-time finite element method, space-time multigrid preconditioner, topology optimisation formulation, and implementation; Section \ref{sec:result_comparison} compares the presented space-time approach with traditional time-stepping through scaling studies and relative comparisons; Section \ref{sec:result_topopt} presents topology optimisation results for the two examples using a range of conditions and explores the performance of the space-time approach; Section \ref{sec:discussion} discusses the results, limitations, future work, and provides conclusions.

\section{Physical problem} \label{sec:physics}

This section describes the governing equations, the design representation, and the two example problems used throughout the paper.

\subsection{Governing equations}

In this work, we consider the transient thermal diffusion equation as defined below:
\begin{equation}
    \mathcal{C} \dfrac{\partial T}{\partial t} - \dfrac{\partial}{\partial x_i} \left( k \dfrac{\partial T}{\partial x_i}\right) = Q \textrm{ for } \x \in \Omega \textrm{ and } t \in \mathcal{T}
\end{equation}
where $T$ is the temperature, $t$ is time, $x_{i}$ is the \textit{i}'th spatial direction, $\mathcal{C} = \rho c_{p}$ is the volumetric heat capacity (the product of the mass density $\rho$ and the specific heat capacity $c_{p}$), $k$ is the thermal conductivity coefficient, $Q$ is the volumetric heat generation, $\Omega$ is the two-dimensional spatial domain of interest, and $\mathcal{T}=\left[ 0;\tau \right]$ is the time frame of interest.

In order to provide a consistent scaling of the final discrete equations for various problems, we rescale the equations by non-dimensionalising the following quantities:
\begin{subequations}
    \begin{align}
        T &=  \Delta T \cdot T^* \\
        t &=  \tau \cdot t^* \\
        x_i &=  L \cdot {x_i}^{*}
    \end{align}
\end{subequations}
where $\Delta T$ is a characteristic temperature difference, $\tau$ is the final time under consideration, and $L$ is a characteristic spatial dimension. Throughout the presented work, the characteristic temperature difference is set to $\Delta T = 1$ without loss of generality.
This process yields the rescaled governing equations:
\begin{equation}
    \mathcal{C} \dfrac{\partial T^{*}}{\partial t^{*}} - \dfrac{\partial}{\partial {x_i}^{*}} \left( \Tilde{k} \dfrac{\partial T^{*}}{\partial {x_i}^{*}}\right) = \Tilde{Q}
\end{equation}
where:
\begin{equation}
    \Tilde{k} = \frac{k \tau}{L^{2}}
\end{equation}
and:
\begin{equation}
    \Tilde{Q} = \frac{Q \tau}{\Delta T}
\end{equation}
Please note that the modified coefficients above are not dimensionless, but are merely a rescaling of the problem to make temperature, space, and time non-dimensional. From hereon, the asterisks $\Box^{*}$ denoting dimensionless quantities are left out for brevity. 

Finally, the governing equation is recast in a space-time continuum, where time is seen as the third spatial direction:
\begin{equation} \label{eq:generalisedCoordinate}
    \xi_i = \left\lbrace \begin{matrix}
        x_1 & \textrm{ if } & i = 1 \\
        x_2 & \textrm{ if } & i = 2 \\
        t & \textrm{ if } & i = 3
    \end{matrix} \right.
\end{equation}
which yields:
\begin{equation} \label{eq:govequ}
    \Tilde{\mathcal{C}}_i \dfrac{\partial T}{\partial {\xi_i}} - \dfrac{\partial}{\partial {\xi_i}} \left( \Tilde{k}_{ij} \dfrac{\partial T}{\partial {\xi_j}}\right) = \Tilde{Q} \textrm{ for } \xiv \in \Pi
\end{equation}
where $\Pi = \Omega \times \mathcal{T}$ is the (2+1)D space-time domain of interest and the coefficients are given by:
\begin{equation}
    \Tilde{\mathcal{C}}_i = \left\lbrace \begin{matrix}
        0 & \textrm{ if } & i = 1 \\
        0 & \textrm{ if } & i = 2 \\
        \mathcal{C} & \textrm{ if } & i = 3
    \end{matrix} \right.
\end{equation}
and:
\begin{equation} \label{eq:originalConductivity}
    \Tilde{k}_{ij} = \left\lbrace \begin{matrix}
        \Tilde{k} & \textrm{ if } & i = j = 1 \\
        \Tilde{k} & \textrm{ if } & i = j = 2 \\
        0 & \multicolumn{2}{c}{\textrm{otherwise}}
    \end{matrix} \right.
\end{equation}
The discretisation of the above governing equation in space-time is detailed in Section \ref{sec:method_cgfem}.

\subsection{Design representation} \label{sec:method_desrep}

\begin{figure}
    \centering
    \includegraphics[width=0.45\linewidth]{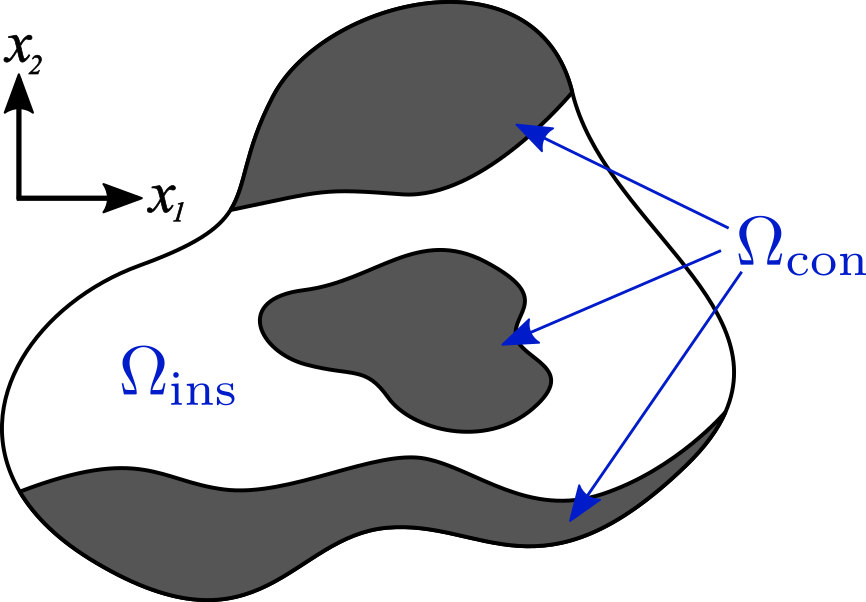}
    \caption{Two-dimensional design representation showing the areas of conductive material, $\Omega_\text{con}$, and insulative material, $\Omega_\text{ins}$.}
    \label{fig:designRep_2D}
\end{figure}
Figure \ref{fig:designRep_2D} shows an arbitrary two-dimensional domain, $\Omega = \Omega_\textrm{ins} \cup \Omega_\textrm{con}$, consisting of areas of conductive material, $\Omega_\textrm{con}$, and insulating material, $\Omega_\textrm{ins}$. We parametrise the design through a spatially-varying field called the design field, basically the characteristic function of the conductive domain:
\begin{equation}
    \gamma\fofx = 
        \begin{cases} 
            1 & \textrm{if } \vecr{x} \in \Omega_\textrm{con} \\
            0 & \textrm{if } \vecr{x} \in \Omega_\textrm{ins}
   \end{cases}
\end{equation}
To allow for density-based topology optimisation \citep{Sigmund2013}, the design field is then allowed to vary continuously between 0 and 1:
\begin{equation}
    0 \leq \gamma\fofx \leq 1   \textrm{   for } \x\in \Omega
\end{equation}
To couple the design representation to the governing equation, Equation \ref{eq:govequ}, the volumetric heat capacity and thermal conductivity are then coupled to the design field through interpolation functions:
\begin{subequations} \label{eq:interpolation_space}
    \begin{align}
        \mathcal{C}(\x) &= \mathcal{C}_\mathrm{SIMP}(\gamma(\x)) \\
        \Tilde{k}(\x) &= \Tilde{k}_\mathrm{SIMP}(\gamma(\x))
    \end{align}
\end{subequations}
where the functions $\Tilde{k}_\mathrm{SIMP}$ and $\mathcal{C}_\mathrm{SIMP}$ are defined using the modified SIMP approach \citep{Bendsoee2004}:
\begin{subequations} 
    \begin{align}
        \label{eq: SIMP c}
        \mathcal{C}_\mathrm{SIMP}(\gamma) = \mathcal{C}_\mathrm{ins} + (\mathcal{C}_\mathrm{con} - \mathcal{C}_\mathrm{ins}) \gamma^{p_c} \\
        \label{eq: SIMP k}
        \Tilde{k}_\mathrm{SIMP}(\gamma) = \Tilde{k}_\mathrm{ins} + (\Tilde{k}_\mathrm{con} - \Tilde{k}_\mathrm{ins}) \gamma^{p_k}
    \end{align}
\end{subequations}
with $p_c$ and $p_k$ being the interpolation parameters for the capacity and conductivity, respectively. 

To ensure direct comparison, the default values in this work are as defined by \citet{Appel2024}: $\mathcal{C}_\mathrm{ins} = 0.5$, $\mathcal{C}_\mathrm{con} = 1.0$, $k_\mathrm{ins} = 0.03$, $k_\mathrm{con} = 3.0$, $p_c = 2$, and $p_k=3$. These values are used throughout unless otherwise stated.

\subsubsection{Space-time design representation}

\begin{figure}
    \centering
    \subfloat[Time-constant design]{
        \includegraphics[width=0.45\linewidth]{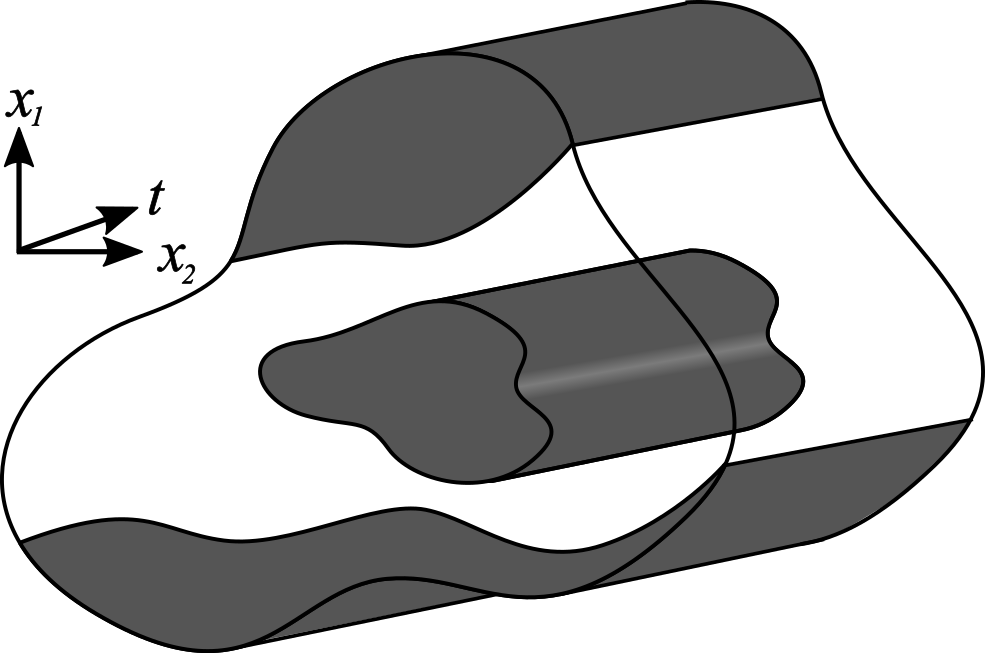}
        \label{fig:designRep_2p1D}
    }
    \hfill
    \subfloat[Space-time design]{
        \includegraphics[width=0.45\linewidth]{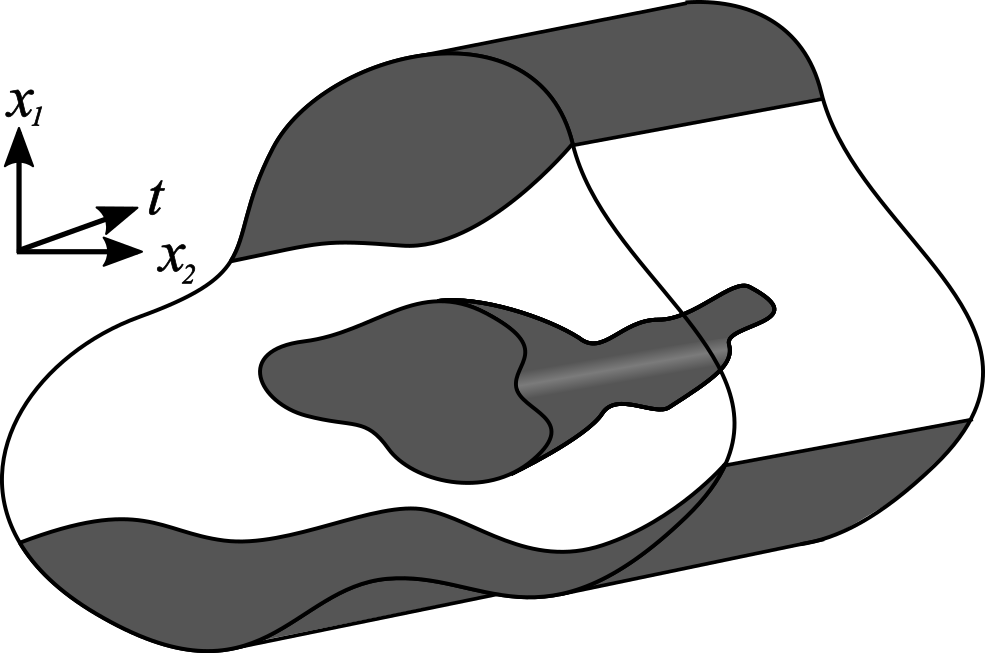}
        \label{fig:designRep_2p1D_st}
    }
    \caption{(2+1)D space-time design representation showing the areas of conductive material, $\Pi_\text{con}$, and insulative material, $\Pi_\text{ins}$.}
    \label{fig:designRep}
\end{figure}
Figure \ref{fig:designRep} shows two possible (2+1)D space-time design representations. For most physical problems, where the design is usually constant over time, the arbitrary design from Figure \ref{fig:designRep_2D} is simply extruded in the time direction as shown in Figure \ref{fig:designRep_2p1D}. This yields the space-time domain $\Pi = \Pi_\textrm{ins} \cup \Pi_\textrm{con}$ consisting of domains of conductive material, $\Pi_\textrm{con} = \Omega_\textrm{con} \times \mathcal{T}$, and insulating material, $\Pi_\textrm{ins} = \Omega_\textrm{ins} \times \mathcal{T}$. 

However, the space-time formulation in Equation \ref{eq:govequ} easily allows for time-dependent variation in the design, as illustrated in Figure \ref{fig:designRep_2p1D_st}. Here the design field becomes a function of the generalised coordinate defined in Equation \ref{eq:generalisedCoordinate}:
\begin{equation}
    0 \leq \gamma\fof{\xiv} \leq 1   \textrm{   for } \xiv \in \Pi
\end{equation}
Thus, the volumetric heat capacity and thermal conductivity coefficients also vary in space-time:
\begin{subequations} \label{eq:interpolation_spacetime}
    \begin{align}
        \mathcal{C}(\xiv) &= \mathcal{C}_\mathrm{SIMP}(\gamma(\xiv)) \\
        \Tilde{k}(\xiv) &= \Tilde{k}_\mathrm{SIMP}(\gamma(\xiv))
    \end{align}
\end{subequations}
This idea was first explored by \citet{Jensen2009,Jensen2010} for one-dimensional wave propagation. Sadly, this has limited applications in practise, unless the time-variation is strictly constrained such as for manufacturing process \citep{Wang2020}. Nonetheless, examples of morphing structures will be demonstrated in Sections \ref{sec:results_exp2_timevar} and \ref{sec:results_exp2_highres}.

\subsection{Example problems} \label{sec:exampleDefs}

Two example problems are introduced and used throughout the paper, referred to as Example 1 and Example 2 from hereon out. The following subsections detail their boundary conditions and the applied heat sources.

\subsubsection{Example 1: Oscillating heat source} \label{sec:example1def}

\begin{figure}
    \centering
    \includegraphics[height=0.5\linewidth]{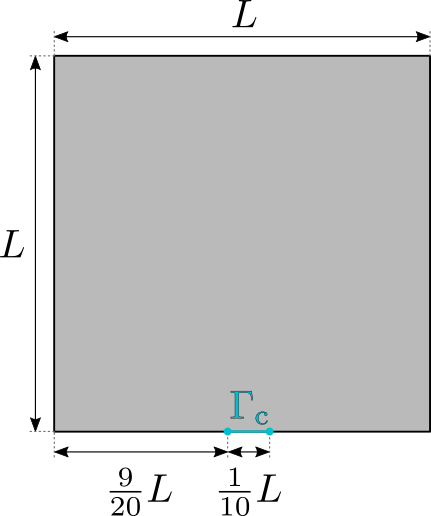}
    \caption{Sketch of Example 1: square domain with distributed time-dependent heat source and a small heat sink area at the centre of the bottom boundary.}
    \label{fig:sketch_oscHeat}
\end{figure}
Figure \ref{fig:sketch_oscHeat} shows the schematic layout of the first example with a distributed time-dependent heat source, the same as treated by \citet{Appel2024}. All outer boundaries have homogenous Neumann conditions (thermally insulated), $\dpart{T}{x_{i}}n_{i} = 0$, except a small region of width $\frac{1}{10}L$ in the centre of the lower boundary, $\Gamma_\textrm{c}$, where a homogenous Dirichlet condition (heat sink) is imposed, $T=0$. The heat source is equally-distributed across the spatial domain with dimension $L=1$, but varies in time according to:
\begin{equation}
    \Tilde{Q}(x_{1},x_{2},t) = \frac{1}{2}\Tilde{Q}_{0}\bof{1-t}\bof{1+\cos\fof{50t}}
\end{equation}
which yields a heat source that oscillates with an angular frequency of 50 and falls from a value of $\Tilde{Q}_{0} = 100$ over a span of 1 unit of time. Thus, the final time is set to $\tau = 1$.

\subsubsection{Example 2: Moving heat source} \label{sec:example2def}

\begin{figure}
    \centering
    \includegraphics[height=0.5\linewidth]{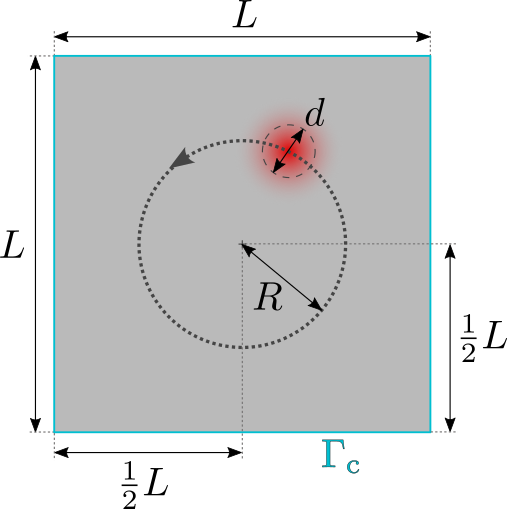}
    \caption{Sketch of Example 2: square domain with a moving heat source and cold walls around the entire outer boundary.}
    \label{fig:sketch_movingHeat}
\end{figure}
Figure \ref{fig:sketch_movingHeat} shows the schematic layout of the second example with a moving heat source. A homogenous Dirichlet condition is imposed on all outer boundaries, $T=0$ on $\Gamma_\textrm{c}$. The heat source is a Gaussian function with a peak value of $\Tilde{Q}_{0}=1$, the centre of which moves in a circle around the centre of the domain. The size of the domain is set to $L=1$ and the radius of the movement is set to $R=0.25$. The approximate diameter of the Gaussian heat source is $d\approx0.3$, which results from the definition:
\begin{equation}
    \tilde{Q}(x_{1},x_{2},t) = \tilde{Q}_{0} \cdot \exp \fof{ -\frac{\bof{x_{1}-x_{1}^{0}(t)}^{2} + \bof{x_{2}-x_{2}^{0}(t)}^{2}}{2\sigma^{2}} }
\end{equation}
with the width parameter set to $\sigma=0.05$. The centre position of the heat source are defined by the $x_{1}$- and $x_{2}$-coordinates, respectively:
\begin{subequations}
    \begin{align}
        x_{1}^{0}(t) &= \frac{L}{2} + R \cos\fof{\omega t + \frac{\pi}{2}} \\
        x_{2}^{0}(t) &= \frac{L}{2} + R \sin\fof{\omega t + \frac{\pi}{2}}
    \end{align}
\end{subequations}
where $\omega$ is the angular frequency of the rotation. The angular frequency is herein set to $\omega = \pi$ meaning that one rotation will occur per 2 units of time. The problem is solved for 3 rotations, that is $\tau = 6$. The initial location of the source at $t=0$, is $x_{1}^{0}(0) = \frac{1}{2}L$ and $x_{2}^{0}(0) = \frac{3}{4}L$.

\section{Methodology} \label{sec:method}

This section introduces the methodological and implementation details for the space-time finite element method, the space-time multigrid preconditioner, topology optimisation formulation, as well as the computational resources used.

\subsection{Continuous-Galerkin (CG) Space-Time Finite Element (STFE) Method} \label{sec:method_cgfem}

A thorough summary of work applying Galerkin finite element methods to space-time domains is beyond the scope of the present article. Early works include the work by \citet{Hulme1972}, \citet{Winther1981}, \citet{Aziz1989} and \citet{Hulbert1990}, with the interested reader being referred to the review by \citet{Steinbach2019}. Notable recent publications deserving to be highlighted is the work by: \citet{Langer2019} presents mathematical analysis of a stabilised STFE for diffusion with varying coefficients and results using algebraic multigrid; and \citet{Danwitz2023} presents a stabilised STFE using both continuous and discontinuous time spaces for advection-diffusion problems applied to (3+1)D problems with time-dependent geometry.

\subsubsection{Continuous weak form}

To discretise the governing equation (Equation \ref{eq:govequ}), we apply the Galerkin method using continuous test and trial functions. Equation \ref{eq:govequ} is multiplied by a test function $v$ and integration-by-parts is applied on the conductivity term to yield the weak form:
\begin{equation}
    \int_\Pi v\,\Tilde{\mathcal{C}}_i \dfrac{\partial T}{\partial {s_i}}\,dV + \int_\Pi \dfrac{\partial v}{\partial {s_i}} \Tilde{k}_{ij} \dfrac{\partial T}{\partial {s_j}}\,dV - \int_{\delta\Pi} v\,\Tilde{k}_{ij} \dfrac{\partial T}{\partial {s_j}} n_i \,dS = \int_\Pi v\, \Tilde{Q} \,dV
\end{equation}
Due to the time-derivative term being purely convective in the time-direction (having no conductivity defined in Equation \ref{eq:originalConductivity}), the standard continuous Bubnov-Galerkin method is unstable if large gradients occur. To stabilise the formulation, artificial diffusion is added in the time-direction by modifying the conductivity tensor to:
\begin{equation}
    \Tilde{k}_{ij} = \left\lbrace \begin{matrix}
        \Tilde{k} & \textrm{ if } & i = j = 1 \\
        \Tilde{k} & \textrm{ if } & i = j = 2 \\
        \hat{k}_{ad} & \textrm{ if } & i = j = 3 \\
        0 & \multicolumn{2}{c}{\textrm{otherwise}}
    \end{matrix} \right.
\end{equation}
where:
\begin{equation}
    \hat{k}_{ad} = \frac{1}{2} \Tilde{\mathcal{C}}\Delta t
\end{equation}
This artificial diffusion ensures that the element ``P\'eclet number'' for the time-direction is unity according to the definition by \citet{Brooks1982}.

It is known that continuous Galerkin (CG) finite elements in space give rise to schemes similar to central-difference methods under certain conditions, such as regular meshes of piecewise linear elements \citep{Brooks1982,Strang1973,Hughes1987}. This is also the case when using CG finite elements in time, where using linear elements in time can yield schemes similar to the Crank-Nicholson method \citep{Aziz1989,Hansbo2000}.
This means that the method using trilinear basis functions gives second-order accuracy in both time and space. Unfortunately, when adding the artificial diffusion to ensure stability, this lowers the accuracy in time to only first-order, resembling more a backward-difference type scheme.

\subsubsection{Discretised system}

It is important to note that the chosen discretisation is non-causal, meaning that the connectivity is not only forward in time and that the system cannot be posed as a time-stepping method. Thus, the system is solved all-at-once, with all degrees-of-freedom (DOFs) in space and time put into one large system of equations. This vastly increases the number of DOFs, computational cost and memory requirements, but allows for full space-time parallelisation to reduce the computational time.

The computational domains are restricted to cuboidal space-time domains, which are discretised using a regular mesh of cuboidal elements with piecewise trilinear basis functions. We restrict ourselves to square-in-space elements, where $\Delta x = \Delta x_{1} = \Delta x_{2}$ is defined as the element spatial size. The element time size is defined as $\Delta t$, such that an element becomes $\Delta x \times \Delta x \times \Delta t$ in dimension. The space-time mesh is described as $n_{x_1} \times n_{x_2} \times n_{t}$, with a total number of elements $N_{e} = n_{x_1}  n_{x_2}  n_{t}$ where $n_{x_1}$ is the number of elements in the $x_{1}$-direction, $n_{x_2}$ is the number of elements in the $x_{2}$-direction, and $n_{t}$ is the number of elements in the time-direction. The total number of nodes is defined as $N_{n} = (n_{x_1}+1)( n_{x_2}+1)(n_{t}+1)$.

The full discretised space-time system can be assembled to be:
\begin{equation}
    \matx{J}\s=\vecr{f}
\end{equation}
where $\s$ is the state vector containing all nodal temperature DOFs:
\begin{equation}
   \s = \begin{Bmatrix} 
   T_0 \\ 
   T_1 \\
   T_2 \\ 
   \vdots \\ 
   T_{N_{n}}
   \end{Bmatrix}
\end{equation}
and $\matx{J}$ is the system matrix and $\vecr{f}$ is the source vector, assembled by element-wise integration of the left- and right-hand side of Equation \ref{eq:govequ}, respectively.

\subsection{Space-time multigrid preconditioner} \label{sec:method_mg}

To solve the large system of equations resulting from the CG-STFE discretisation, the Flexible Generalised Minimal Residual (FGMRES) method \citep{Saad1993} is used, preconditioned using a space-time multigrid (STMG) method \citep{Hackbusch1983}. The Jacobi-preconditioned Generalised Minimal Residual (GMRES) method \citep{Saad1986} is used both as a solver for the coarsest grid and as a smoother for the remaining grids. We use a classical V-cycle on a hierarchy of grids defined by a semi-coarsening strategy defined in the next subsection. 
The solver parameters are kept constant throughout: a relative tolerance of the outer F-GMRES solver of $10^{-5}$; a relative tolerance of the GMRES smoothers and coarse solver of $10^{-6}$; at most 10 smoothing iterations; and at most 200 coarse solver iterations. We employ a warm start of the F-GMRES solver by using the solution from the previous design iteration as the initial vector for the new solution.

To define the system matrices on the grids of the hierarchy, Galerkin projection is used, where the system matrices are computed using the following:
\begin{equation}
    \matx{J}_{l+1} = {\matx{P}_{l}}^{\intercal} \matx{J}_{l} {\matx{P}_{l}}
\end{equation}
where $\matx{J}_{l}$ is the system matrix on the $l$\textsuperscript{th} grid and $\matx{P}_{l}$ is the prolongation operator which interpolates from the $(l+1)$\textsuperscript{th} grid to the $l$\textsuperscript{th} grid. The interpolation method used for the prolongation operators is trilinear space-time interpolation, meaning it is non-causal and transfers information both forwards and backwards in time. The restriction operator for going from the $l$\textsuperscript{th} grid to the $(l+1)$\textsuperscript{th} grid is simply defined as the transposed prolongation operator, ${\matx{P}_{l}}^{\intercal}$.

\subsubsection{Semi-coarsening strategy}

When using pointwise smoothers combined with full space-time coarsening, STMG methods often exhibit poor convergence rates \citep{Horton1995}. This can be avoided by using other types of smoothers \citep{Gander2016_stmg_br, franco2018_stmg_with_CN} or by using semi-coarsening strategies \citep{Horton1995}.
Despite Jacobi-preconditioned GMRES as a smoother not being truly pointwise, it acts like a pointwise smoother when using only a few smoothing iterations. Thus, we opt to use a semi-coarsening strategy. The chosen semi-coarsening strategy uses two different kinds of coarsening:
\begin{itemize}
    \item Semi-coarsening in space, where $\Delta x$ is doubled on the next coarse level while $\Delta t$ is unchanged. 
    \item Semi-coarsening in time, where $\Delta t$ is doubled on the next coarse level while $\Delta x$ is unchanged. 
\end{itemize}
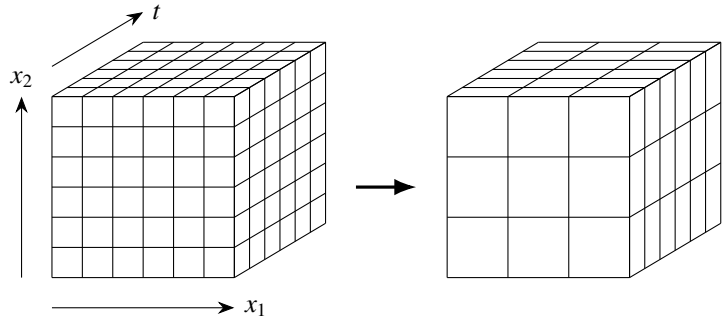
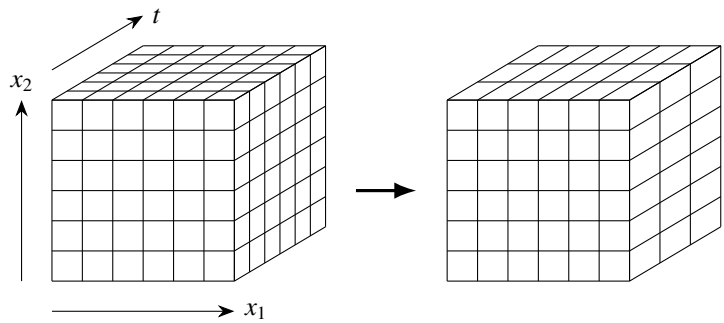
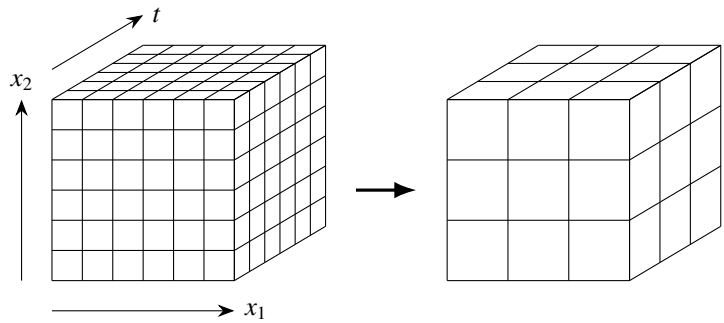
\begin{figure} 
\def\dx{0.5}
\def\dy{0.3}
\centering
\subfloat[Semi-coarsening in space (spatial coarsening).]{
\begin{tikzpicture}[scale=0.4]
\foreach \i in {0,...,6}  \draw[thin, solid, black] (0, \i) -- (6, \i); 
\foreach \i in {0,...,6}  \draw[thin, solid, black] (\i, 0) -- (\i, 6); 
\foreach \i in {0,...,6}  \draw[thin, solid, black] (\dx*\i, 6+\dy*\i) -- (6+\dx*\i, 6+\dy*\i); 
\foreach \i in {0,...,6}  \draw[thin, solid, black] (\i, 6) -- (\i+6*\dx, 6+6*\dy); 
\foreach \i in {0,...,6}  \draw[thin, solid, black] (6+\dx*\i, \dy*\i) -- (6+\dx*\i, \dy*\i+6); 
\foreach \i in {0,...,6}  \draw[thin, solid, black] (6, \i) -- (6+6*\dx, \i+6*\dy); 
\draw[black, thin, {}-{Stealth[scale=1.3]} ] (0,-1)  -- (6,-1) node[anchor=west] {$x_{1}$}; 
\draw[black, thin, {}-{Stealth[scale=1.3]} ] (-1,0)  -- (-1,6) node[anchor=south] {$x_{2}$}; 
\draw[black, thin, {}-{Stealth[scale=1.3]} ] (0,7)  -- (0+6*\dx,7+6*\dy) node[anchor=west] {$t$}; 
\draw[black, very thick, {}-{Latex[scale=1]} ] (10,3)  -- (12,3); 
\foreach \i in {0,...,3}  \draw[thin, solid, black] (13+0, 2*\i) -- (13+6, 2*\i); 
\foreach \i in {0,...,3}  \draw[thin, solid, black] (13+2*\i, 0) -- (13+2*\i, 6); 
\foreach \i in {0,...,6}  \draw[thin, solid, black] (13+\dx*\i, 6+\dy*\i) -- (13+6+\dx*\i, 6+\dy*\i); 
\foreach \i in {0,...,3}  \draw[thin, solid, black] (13+2*\i, 6) -- (13+2*\i+6*\dx, 6+6*\dy); 
\foreach \i in {0,...,6}  \draw[thin, solid, black] (13+6+\dx*\i, \dy*\i) -- (13+6+\dx*\i, \dy*\i+6); 
\foreach \i in {0,...,3}  \draw[thin, solid, black] (13+6, 2*\i) -- (13+6+6*\dx, 2*\i+6*\dy); 
\end{tikzpicture}
} \\
\subfloat[Semi-coarsening in time (time coarsening).]{
\begin{tikzpicture}[scale=0.4]
\foreach \i in {0,...,6}  \draw[thin, solid, black] (0, \i) -- (6, \i); 
\foreach \i in {0,...,6}  \draw[thin, solid, black] (\i, 0) -- (\i, 6); 
\foreach \i in {0,...,6}  \draw[thin, solid, black] (\dx*\i, 6+\dy*\i) -- (6+\dx*\i, 6+\dy*\i); 
\foreach \i in {0,...,6}  \draw[thin, solid, black] (\i, 6) -- (\i+6*\dx, 6+6*\dy); 
\foreach \i in {0,...,6}  \draw[thin, solid, black] (6+\dx*\i, \dy*\i) -- (6+\dx*\i, \dy*\i+6); 
\foreach \i in {0,...,6}  \draw[thin, solid, black] (6, \i) -- (6+6*\dx, \i+6*\dy); 
\draw[black, thin, {}-{Stealth[scale=1.3]} ] (0,-1)  -- (6,-1) node[anchor=west] {$x_{1}$}; 
\draw[black, thin, {}-{Stealth[scale=1.3]} ] (-1,0)  -- (-1,6) node[anchor=south] {$x_{2}$}; 
\draw[black, thin, {}-{Stealth[scale=1.3]} ] (0,7)  -- (0+6*\dx,7+6*\dy) node[anchor=west] {$t$}; 
\draw[black, very thick, {}-{Latex[scale=1]} ] (10,3)  -- (12,3); 
\foreach \i in {0,...,6}  \draw[thin, solid, black] (13+0, \i) -- (13+6, \i); 
\foreach \i in {0,...,6}  \draw[thin, solid, black] (13+\i, 0) -- (13+\i, 6); 
\foreach \i in {0,...,3}  \draw[thin, solid, black] (13+\dx*2*\i, 6+\dy*2*\i) -- (13+6+\dx*2*\i, 6+\dy*2*\i); 
\foreach \i in {0,...,6}  \draw[thin, solid, black] (13+\i, 6) -- (13+\i+6*\dx, 6+6*\dy); 
\foreach \i in {0,...,3}  \draw[thin, solid, black] (13+6+\dx*2*\i, \dy*2*\i) -- (13+6+\dx*2*\i, \dy*2*\i+6); 
\foreach \i in {0,...,6}  \draw[thin, solid, black] (13+6, \i) -- (13+6+6*\dx, \i+6*\dy); 
\end{tikzpicture}
} \\
\subfloat[Full space-time coarsening.]{
\begin{tikzpicture}[scale=0.4]
\foreach \i in {0,...,6}  \draw[thin, solid, black] (0, \i) -- (6, \i); 
\foreach \i in {0,...,6}  \draw[thin, solid, black] (\i, 0) -- (\i, 6); 
\foreach \i in {0,...,6}  \draw[thin, solid, black] (\dx*\i, 6+\dy*\i) -- (6+\dx*\i, 6+\dy*\i); 
\foreach \i in {0,...,6}  \draw[thin, solid, black] (\i, 6) -- (\i+6*\dx, 6+6*\dy); 
\foreach \i in {0,...,6}  \draw[thin, solid, black] (6+\dx*\i, \dy*\i) -- (6+\dx*\i, \dy*\i+6); 
\foreach \i in {0,...,6}  \draw[thin, solid, black] (6, \i) -- (6+6*\dx, \i+6*\dy); 
\draw[black, thin, {}-{Stealth[scale=1.3]} ] (0,-1)  -- (6,-1) node[anchor=west] {$x_{1}$}; 
\draw[black, thin, {}-{Stealth[scale=1.3]} ] (-1,0)  -- (-1,6) node[anchor=south] {$x_{2}$}; 
\draw[black, thin, {}-{Stealth[scale=1.3]} ] (0,7)  -- (0+6*\dx,7+6*\dy) node[anchor=west] {$t$}; 
\draw[black, very thick, {}-{Latex[scale=1]} ] (10,3)  -- (12,3); 
\foreach \i in {0,...,3}  \draw[thin, solid, black] (13+0, 2*\i) -- (13+6, 2*\i); 
\foreach \i in {0,...,3}  \draw[thin, solid, black] (13+2*\i, 0) -- (13+2*\i, 6); 
\foreach \i in {0,...,3}  \draw[thin, solid, black] (13+\dx*2*\i, 6+\dy*2*\i) -- (13+6+\dx*2*\i, 6+\dy*2*\i); 
\foreach \i in {0,...,3}  \draw[thin, solid, black] (13+2*\i, 6) -- (13+2*\i+6*\dx, 6+6*\dy); 
\foreach \i in {0,...,3}  \draw[thin, solid, black] (13+6+\dx*2*\i, \dy*2*\i) -- (13+6+\dx*2*\i, \dy*2*\i+6); 
\foreach \i in {0,...,3}  \draw[thin, solid, black] (13+6, 2*\i) -- (13+6+6*\dx, 2*\i+6*\dy); 
\end{tikzpicture}
}
\caption{Illustration of: (a-b) the two semi-coarsening operations available in the semi-coarsening strategy defined in Algorithm \ref{alg:coarstrat}; (c) suboptimal full space-time coarsening.}
\label{fig:semicoarsening}
\end{figure}
These coarsening types are illustrated in Figure \ref{fig:semicoarsening}, along with full space-time coarsening.
\begin{algorithm}
\caption{Coarsening strategy.}
\label{alg:coarstrat}
\begin{algorithmic}[1]
    \STATE Construct level 0 (the finest level)
    \STATE $D_\mathrm{eff} \gets \sqrt{ \dfrac{ \Tilde{k}_\mathrm{con} \Tilde{k}_\mathrm{ins} }{ \Tilde{\mathcal{C}}_\mathrm{con} \Tilde{\mathcal{C}}_\mathrm{ins} } } $
    \FOR {$l = 1, \hdots, N_l-1$}
        \STATE $\lambda_{\mathrm{eff},l-1} \gets D_\mathrm{eff} \dfrac{\Delta t_{l-1}}{  {\Delta x_{l-1} }^2  }$
        \IF{$\lambda_{\mathrm{eff},{l-1}} < \lambda_\mathrm{crit}$}
            \STATE Construct level $l$ by semi-coarsening level $l-1$ in time
        \ELSE
            \STATE Construct level $l$ by semi-coarsening level $l-1$ in space
        \ENDIF
    \ENDFOR
\end{algorithmic}
\end{algorithm}
The used coarsening strategy is given in Algorithm \ref{alg:coarstrat} and was proposed by \citet{Appel2025}. In this context: $N_l$ is the number of multigrid levels; $\Delta x_{l-1}$ denotes $\Delta x$ on the $(l-1)$\textsuperscript{th} level; $\Delta t_{l-1}$ denotes $\Delta t$ on the $(l-1)$\textsuperscript{th} level; $D_\mathrm{eff}$ is the effective diffusivity of the problem, $\lambda_{\mathrm{eff},{l-1}}$ is the effective anisotropy parameter on the $(l-1)$\textsuperscript{th} level, and $\lambda_\mathrm{crit}$ is a user-defined parameter.
Through numerical experiments, it was found that the formulae defining $D_\mathrm{eff}$ and $\lambda_{\mathrm{eff},{l-1}}$ resulted in reliable indicators for whether it is optimal to apply semi-coarsening in space or time \citep{Appel2025}. These experiments focused on a discretisation using finite elements in space and the backward Euler method in time. However, preliminary tests show that these formulae also work well for the STFE discretisation described in Section \ref{sec:method_cgfem} and, thus, they are applied here. 
In the present study, $\lambda_\mathrm{crit}$ is set to 0.5 since this has been observed to work decently in practise.

\subsection{Topology optimisation} \label{sec:method_topopt}

Topology optimisation is carried out using the density-based method \citep{Sigmund2013} based on the design representation described in Section \ref{sec:method_desrep}. This subsection describes the optimisation problem, filtering and projection, sensitivity analysis, and optimisation solver.

\subsubsection{Optimisation problem}

The optimisation problem is formulated in terms of the discretised design field, resulting in the design variable vector $\g$:
\begin{equation}
\begin{aligned}
    \min_{\g}\!.& \quad \phi(\s, \g) \\
    \text{s.t.}& \quad \chi(\g) \leq 0 \\
    & \quad 0 \leq \gamma_e \leq 1 \quad \text{for} \quad e=1,...,N_{e}
\end{aligned}
\label{eq:optproblem}
\end{equation}
where $\s$ is implicitly given by the solution of the discretised system of equations, given in residual form as:
\begin{equation} \label{eq:govequ_residual}
    \vecr{r} = \matx{J}\vecr{s} - \vecr{f} = \vecr{0}
\end{equation}
Thus, the optimisation problem is solved in a nested formulation, where for each design iteration the system of equations is solved for the current state vector $\s$.

The objective functional $\phi(\s, \g)$ is the p-norm of the element-wise averaged temperature:
\begin{equation}
    \phi(\s, \g) = \left( \sum_{e=1}^{N_e} {T_{\text{avg}, e}}^P \right)^{1/P}
\end{equation}
where $P$ is set to 20 to remain consistent with our previous work \citep{Appel2024}. The element-wise averaged temperature is defined as:
\begin{equation}
    T_{\text{avg},e} = \frac{1}{8} \sum_{n=1}^{8} T_{e,n}
\end{equation}
where $T_{e,n}$ is the temperature DOF of the $n$'th node of the $e$'th element. 
The volume constraint functional $\chi(\g)$ is defined as:
\begin{equation}
    \chi(\g) = \frac{1}{v_{f}\left| \Pi \right|}\sum_{e=1}^{N_e} \gamma_{e}v_{e} - 1 
\end{equation}
where $v_f\in\left] 0;1 \right[$ is the so-called volume fraction, denoting the fraction of the total space-time volume $\left| \Pi \right|$ that can be occupied by conductive material. Unless otherwise specified, the volume fraction is set to $v_f = 0.3$ throughout.

\subsubsection{Filtering and projection}

To regularise the optimisation problem, we apply the ``PDE filtering method'' by solving a reaction-diffusion equation \citep{Lazarov2011} posed in the full space-time domain:
\begin{equation}
   - \dfrac{\partial}{\partial \xi_i} \left( d_{ij} \dfrac{\partial \tilde{\gamma}}{\partial \xi_j}\right) + \tilde{\gamma} = \gamma \textrm{ for } \xiv \in \Pi
\end{equation}
where $\tilde{\gamma}$ is the filtered design field and $d_{ij}$ is the anisotropic diffusion tensor \citep{Lazarov2011,Wang2020b} defined in our case as:
\begin{equation}
    d_{ij} = \left\lbrace \begin{matrix}
        \tfrac{{r_x}^{2}}{12} & \textrm{ if } & i = j = 1 \\
        \tfrac{{r_x}^{2}}{12} & \textrm{ if } & i = j = 2 \\
        \tfrac{{r_t}^{2}}{12} & \textrm{ if } & i = j = 3 \\
        0 & \multicolumn{2}{c}{\textrm{otherwise}}
    \end{matrix} \right.
\end{equation}
with $r_{x}$ and $r_{t}$ being the filter radii in the spatial directions, $x_{1}$ and $x_{2}$, and time direction, $t$, respectively. Unless otherwise defined, these are set to the minimum value to ensure stability \citep{Lazarov2011}: $r_{x} = 2.4\Delta x$ and $r_{t} = 2.4\Delta t$.

Filtering introduces a lot of intermediate design field values, which are undesirable from a physical point of view. The goal is to end up with a design field with binary values, indicating only purely conductive or insulative material. Thus, we apply ``Heaviside projection'' \citep{Guest2004} using the projection function proposed by \citet{Wang2011}:
\begin{equation}
    \bar{\tilde{\gamma}}\fof{\xiv} = \frac{ \tanh\fof{\beta\eta} + \tanh\fof{\beta\bof{\tilde{\gamma}\fof{\xiv} - \eta }} }{ \tanh\fof{\beta\eta} + \tanh\fof{ \beta \bof{1-\eta}  } }
\end{equation}
where $\eta=0.5$ and $\beta=32$ are the chosen values for the threshold and projection sharpness, respectively.

The results from the projection, $\bar{\tilde{\gamma}}$, is termed the physical design field and is passed to the governing equations through the interpolation functions, Equations \ref{eq:interpolation_space} and \ref{eq:interpolation_spacetime}. The sensitivities are corrected according to the chain rule, effectively meaning that the sensitivities are also passed through another reaction-diffusion equation \citep{Lazarov2011}, one for each functional.

\subsubsection{Optimisation solver and sensitivity analysis}

The Method of Moving Asymptotes (MMA) \citep{Svanberg1987} is used to solve the optimisation problem (Equation \ref{eq:optproblem}). We use a modified version of the initial asymptotes as proposed by \citet{Guest2011} to allow for continuation-free optimisation using a constant high value of $\beta$. MMA is a gradient-based method, meaning that it need the gradients, or sensitivities, of the objective and constraint functionals, $\phi(\s,\g)$ and $\chi(\g)$, with respect to the design variables, $\g$. As mentioned above, the sensitivities are found with respect to the physical design variables, $\bar{\tilde{\g}}$, and then corrected according to the chain rule. Below the sensitivities with respect to the physical design variables are shown.

The volume constraint only depends explicitly on the design variables and thus the sensitivities are easily computed as:
\begin{equation}
    \dpart{\chi}{\bar{\tilde{\gamma_{e}}}} = \frac{v_{e}}{v_{f}\left| \Pi \right|}
\end{equation}
Since the objective functional is a function of the state vector, $\s$, which itself depends implicitly on the design variables, $\g$, we use adjoint sensitivity analysis for computing the gradients. The total derivative is given by:
\begin{equation}
    \frac{d\phi}{\bar{\tilde{\gamma_{e}}}} = \dpart{\phi}{\bar{\tilde{\gamma_{e}}}} - \L^\intercal \dpart{\vecr{r}}{\bar{\tilde{\gamma_{e}}}}
\end{equation}
where $\L$ is the collected vector of all adjoint DOFs, similar to $\s$:
\begin{equation}
   \L = \begin{Bmatrix} 
   \lambda_0 \\ 
   \lambda_1 \\
   \lambda_2 \\ 
   \vdots \\ 
   \lambda_{N_{n}}
   \end{Bmatrix}
\end{equation}
This vector is given as the solution to the adjoint problem:
\begin{equation}
    {\dpart{\vecr{r}}{\s}}^\intercal \L = {\dpart{\phi}{\s}}^\intercal
\end{equation}
For our particular problem and model, the sensitivities boil down to:
\begin{equation}
    \frac{d\phi}{d\bar{\tilde{\gamma_{e}}}} = - \L^\intercal \dpart{\vecr{r}}{\bar{\tilde{\gamma_{e}}}}
\end{equation}
where $\L$ is the solution to:
\begin{equation}
    {\matx{J}}^\intercal \L = {\dpart{\phi}{\s}}^\intercal
\end{equation}
with the right-hand side given by the finite element assembly of:
\begin{equation}
    \dpart{\phi}{\s_{e}} = \frac{1}{P} \theta^{\frac{1-P}{P}} \cdot P \cdot {T_{avg,e}}^{P-1} \left\lbrace \tfrac{1}{8}\,\,\tfrac{1}{8}\,\,\tfrac{1}{8}\,\,\tfrac{1}{8}\,\,\tfrac{1}{8}\,\,\tfrac{1}{8}\,\,\tfrac{1}{8}\,\,\tfrac{1}{8} \right\rbrace
\end{equation}
where $\theta = \sum_{e=1}^{N_e} {T_{\text{avg}, e}}^P$.

\subsection{Implementation} \label{sec:method_implementation}

The methodology is implemented in an in-house version of the ``TopOpt in PETSc'' code by \citet{Aage2015}. The code uses the `Portable, Extensible Toolkit for Scientific Computation' (PETSc) \citep{petsc-user-ref} to handle the mesh, multigrid hierachy, and linear solvers. Specifically, the code uses KSP for linear solvers, PCMG for multigrid, and DMDA for regular mesh handling and semi-coarsening hierarchy. To solve the optimisation problem, a fully parallel version of MMA is used as provided by \citet{Aage2015}. 

\subsubsection{Time-constant design} \label{sec:method_extrusion}

In order to force the space-time design field to represent a constant design through time, as illustrated in Figure \ref{fig:designRep_2p1D}, we couple the discretised design variables together using an extrusion operator applied through a matrix-vector product. It is possible to emulate extrusion implicitly by setting a very high filter radius in the time-direction, as has previously been done for structural mechanics in three-dimensions \citep{Wang2020b}. However, in a large-scale framework it is necessary to solve the filter equations using an iterative solver and the performance of this solver breaks down for very high anisotropy. Therefore, we choose to couple the design variables explicitly together.

First, the driving design variables are identified on the front face of the space-time domain by finding element numbers for the first time slab. Secondly, the element numbering for the equivalent elements are found for all subsequent time slabs. This is done through assuming a regular domain partitioning in the $(x_1,x_2)$-plane and the numbering scheme inherent to the regular meshes used. Thirdly, a sparse tall-and-skinny rectangular matrix is built that couples driving design variables (columns) to driven design variables (rows). The number of design variables passed to MMA is reduced to only the driving design variables in the first time slab and the sensitivities are found by summing up through time (using the transposed extrusion operator).

It should be noted that the current implementation sets certain requirements to the combination of the coarsest level discretisation and the partitioning, since it assumes regular domain partitioning. Due to the specifics of the DMDA refinement structure and partitioning, a regular domain partitioning cannot always be ensured - especially for very coarse discretisation of the coarsest level and for very high number of cores/partitions.

\subsection{Computational resources} \label{sec:compresources}

All presented results were run on the LUMI supercomputer, hosted by CSC (Finland) and owned by the EuroHPC Joint Undertaking and the LUMI consortium.
Specifically the LUMI-C partition is used consisting of 2048 CPU-based compute nodes. Each node is equipped with 256GiB memory and 128 physical cores total distributed on two AMD EPYC 7763 processors (2.4GHz base, 3.5GHz boost) \citep{LUMI_specs}. The queuing system limited each job to at most 512 nodes (65536 cores).

\section{Comparisons} \label{sec:result_comparison}

This section compares the parallel performance of a traditional time-stepping approach and the proposed space-time approach through strong scaling, computational time and computational cost studies. Example 1 is used and is discretised using a $640\times 640$ mesh for time-stepping and a $640\times 640\times 1280$ mesh for the space-time approach. A first-order backward difference scheme is used for the time-stepping approach and the optimisation problem and adjoint sensitivities are as described by \citet{Appel2024}. The material parameters are also as defined by \citet{Appel2024}.
For both methods, the output of data is turned off to make comparisons only about the computational cost and scalability of the methods themselves.

In order to keep the computational time tractable when using few cores and keep the resource requirements generally low (available computational resources on LUMI were limited by the project allocation), only 10 design iterations are performed for both methods and only 10 time steps are used for the time-stepping method. The computational time for time-stepping is then scaled up to 1280 time steps, based on the fact that the computational time is observed to scale more or less linearly with the number of time steps. However, this is actually an unfair advantage for the time-stepping approach from a memory perspective, since storing the 10 time steps of the forward and adjoint solutions in memory is much less intensive than storing 1280 time steps. In fact, the time-stepping approach for 1280 time steps reached memory limitations on a single node and actually requires substantially more nodes to solve the problem without memory issues. However, computational time is the main measure of interest, not memory usage, and both codes have in no way been memory-optimised. Lastly, only running 10 design iterations may underestimate the computational work needed to solve the space-time problem, since generally it is observed that the number of linear iterations increases towards the end of the optimisation process. However, this is also observed for the time-stepping approach but to a lesser extent. Therefore, the speed-up estimates should be taken with some uncertainty and it is unlikely that the exact numbers can be generalised to other problems.

The design extrusion operation is turned off for the space-time approach, since the current implementation requires a regular partitioning which the semi-coarsening scheme does not always yield for large numbers of cores. Turning it off will allow us to explore scaling up to $65,\!536$ cores without issues related to the extrusion operator.

The solver parameters are kept identical for both methods and are as specified in Section \ref{sec:method_mg}. The number of multigrid levels is $N_l = 6$ using full spatial coarsening for the time-stepping approach and $N_l = 8$ using semi-coarsening for the space-time approach, yielding the multigrid hierarchies listed in Table \ref{tab:scaling_multigrid_hierarchies}.
\begin{table}
    \centering
    \begin{tabular}{ccc}
        Level & Space-time & Time-stepping \\ \hline
        0 & $640\times 640\times 1280$ & $640\times 640$ \\
        1 & $320\times 320\times 1280$ & $320\times 320$ \\
        2 & $160\times 160\times 1280$ & $160\times 160$ \\
        3 & $80\times 80\times 1280$ & $80\times 80$ \\
        4 & $40\times 40\times 1280$ & $40\times 40$ \\
        5 & $20\times 20\times 1280$ & $20\times 20$ \\
        6 & $20\times 20\times 640$ & - \\
        7 & $20\times 20\times 320$ & -\\
    \end{tabular}
    \caption{Multigrid hierarchies for the (2+1)D space-time approach and the time-stepping approach for a two-dimensional spatial domain.}
    \label{tab:scaling_multigrid_hierarchies}
\end{table}
It can be seen that the coarsening strategy prefers semi-coarsening in space for the first five levels and then switches to semi-coarsening in time for the last two.

\subsection{Strong scaling} \label{sec:result_scaling}

\begin{figure}
    \centering
    \subfloat[]{\includegraphics[height=0.45\linewidth]{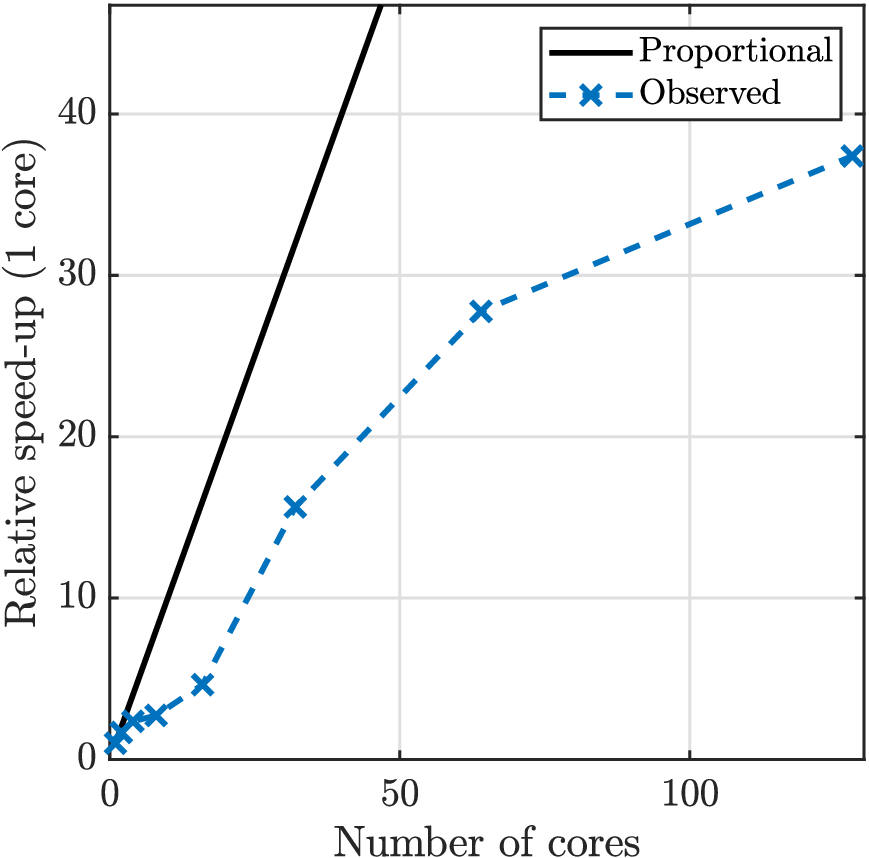}\label{fig:TS_scaling-a}}
    \hfill
    \subfloat[]{\includegraphics[height=0.45\linewidth]{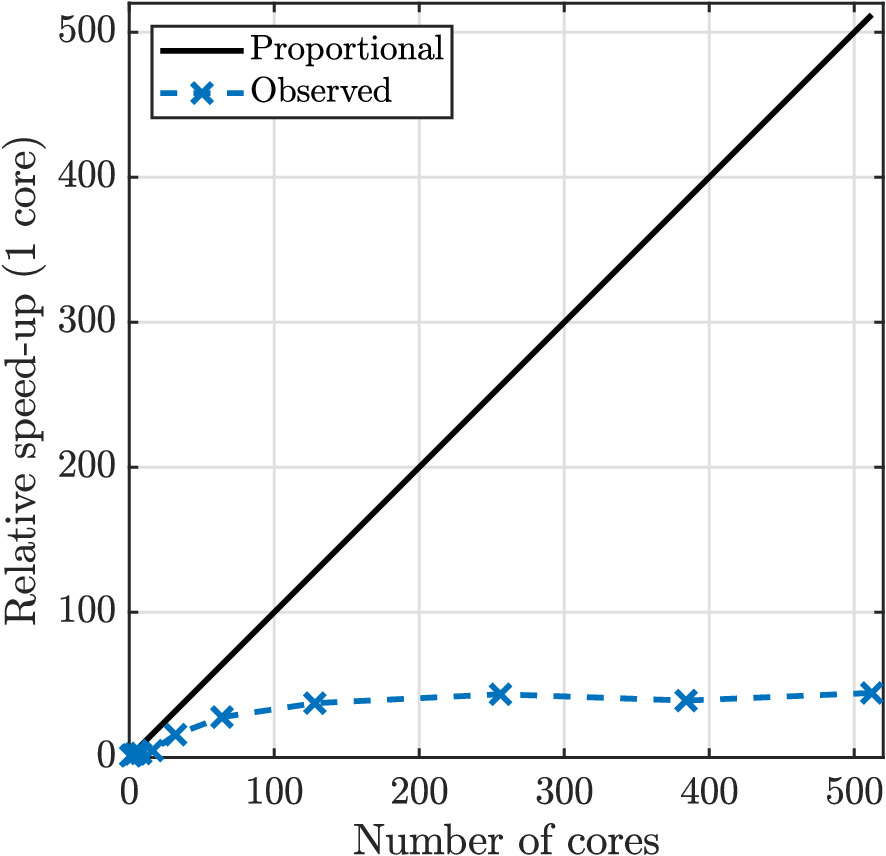}\label{fig:TS_scaling-b}}
    \caption{Relative speed-up of the time-stepping approach for Example 1, as a function of number of cores on: (a) a single node; (b) multiple nodes. The reference for the speed-up is execution performed using a single core.}
    \label{fig:TS_scaling}
\end{figure}
Figure \ref{fig:TS_scaling} shows strong scaling results of the time-stepping approach. It can be seen that the scaling is quite poor, even for a low number of cores on a single node, as seen in Figure \ref{fig:TS_scaling-a}, and especially for multiple nodes, as seen in Figure \ref{fig:TS_scaling-b}. This is because the two-dimensional resolution of $640\times 640$ elements only yields $410,\!881$ DOFs. This is actually a very small computational problem by modern standards and, thus, the parallel performance is quickly polluted by communication cost. 

For the proposed space-time approach, the problem becomes three-dimensional (or rather (2+1)D) and, therefore, significantly larger. The computational mesh consists of $640\times 640\times 1280$ elements, which yields a total of $526,\!338,\!561$ DOFs. This is already a very large computational problem, even for a relatively coarse two-dimensional spatial resolution. This requires significantly more memory to store all the necessary data and the full linear system. In practice, the minimum number of computational nodes necessary to solve the problem is observed to be 8 nodes (2048 GiB of RAM).
\begin{figure}
    \centering
    \subfloat[]{\includegraphics[height=0.45\linewidth]{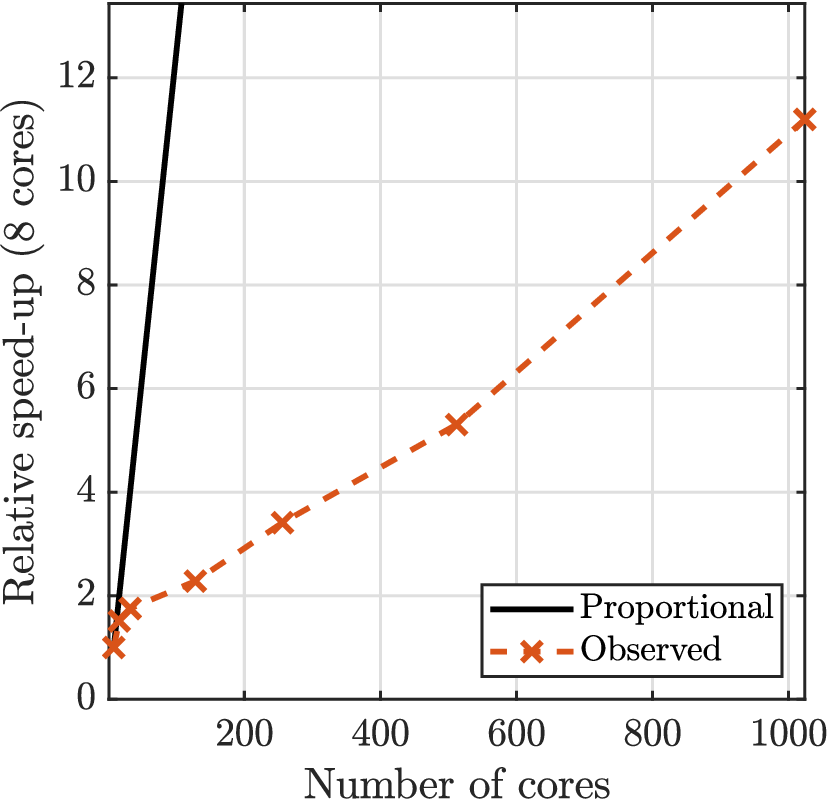}\label{fig:ST_scaling-a}}
    \hfill
    \subfloat[]{\includegraphics[height=0.45\linewidth]{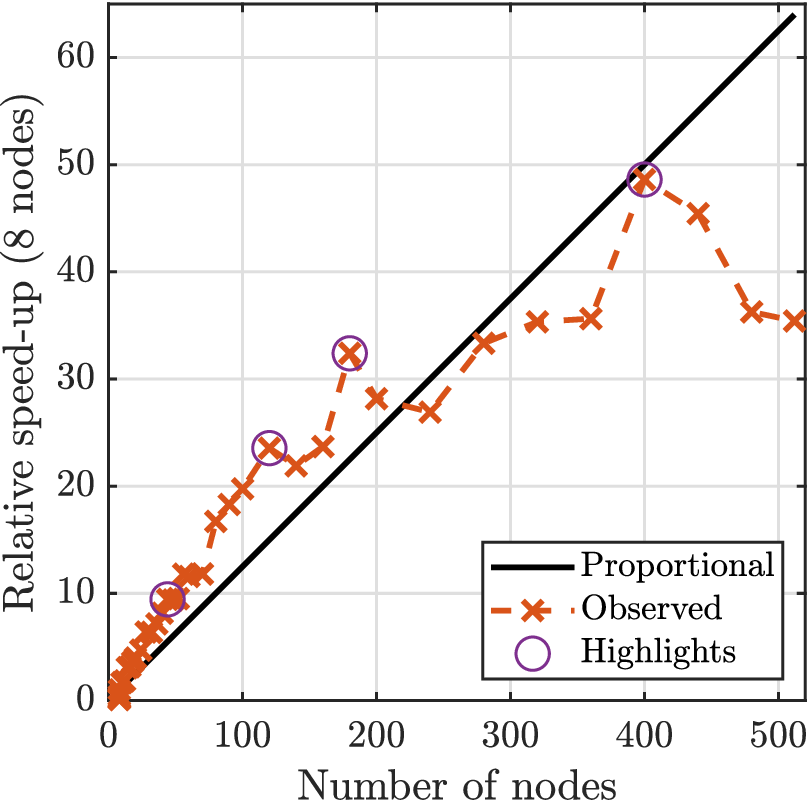}\label{fig:ST_scaling-b}}
    \caption{Relative speed-up of the space-time approach for Example 1, as a function of: (a) number of cores on 8 nodes; (b) number of full nodes with 128 cores each. The reference for the speed-up and starting point of the graphs are execution performed using: (a) 1 core on 8 nodes; (b) 8 full nodes (1024 cores). The circled points are discussed in Section \ref{sec:result_compcost}.}
    \label{fig:ST_scaling}
\end{figure}
Figure \ref{fig:ST_scaling} shows strong scaling results when using the proposed space-time approach. 
Figure \ref{fig:ST_scaling-a} shows results from using 8 nodes with less than 128 cores, relative to using 1 core per node. An unusual trend is observed, where first the scaling bends off at $8\times 4$ cores but then starts a more or less linear scaling after that. In general, sub-node scaling is observed to be far from optimal for the proposed approach. However, when looking at scaling by increasing the number of full nodes, excellent scaling is observed in Figure \ref{fig:ST_scaling-b} up to 400 nodes, where $\sim50\times$ speed-up is observed for a $50\times$ increase in nodes (from 8 to 400).
It can also be seen that the speed-up scales better than directly-proportional, which may be due to cache effects.

Lastly, it should be noted that some odd drops in performance are seen for some configurations. For instance, 140 and 160 nodes yield longer computational times than expected based on 120 and 180 nodes, and the same is seen up 400 nodes. We do not know exactly why this is and finding this reason is beyond the scope of the current paper, which presents a proof-of-concept. However, our current hypothesis is that it may be due to irregular partitioning of the mesh and thus poor load balancing on the coarser levels of the semi-coarsened hierarchy. Four of the best performers are highlighted in Figure \ref{fig:ST_scaling-b} and will be discussed further in Section \ref{sec:result_compcost}.

\subsection{Computational time} \label{sec:result_comptime}

\begin{figure}
    \centering
    \includegraphics[height=0.45\linewidth]{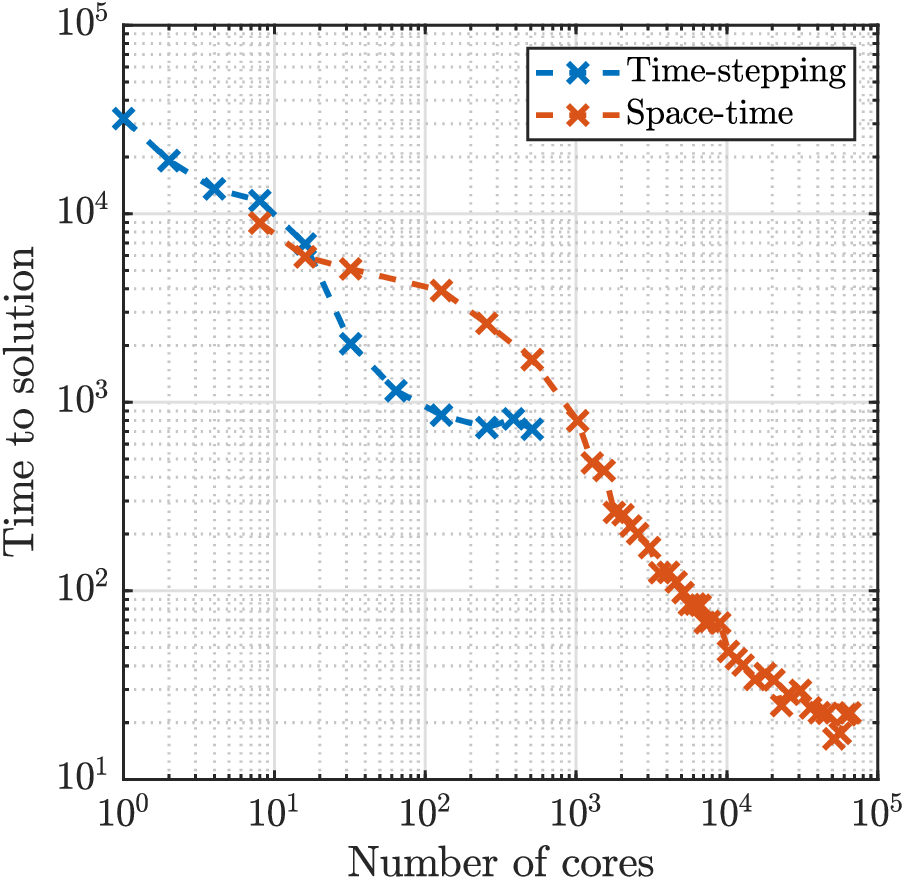}
    \caption{Computational time-to-solution, as a function of number of cores for both methods applied to Example 1.}
    \label{fig:TSvsST_timing}
\end{figure}
Figure \ref{fig:TSvsST_timing} compares the computational time-to-solution. For a relatively small number of cores, the time-stepping approach is the only viable option, since the space-time approach requires a large amount of memory. For a medium amount of cores, the time-stepping approach appears to be the fastest, at least for this particular configuration for the space-time method (less than full nodes, but 8 nodes at minimum). However, a crossing point is seen at around 1000 cores, where the time-stepping scaling has plateaued (see Figure \ref{fig:TS_scaling}), but the scaling of the space-time approach is just picking up (see Figure \ref{fig:ST_scaling-b}). 
Figure \ref{fig:TSvsST_timing} shows that spatial parallelisation can be used to reduce the time-to-solution of time-stepping by around 1.5 orders of magnitude compared to a single core, from $\sim\! 32000$ seconds to $\sim\! 850$ seconds. However, using the space-time approach, the time-to-solution can be reduced to just $\sim\! 20$ seconds, providing a reduction of over 3 orders of magnitude compared to time-stepping on a single core and almost 3 orders of magnitude compared to the space-time approach on 8 cores (distributed over 8 nodes).
Obviously, this reduction in computational time comes at a cost, which will be discussed in the following subsection.

\subsection{Computational cost} \label{sec:result_compcost}

\begin{figure}
    \centering
    \subfloat[]{\includegraphics[height=0.45\linewidth]{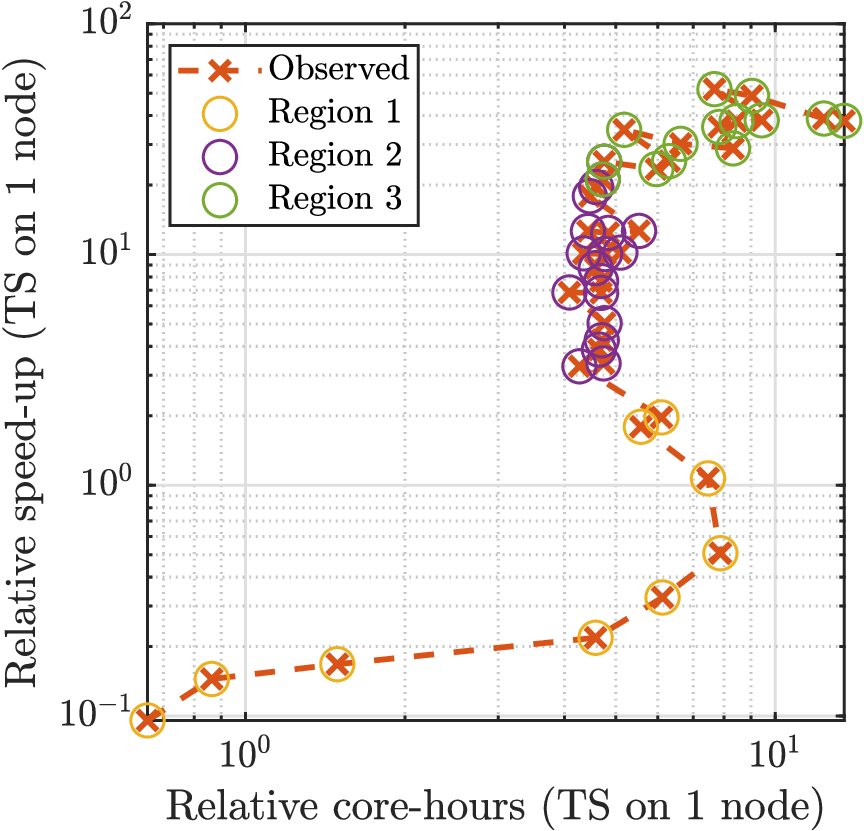}\label{fig:TSvsST_cost-a}}
    \hfill
    \subfloat[]{\includegraphics[height=0.45\linewidth]{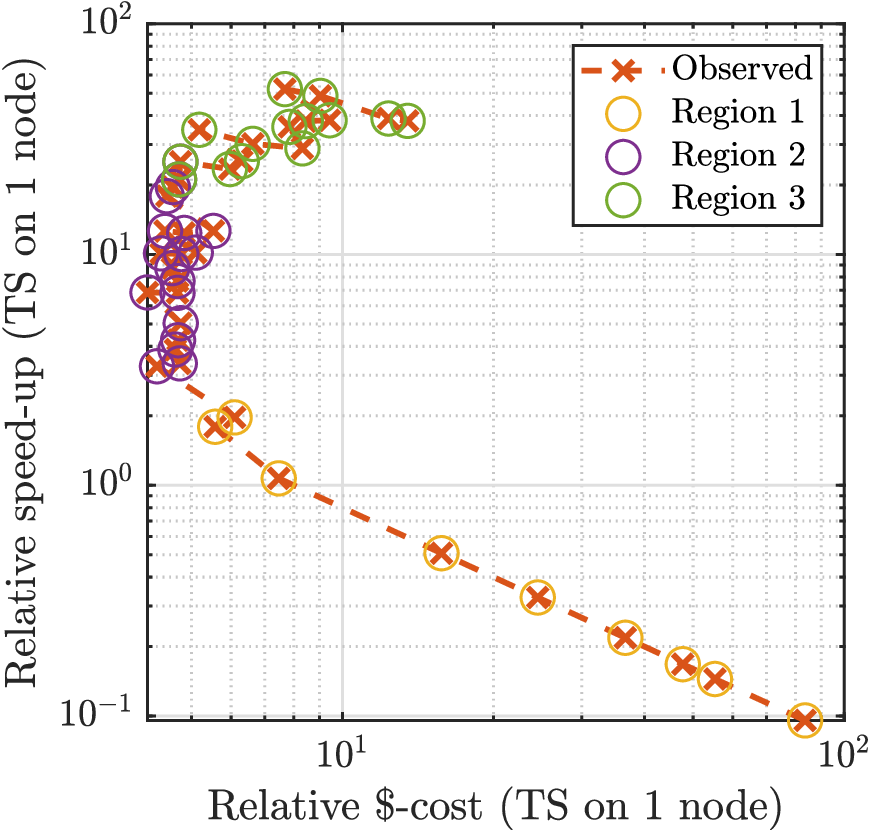}\label{fig:TSvsST_cost-b}}
    \caption{Speed-up for the proposed space-time approach relative to time-stepping on 1 full node, as a function of: (a) relative number of core-hours; (b) relative monetary cost. The different coloured circles group the points into three regions as discussed in Section \ref{sec:result_compcost}.}
    \label{fig:TSvsST_cost}
\end{figure}
Figure \ref{fig:TSvsST_cost} shows the relative speed-up of the proposed space-time approach compared to time-stepping on 1 full node (128 cores). The oscillations in the data are due to the intermediate drops in performance observed in Figure \ref{fig:ST_scaling-b}.
Figure \ref{fig:TSvsST_cost-a} shows the speed-up as a function of the core-hours, computed naively as the number of cores times computational time in hours. Using this metric, using very few cores distributed over 8 nodes becomes relatively cheap, but also provides really poor speed-up (below 1). However, the fact is that on a system such as LUMI, one pays for the full nodes when all the memory is needed, as is the case here, although very few cores are used. Therefore, Figure \ref{fig:TSvsST_cost-b} shows the speed-up as a function of the actual monetary cost relative to a single computation node, which would be proportional to the actual price paid on a commercial system. It is also worth mentioning that the relative number of core-hours can be translated to a relative energy consumption of running the jobs. 

It seems that the measurements can be divided into three groups, as shown by the three different coloured circles in Figure \ref{fig:TSvsST_cost}. Region 1 shows that the sub-node runs are way too expensive and provide no speed-up, rendering these settings useless in practise. Then after a few increases in the number of nodes, the speed-up becomes vertically increasing in Region 2, where the speed-up is increasing at no additional cost (corresponding to the region of better-than-expected scaling in Figure \ref{fig:ST_scaling-b}). This keeps going up to around a $25\times$ speed-up at just under $5\times$ the cost. Subsequently, the data points start to bend off in Region 3 (similar to the scaling curve in Figure \ref{fig:ST_scaling-b}) when communication costs starts to take over.
The four data points highlighted in Figure \ref{fig:ST_scaling-b} are the following:
\begin{itemize}
    \item $10.1\times$ speed-up at $4.3\times$ the cost using 44 nodes (5632 cores) for 84.5 seconds
    \item $25.2\times$ speed-up at $4.8\times$ the cost using 120 nodes (15360 cores) for 33.9 seconds
    \item $34.7\times$ speed-up at $5.2\times$ the cost using 180 nodes (23040 cores) for 24.6 seconds
    \item $52.1\times$ speed-up at $7.7\times$ the cost using 400 nodes (51200 cores) for 16.4 seconds
\end{itemize}
It should be noted that this is compared to the reference time-stepping solution of 855 seconds.

These estimates should be taken with some uncertainty given the assumptions and simplifications made. As will be seen in the upcoming results, the computational time does not scale directly proportionally with the number of design iterations, meaning that an optimisation with 100 design iterations using 120 nodes does not simply take 339 seconds. This is mainly because the number of linear iterations, to solve the systems of equations, increases beyond the 10 design iterations as the design field begins to converge to binary values - as will be shown in Figure \ref{fig:example2_comparison_linits}.

\section{Topology optimisation} \label{sec:result_topopt}

This section presents the topology optimisation results for the two examples described in Section \ref{sec:exampleDefs}. In the space-time representation, all presentations will be of the thresholded design field - meaning only elements with a physical design variable above 0.5 will be shown.

\subsection{Example 1: Oscillating heat source} \label{sec:result_example1}

The problem definition and boundary conditions for Example 1 are given in Section \ref{sec:example1def} and is based on the problem introduced by \citet{Appel2024}.

\subsubsection{Coarse mesh verification} \label{sec:results_exp1_parareal}

In order to verify the implementation, Example 1A applies the space-time approach to the same computational problem as introduced by \citet{Appel2024}. That is, the space-time mesh is $100\times 100\times 480$ yielding $4,\!906,\!681$ DOFs. 
\begin{figure}
    \centering
    \includegraphics[width=0.5\linewidth]{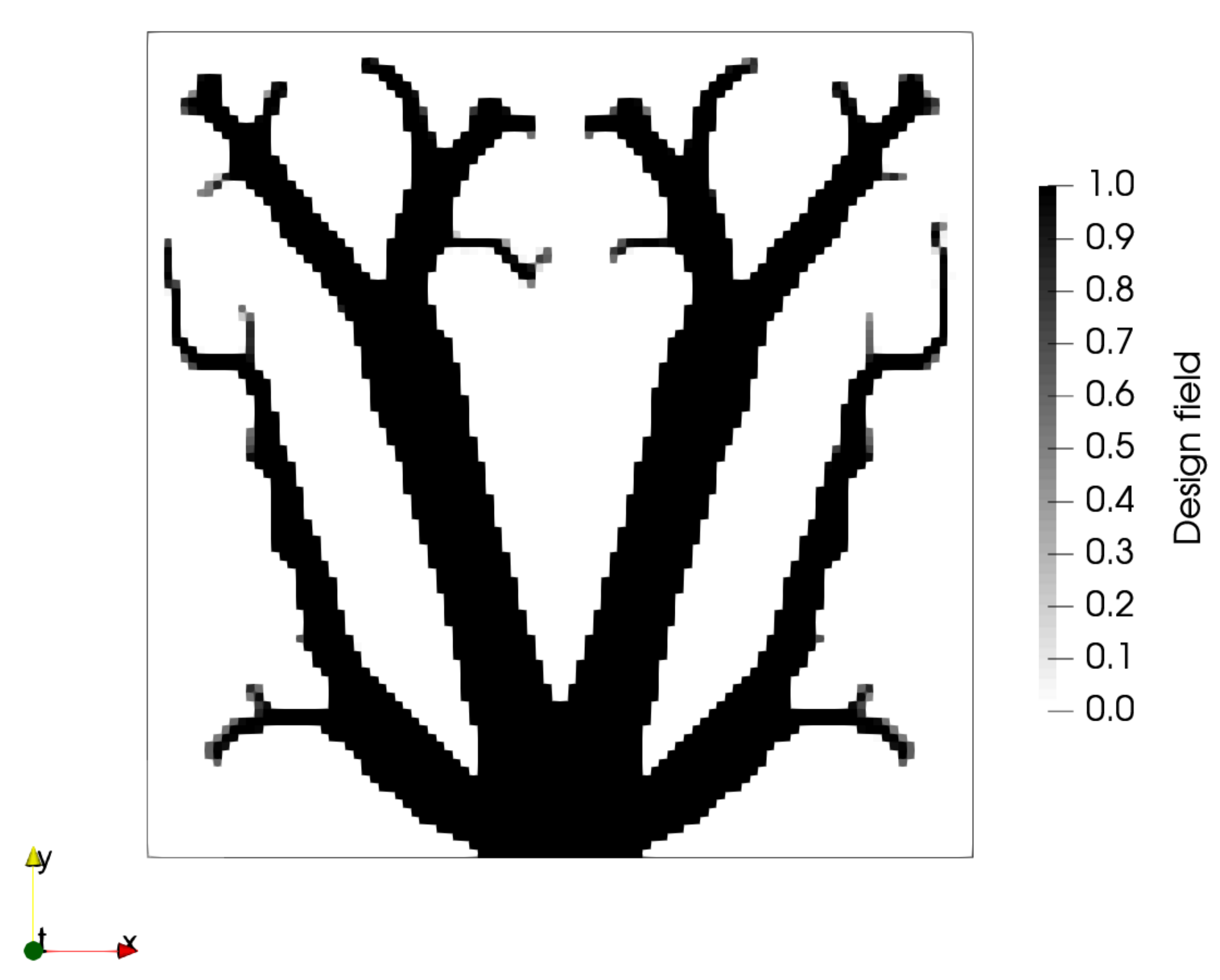}
    \caption{Optimised design field for Example 1A with problem settings defined by \citet{Appel2024} and a bounding box showing the edges of the design domain.}
    \label{fig:example1_pararealSettings}
\end{figure}
Figure \ref{fig:example1_pararealSettings} shows the optimised design field after 100 design iterations. There is qualitative agreement with the results presented by \citet{Appel2024}, but there are minor differences most likely due to different discretisations, implementations, etc. However, this problem is too small to truly observe the benefits of using the proposed space-time approach, so a finer mesh is investigated in the next section.

\subsubsection{Finer mesh} \label{sec:results_exp1_finermesh}

For Example 1B, the space-time mesh is now refined to $640\times 640\times 1280$, which yields $526,\!338,\!561$ DOFs. Due to limitations in the current implementation of the extrusion operator clashing with the partitioning of the coarse grids, we are limited to only $N_l = 7$ multigrid levels (the first 7 meshes listed in Section \ref{sec:result_comparison}). The computational time is affected by the coarsest level being slightly larger, as well as the additional cost of applying the extrusion operator. Lastly, the number of linear iterations increases during the optimisation process as is usual for topology optimisation \citep{Amir2014}.

The problem was solved using 44 nodes (5632 cores) and 120 nodes (15360 cores), which according to the scaling study in Section \ref{sec:result_compcost} should provide approximately $10\times$ and $25\times$ speed-up, respectively, both at  approximately $5\times$ the cost. For 44 nodes, the optimisation took 49 minutes and 12 seconds, and for 120 nodes, it took 43 minutes and 51 seconds. Here it can be seen that the significant increase in computational resources only has a small impact on speed, the reason for which is unknown.

To provide a more realistic estimate of the computational time using the time-stepping approach, we solved 58 iterations using 128 time steps on 1 node and scaled it up to 100 iterations of 1280 time steps (because it was not possible to solve 1280 time steps on 1 node due to the memory-inefficient implementation). The estimated completion time is 15 hours, 13 minutes and 56 seconds, yielding speed-up estimates of $18.6\times$ and $20.8\times$ for 44 nodes and 120 nodes, respectively, at $2.4\times$ and $5.8\times$ the cost. Using 44 nodes, this is better speed-up than expected for cheaper, and using 120 nodes, this is slightly worse than expected. However, we will not go into detail with this, since this work is a proof-of-concept laying the groundwork for future developments and further understanding of the method.
        
\begin{figure}
    \centering
    \subfloat[Spatial representation]{
    \includegraphics[width=0.425\linewidth]{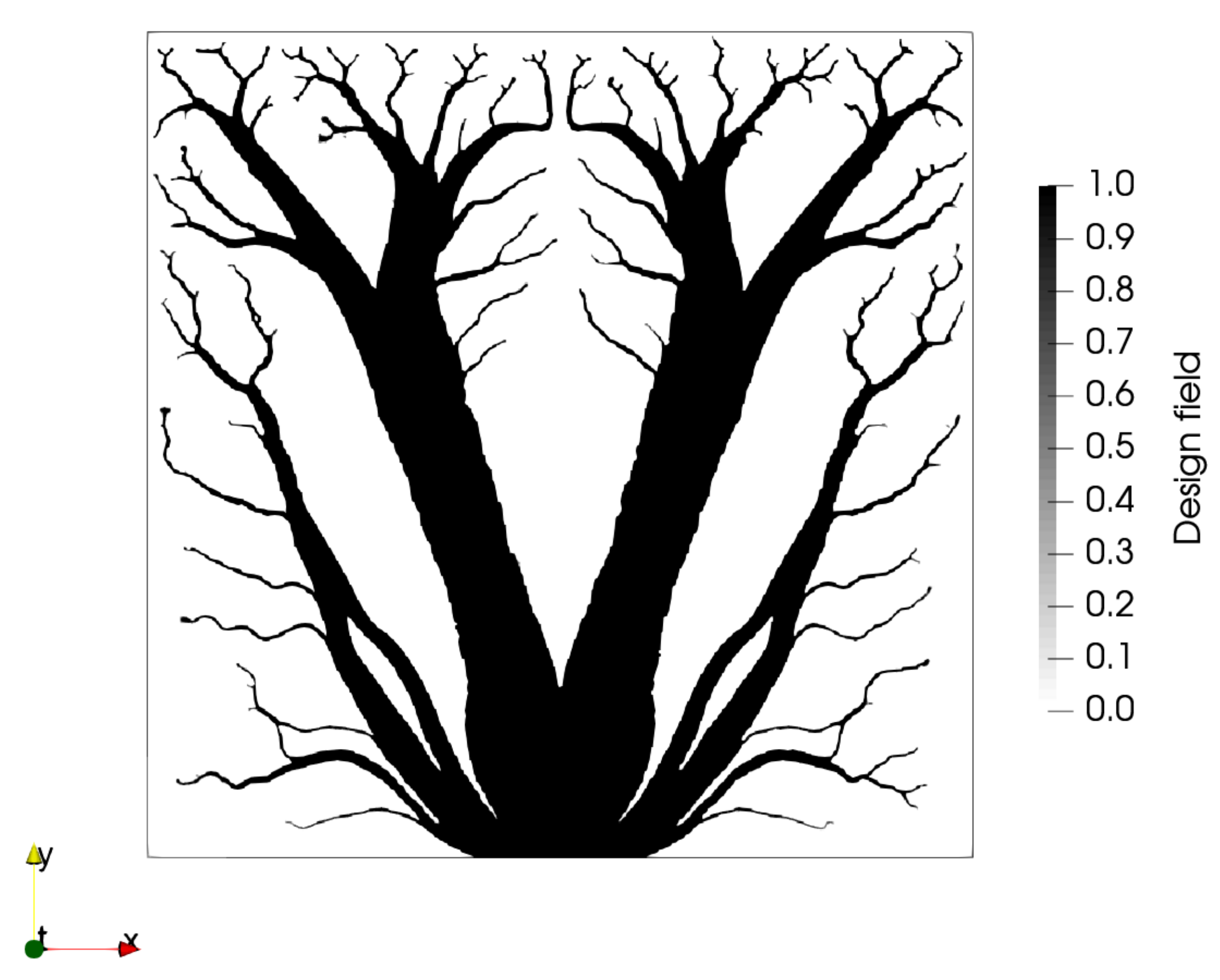}
    \label{fig:example1_finerMesh-a}
    }
    \hfill
    \subfloat[Space-time representation]{
    \includegraphics[width=0.5\linewidth]{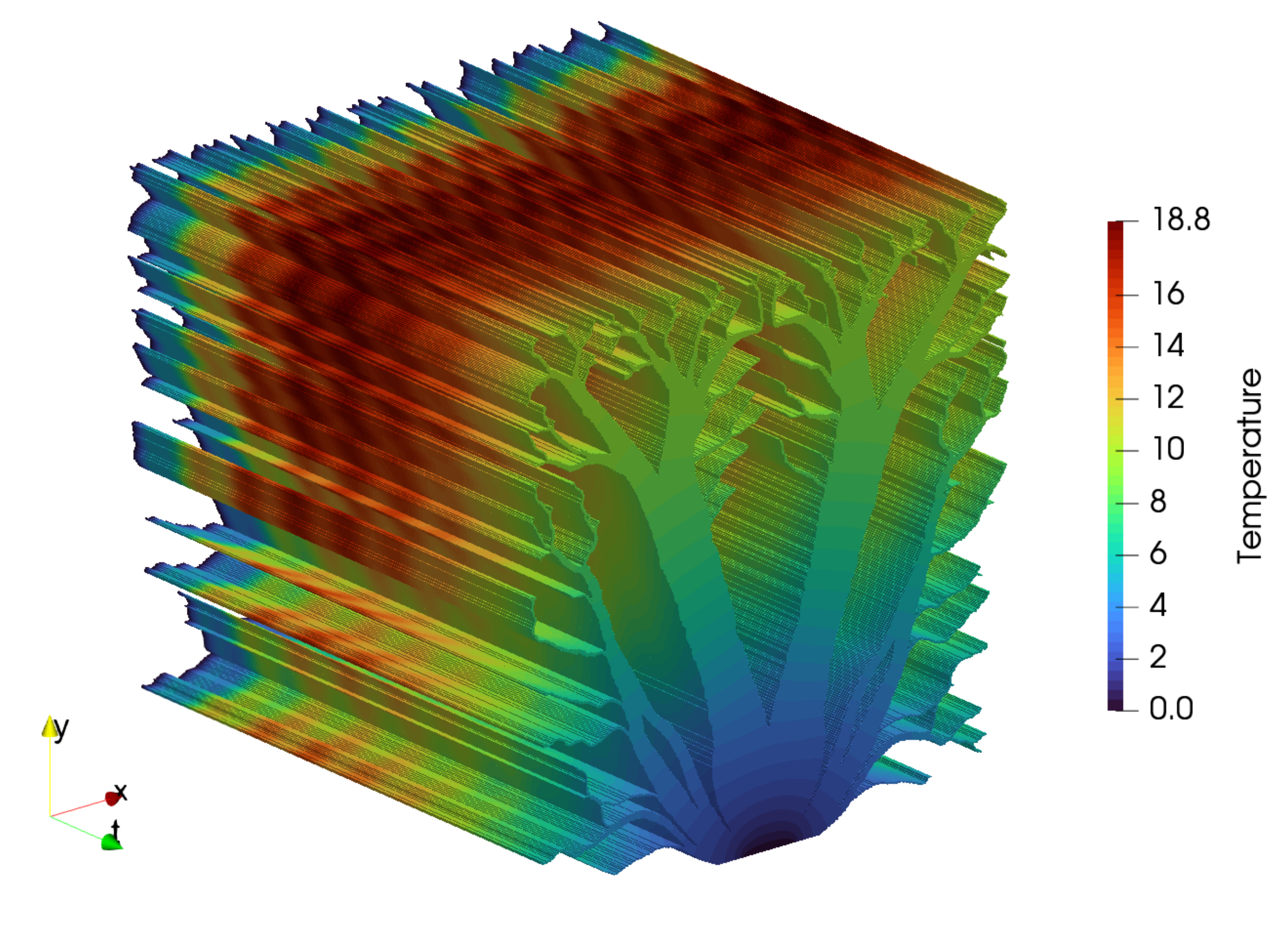}
    \label{fig:example1_finerMesh-b}
    }
    \caption{Optimised design for Example 1B using a $640\times 640\times 1280$ space-time mesh shown in: (a) a two-dimensional spatial representation with a bounding box showing the edges of the design domain; (b) a (2+1)D space-time thresholded representation coloured according to the temperature field.}
    \label{fig:example1_finerMesh}
\end{figure}
Figure \ref{fig:example1_finerMesh} shows the optimised design in two different representations.
Figure \ref{fig:example1_finerMesh-a} shows the design field, which is clearly similar to the coarser result in Figure \ref{fig:example1_pararealSettings}, just with finer features allowed for by the finer mesh. Although it is well-known that conductive branching trees are suboptimal for pure conduction \citep{Yan2018}, they are nonetheless frequently observed in practise.
Figure \ref{fig:example1_finerMesh-b} shows the space-time representation of the optimised design, where the design field has been thresholded at $\bar{\tilde{\chi}} = 0.5$ and is coloured by the temperature field. It can be seen that the design is extruded in the time dimension, as described in Section \ref{sec:method_extrusion}, and that an oscillatory temperature history results from the oscillatory heat source.      
\begin{figure}
    \centering
    \subfloat[$t = 0.25$]{
    \includegraphics[height=0.325\linewidth]{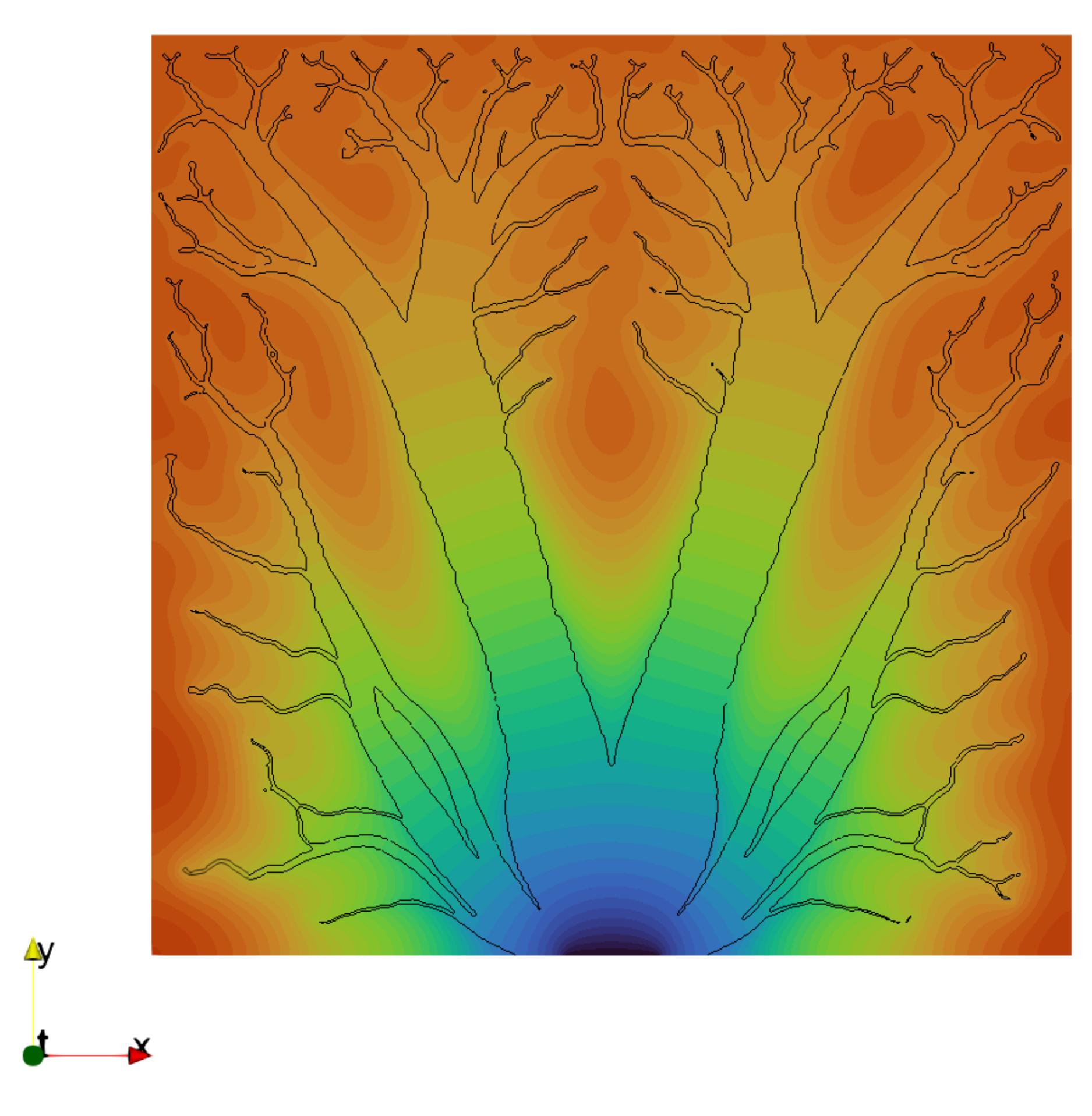}
    \label{fig:example1_finerMesh_temp-a}
    }
    \subfloat[$t = 0.5$]{
    \includegraphics[height=0.325\linewidth]{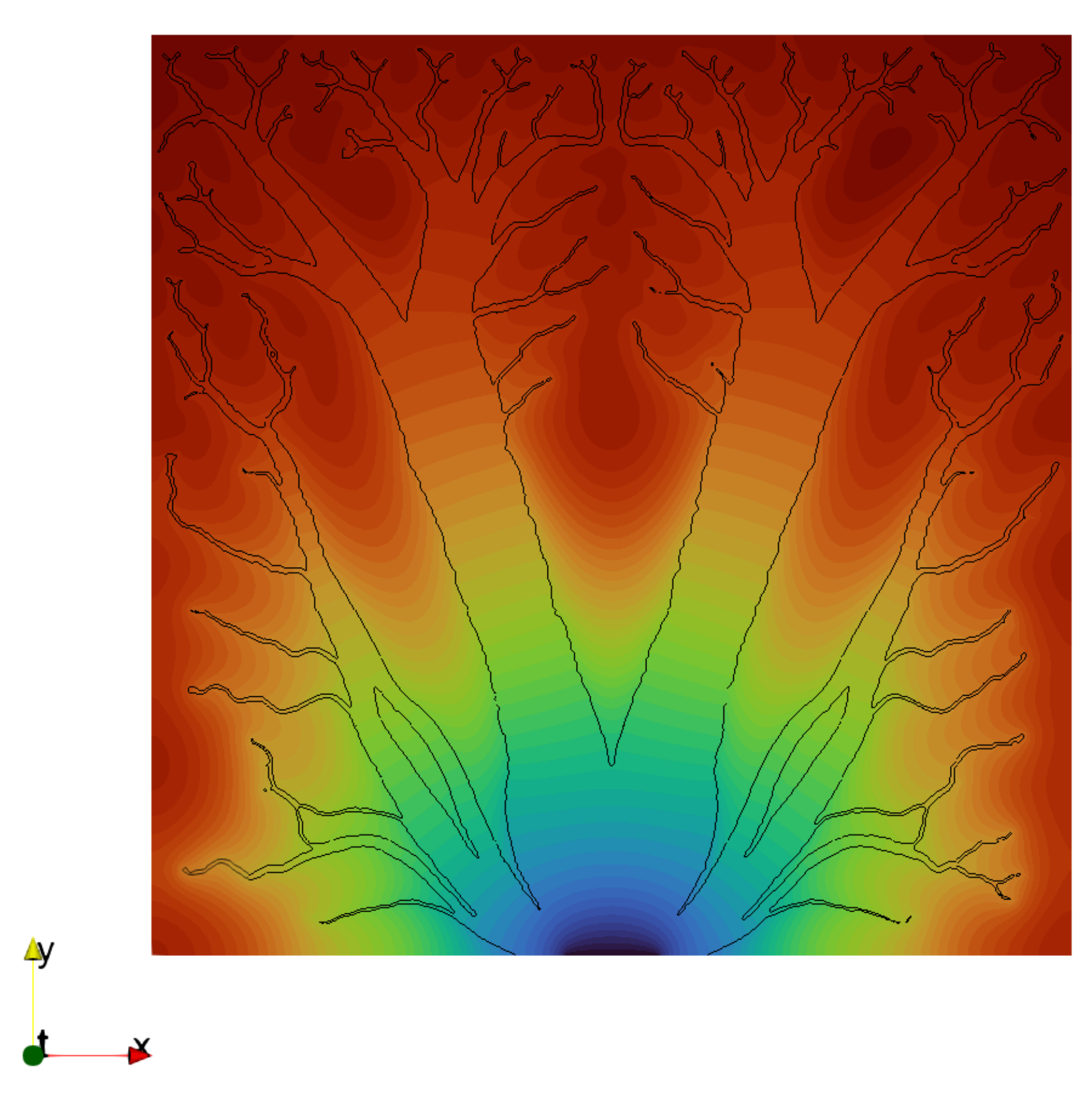}
    \label{fig:example1_finerMesh_temp-b}
    }
    \subfloat[$t = 0.9$]{
    \includegraphics[height=0.325\linewidth]{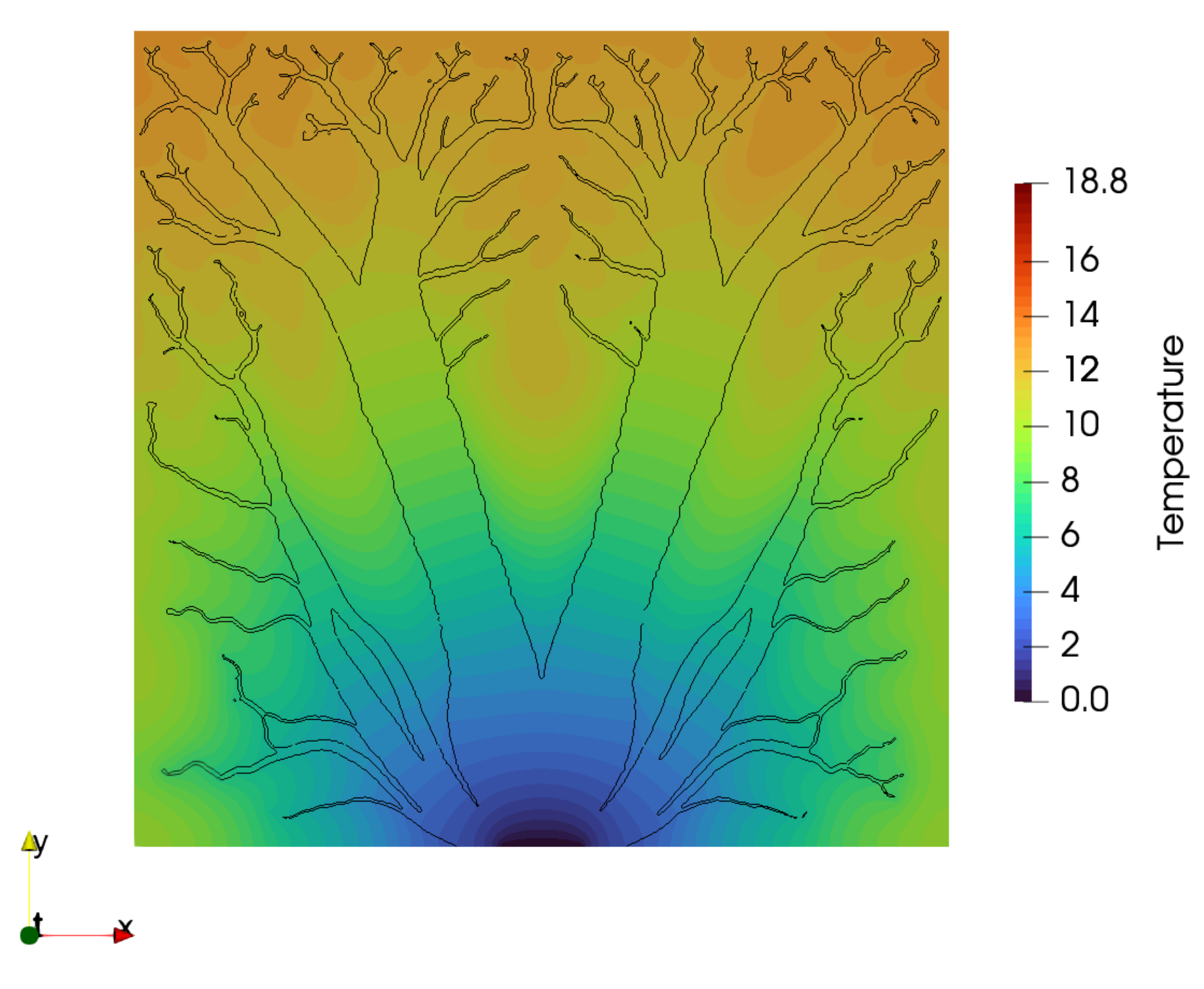}
    \label{fig:example1_finerMesh_temp-c}
    }
    \caption{Time slices of the space-time domain to illustrate the oscillatory temperature history for Example 1B. The colours denote the temperature field and the black lines show the contours of the design.}
    \label{fig:example1_finerMesh_temp}
\end{figure}
The oscillatory temperature history is further illustrated by the time slices presented in Figure \ref{fig:example1_finerMesh_temp}.

To show the benefit of the semi-coarsening scheme, the problem was also solved using full-coarsening only. This is significantly slower and the linear solver requires 9.7 iterations on average for the state problem and a huge 55.6 iterations on average for the adjoint problem, compared to 5.6 and 9.4, respectively, for the semi-coarsening scheme. It might be that the number of iterations could be brought down by increasing the number of smoothing steps or changing the coarse level solver. But for comparable settings, the semi-coarsening scheme has a clear advantage and leads to significantly faster convergence.

\subsubsection{Different material parameters} \label{sec:results_exp1_matprops}

Two different cases are now explored, where minor modifications are made to the material parameters. For Example 1C, the conductivity of the conductive material is increased to $k_\textrm{con} = 10$, while all other parameters are kept as previously. For Example 1D, the volumetric capacity of the conductive material is increased to $\mathcal{C}_\textrm{con} = 100$.
It should be noted that to converge to binary designs, it was necessary to increase $p_\mathcal{C}$ to 4 for the high volumetric capacity case, Example 1D.

The problem is solved using $N_l = 7$ multigrid levels on 44 nodes (5632 cores) due to the good performance from above.
\begin{table}
    \centering
    \begin{tabular}{ccc}
        Level & Ref. (1B) \& High cond. (1C) & High cap. (1D) \\ \hline
        0 & $640\times 640\times 1280$ & $640\times 640\times 1280$ \\
        1 & $320\times 320\times 1280$ & $320\times 320\times 1280$ \\
        2 & $160\times 160\times 1280$ & $160\times 160\times 1280$ \\
        3 & $80\times 80\times 1280$ & $80\times 80\times 1280$ \\
        4 & $40\times 40\times 1280$ & $80\times 80\times 640$ \\
        5 & $20\times 20\times 1280$ & $80\times 80\times 320$ \\
        6 & $20\times 20\times 640$ & $40\times 40\times 320$ \\
    \end{tabular}
    \caption{Multigrid hierarchies for the three material cases of Example 1(B,C,D).}
    \label{tab:example1_multigrid_hierarchies}
\end{table}
The multigrid hierarchies are listed in Table \ref{tab:example1_multigrid_hierarchies} for the reference (Example 1B) and two additional material cases (Examples 1C and 1D). It can be seen that the first 3 coarser levels are the same for all problems, but then the coarsening strategy, outlined in Section \ref{sec:method_mg}, causes the hierarchies to take different paths. It is worth noting that the coarsest grids do not have the same amount of DOFs.

\begin{figure}
    \centering
    \subfloat[Spatial representation]{
    \includegraphics[width=0.425\linewidth]{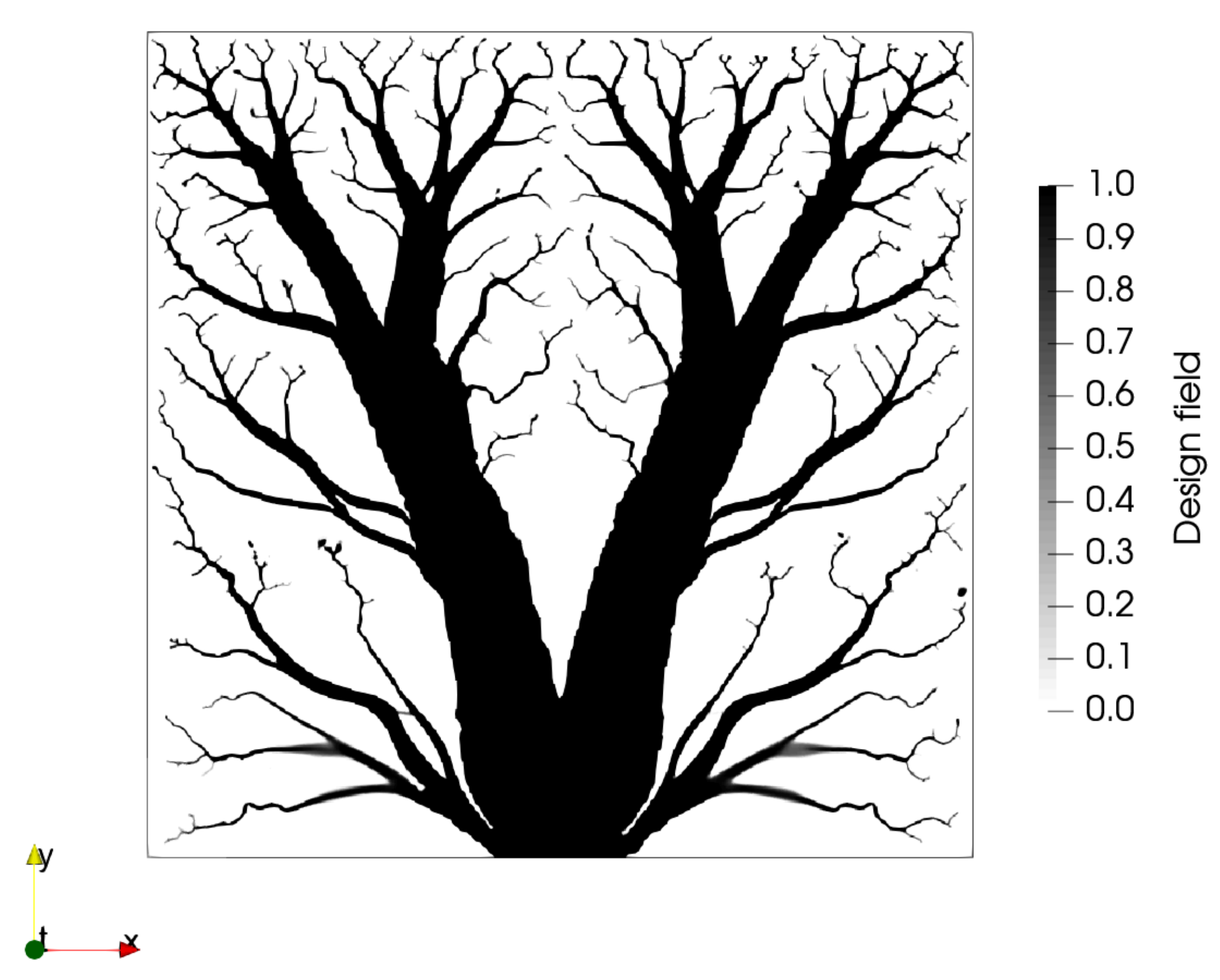}
    \label{fig:example1_higherCond-a}
    }
    \hfill
    \subfloat[Space-time representation]{
    \includegraphics[width=0.5\linewidth]{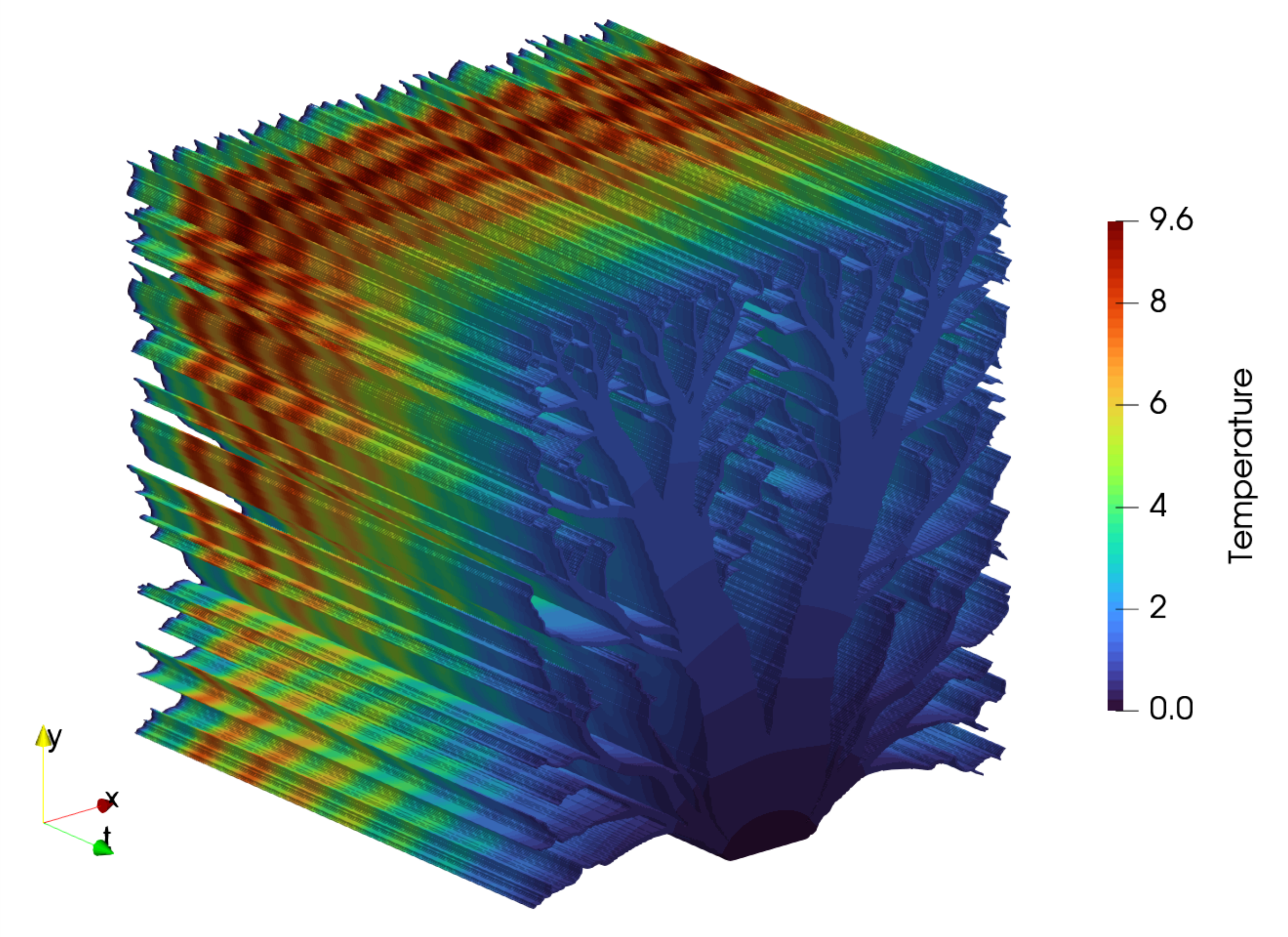}
    \label{fig:example1_higherCond-b}
    }
    \caption{Optimised design for Example 1C with higher conductivity shown in: (a) a two-dimensional spatial representation with a bounding box showing the edges of the design domain; (b) a (2+1)D space-time thresholded representation coloured according to the temperature field.}
    \label{fig:example1_higherCond}
\end{figure}
Figure \ref{fig:example1_higherCond} shows the optimised design for Example 1C, where the conductive material has an increased conductivity. Figure \ref{fig:example1_higherCond-a} shows a design field very similar to the previous design in Figure \ref{fig:example1_finerMesh-a}, but with more fine branching features. Figure \ref{fig:example1_higherCond-b} shows that the design cools down quicker and does not accumulate as much heat as previously, which is due to the higher conductivity allowing it to remove the heat more efficiently.

\begin{figure}
    \centering
    \subfloat[Spatial representation]{
    \includegraphics[width=0.45\linewidth]{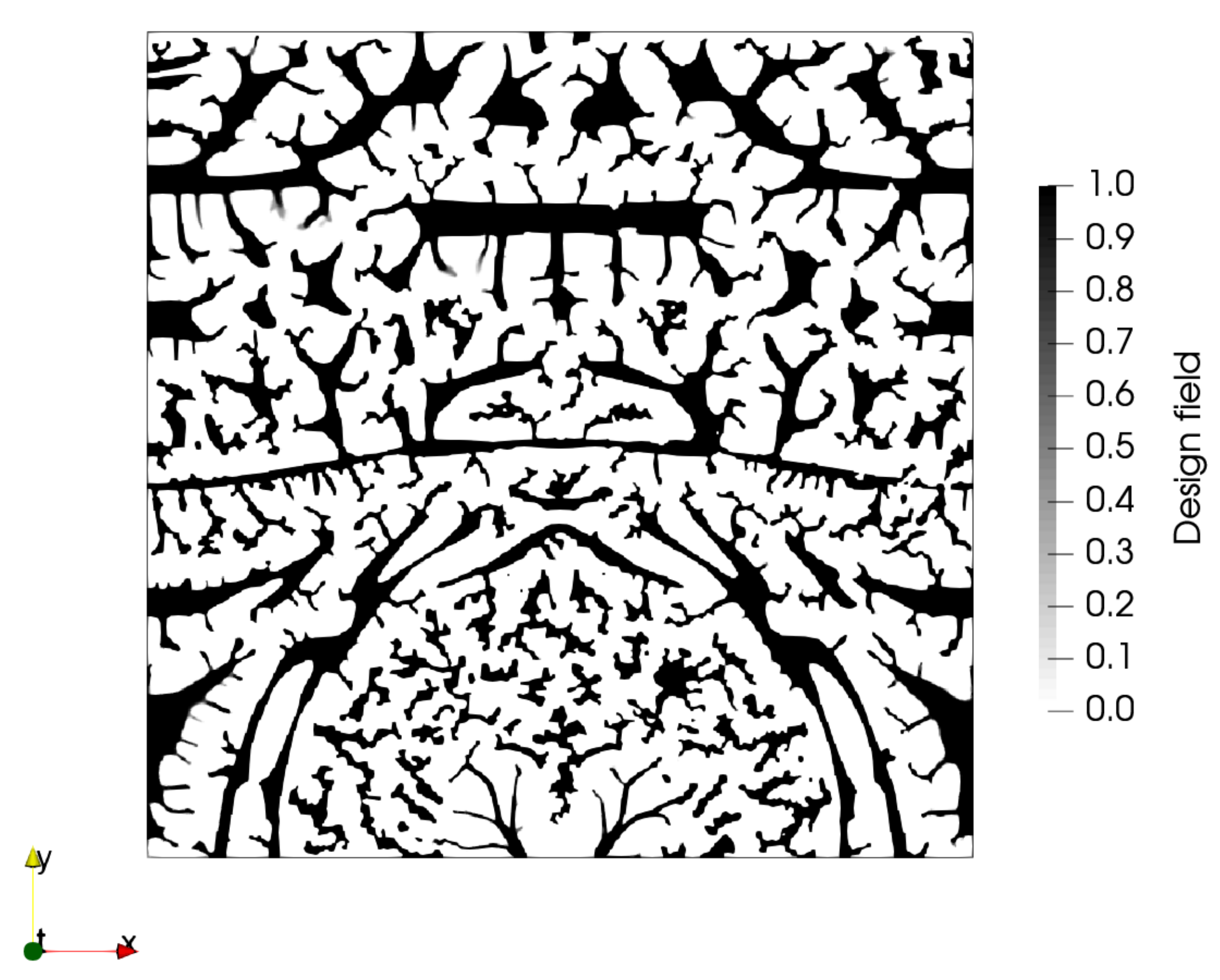}
    \label{fig:example1_higherCap-a}
    }
    \hfill
    \subfloat[Space-time representation]{
    \includegraphics[width=0.5\linewidth]{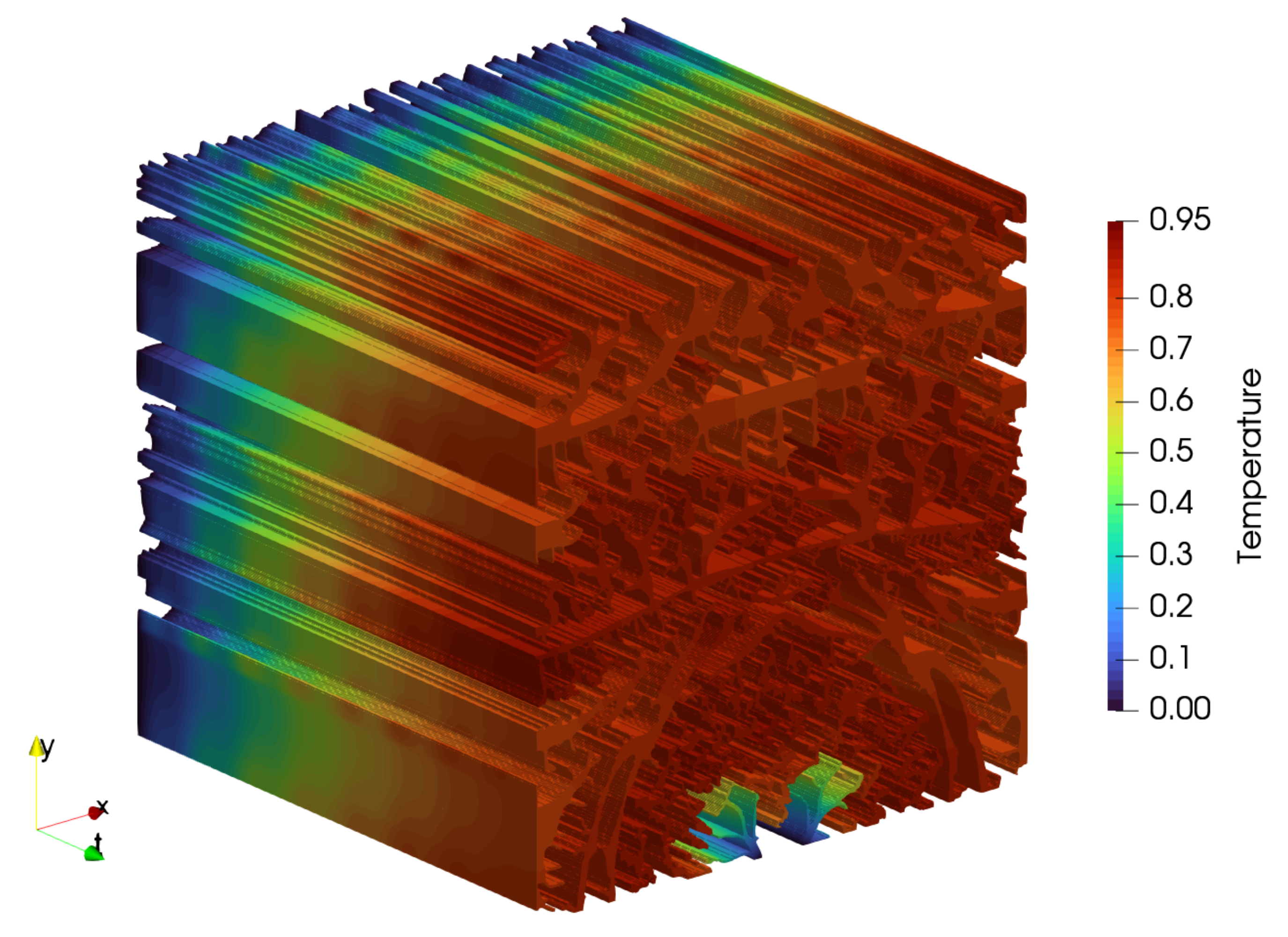}
    \label{fig:example1_higherCap-b}
    }
    \caption{Optimised design for Example 1D with higher volumetric capacity shown in: (a) a two-dimensional spatial representation with a bounding box showing the edges of the design domain; (b) a (2+1)D space-time thresholded representation coloured according to the temperature field. This case required a penalty power for the heat capacity of $p_\mathcal{C}=4$ to converge to binary designs.}
    \label{fig:example1_higherCap}
\end{figure}
Figure \ref{fig:example1_higherCap} shows the optimised design for Example 1D, where the conductive material has an increased volumetric capacity, which in this case could be called the capacitative material. Figure \ref{fig:example1_higherCap-a} shows a design field that is completely different to the conductive tree designs previously observed.
The difference stems from the temperature dynamics being completely different, as shown in Figure \ref{fig:example1_higherCap-b} where it can be seen that the design is accumulating heat throughout the simulation time.
The timescale of diffusion can be computed as:
\begin{equation} \label{eq:diffusionTimescale}
    \tau_\text{diff} = \frac{\mathcal{C}L^{2}}{k}
\end{equation}
which for the conductive material of Example 1B, 1C, and 1D become 0.33, 0.10, and 33.33, respectively. For the high capacity case, this means that the conductive/capacitative material does not react to the fast oscillations of the heat source in the relatively short time frame solved for, $\tau = 1$. Therefore, the best performance is obtained by distributing the capacitative material over the entire domain to evenly accumulate heat and reduce the overall temperature.

\begin{figure}
    \centering
    \subfloat[Space-time domain]{
    \includegraphics[height=0.28\linewidth]{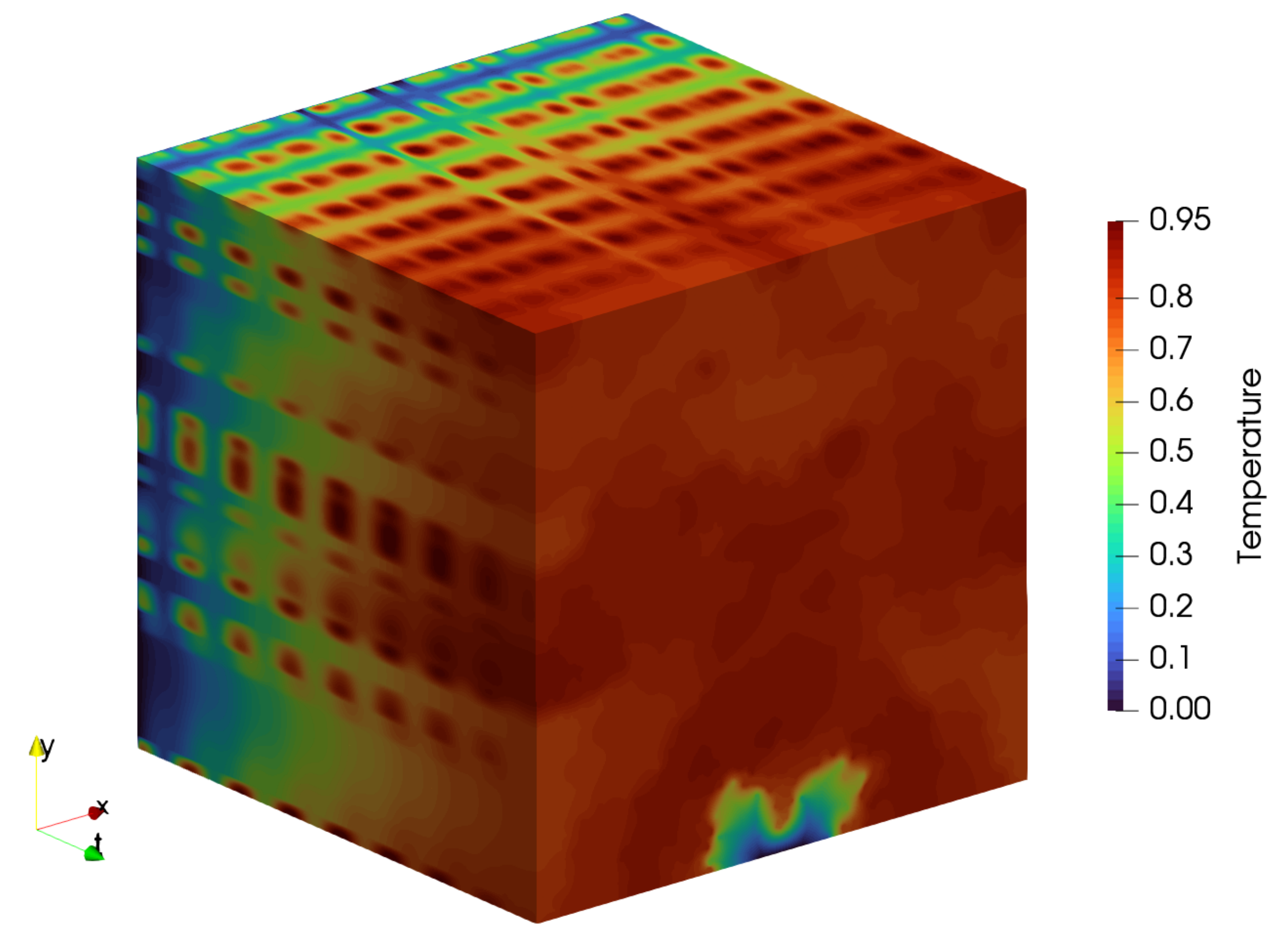}
    \label{fig:example1_higherCap_tempHist-a}
    }
    \subfloat[$t=0.37571$ (peak 4)]{
    \includegraphics[width=0.32\linewidth]{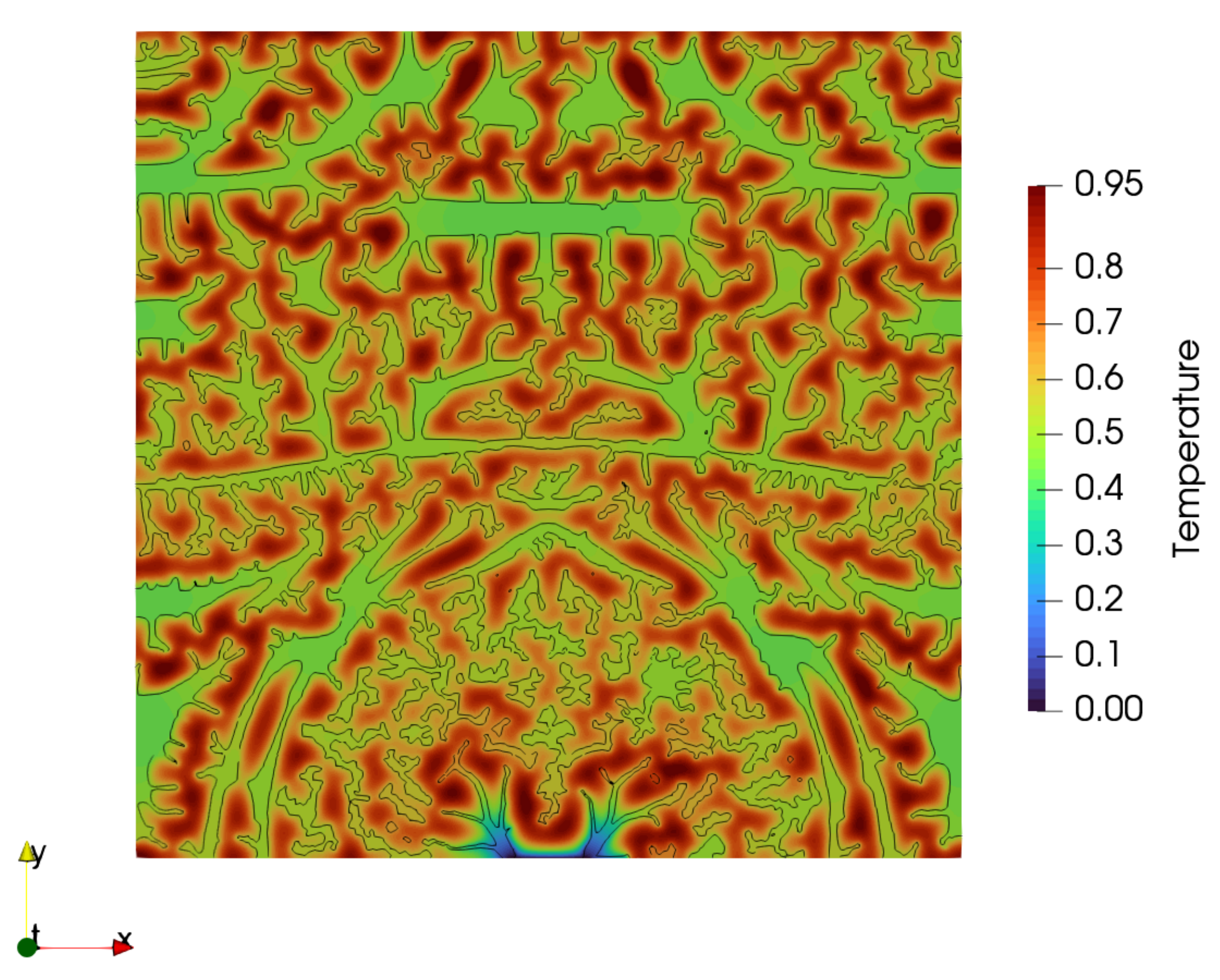}
    \label{fig:example1_higherCap_tempHist-b}
    }
    \subfloat[$t=0.62618$ (peak 6)]{
    \includegraphics[width=0.32\linewidth]{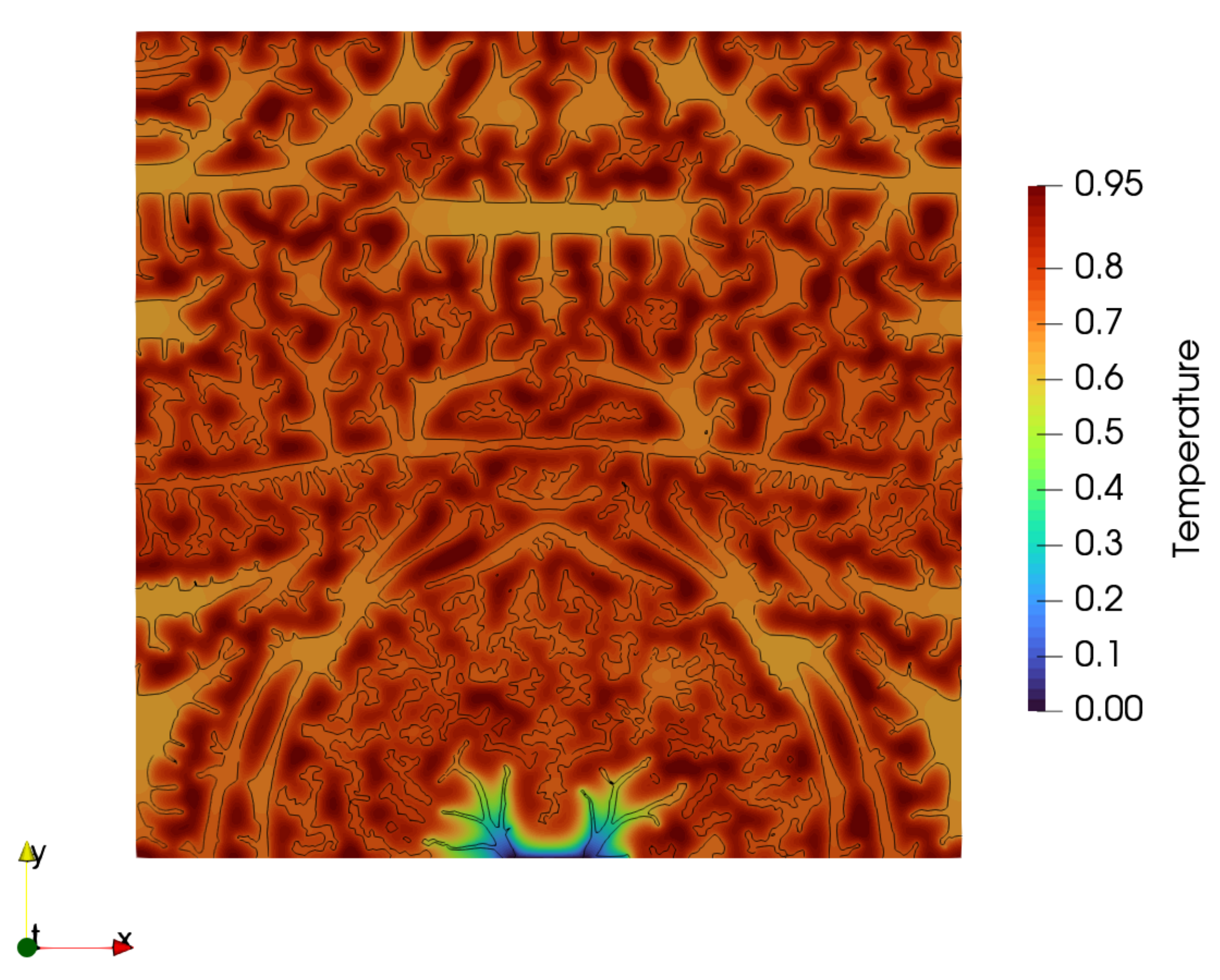}
    \label{fig:example1_higherCap_tempHist-c}
    }
    \caption{Temperature history for the higher volumetric capacity case, Example 1D: (a) the full space-time domain; (b-c) time slices at two peaks of the heat source function. The colours denote the temperature field and the black lines show the contours of the design.}
    \label{fig:example1_higherCap_tempHist}
\end{figure}
To further understand the optimised design, Figure \ref{fig:example1_higherCap_tempHist} shows the temperature history for the entire domain including the insulating material. It can be seen that it increases in temperature significantly more than the capacitative material. The capacitative material has a high thermal inertia, causing it to only slowly react to the peaks of the distributed heat source and slowly absorbing the heat from the insulating material during the valleys of the heat source function.

\begin{table}
    \centering
    \begin{tabular}{ccccc}
         & \multicolumn{2}{c}{Avg. linear it.} &  & \\
        Problem & State & Adjoint & Avg. time per it. [s] & Total time [h:mm:ss] \\ \hline
        Reference (1B) & 5.62 & 9.4 & 26.3 & 0:49:12 \\ 
        High conductivity (1C) & 7.79 & 12.24 & 43.8 & 1:08:09 \\ 
        High capacity (1D) & 8.18 & 14.36 & 43.7 & 1:09:08 
    \end{tabular}
    \caption{Solver performance metrics for the three material cases of Example 1(B,C,D) using 44 nodes (5632 cores) and 7 multigrid levels.}
    \label{tab:example1_multigrid_performance}
\end{table}
Table \ref{tab:example1_multigrid_performance} shows chosen performance metrics for the three cases with the different mesh hierarchies listed previously in Table \ref{tab:example1_multigrid_hierarchies}. It can be seen that the computational performance only slightly varies between the high conductivity and high capacity cases, but that both are harder and, thus, slower to solve than the reference. Although the high capacity case needs slightly higher number of linear iterations than the high conductivity case, especially for the adjoint solver, the average time per iteration and, thus, the total time remains basically the same. This could be due to several factors, such as performance variations across different nodes on LUMI, different number of GMRES iterations on the coarsest levels, or the contrast-dependency of the solver.

\subsection{Example 2: Moving heat source} \label{sec:result_example2}

The problem definition and boundary conditions for Example 2 are given in Section \ref{sec:example2def}. This problem has a moving heat source that is made to test the proposed method for a heat source that varies in both space and time. Unless otherwise stated, the problems are solved using 30 nodes (3840 cores).

\subsubsection{Constant design} \label{sec:results_exp2_constdes}

\begin{figure}
    \centering
    \subfloat[Spatial representation]{
    \includegraphics[width=0.45\linewidth]{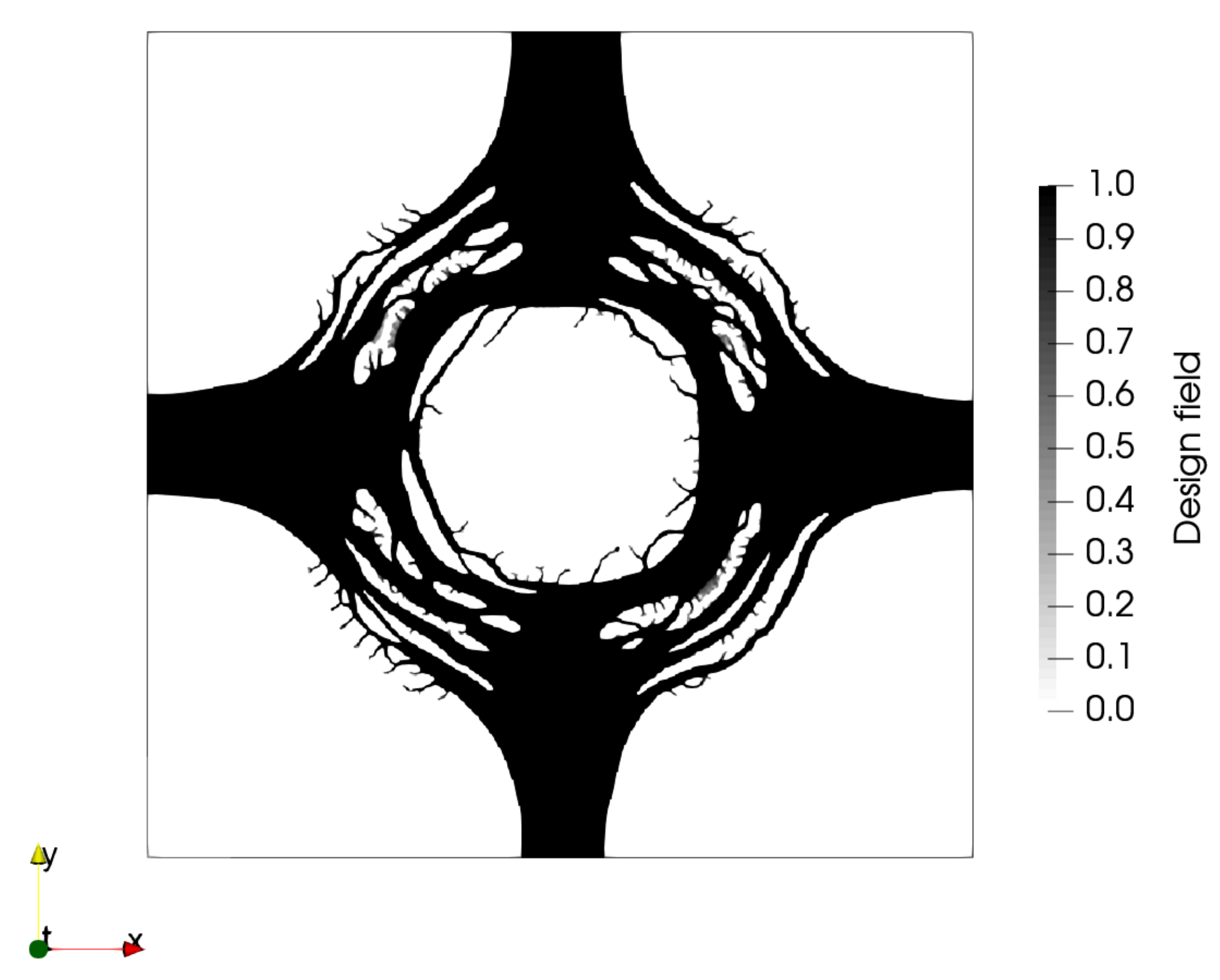}
    \label{fig:example2_constantDesign-a}
    }
    \subfloat[Space-time representation]{
    \includegraphics[width=0.52\linewidth]{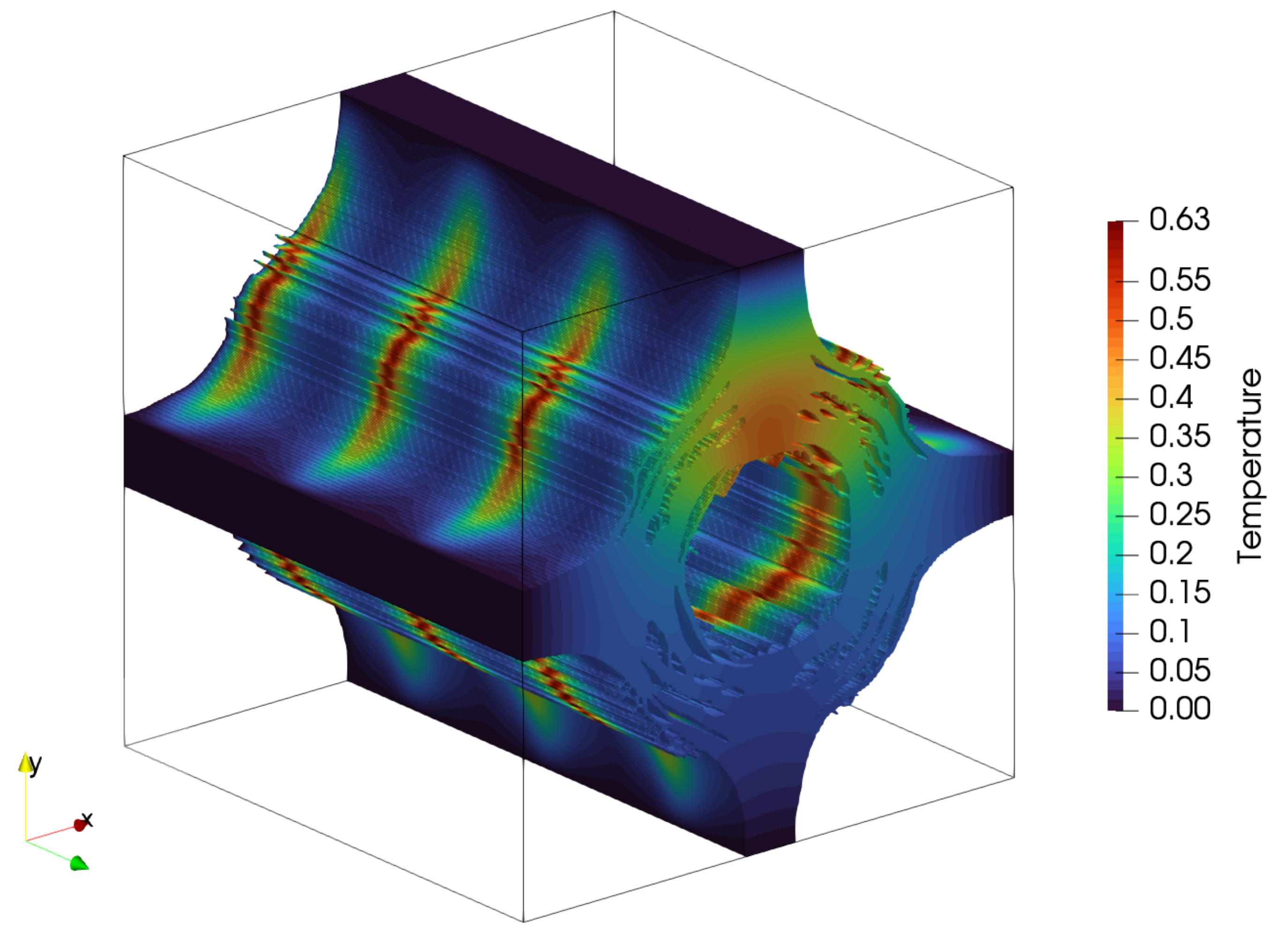}
    \label{fig:example2_constantDesign-b}
    }
    \\
    \subfloat[Time slice ($t=2.4$)]{
    \includegraphics[width=0.45\linewidth]{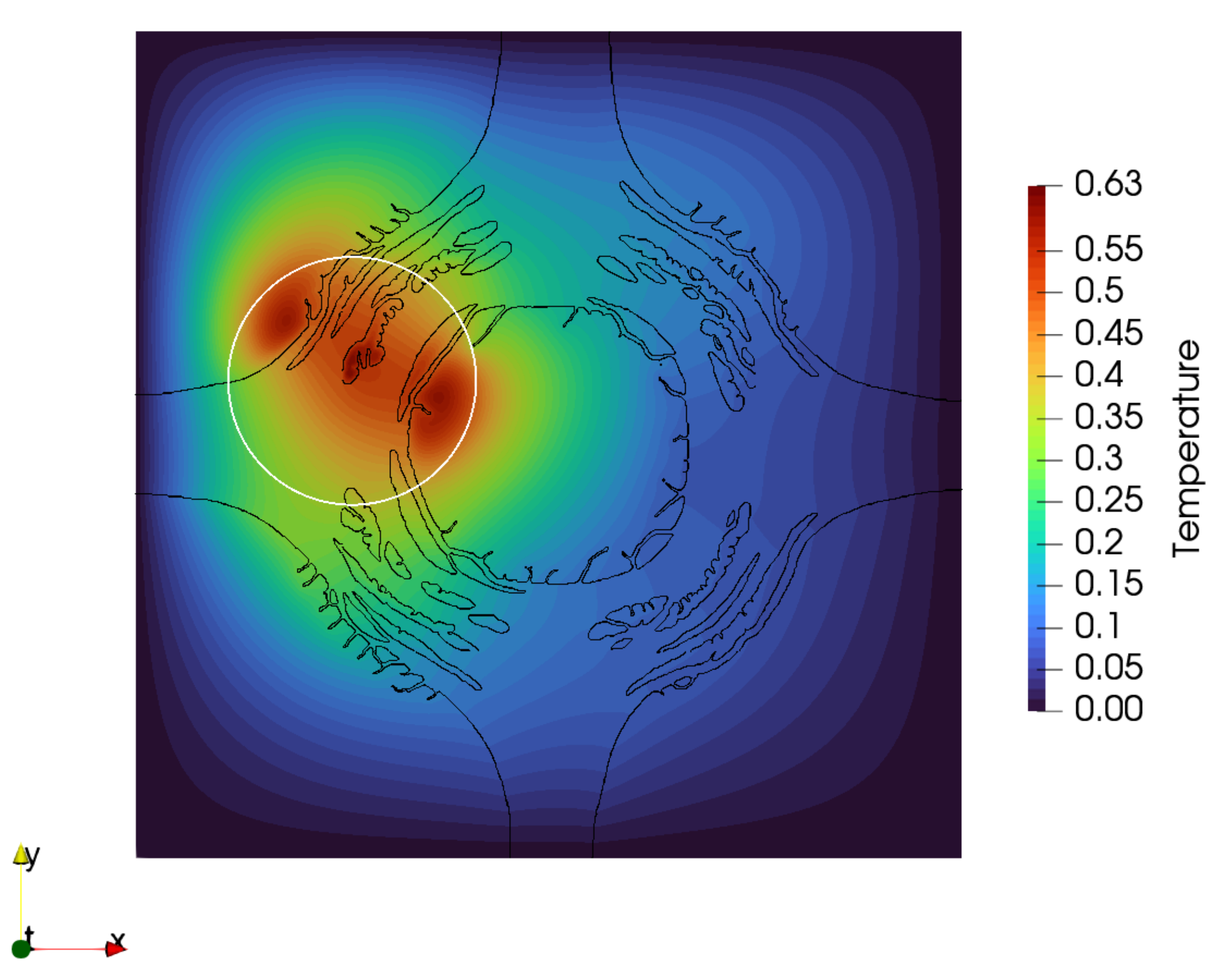}
    \label{fig:example2_constantDesign-c}
    }
    \hfill
    \subfloat[Time slice ($t=3.0$)]{
    \includegraphics[width=0.45\linewidth]{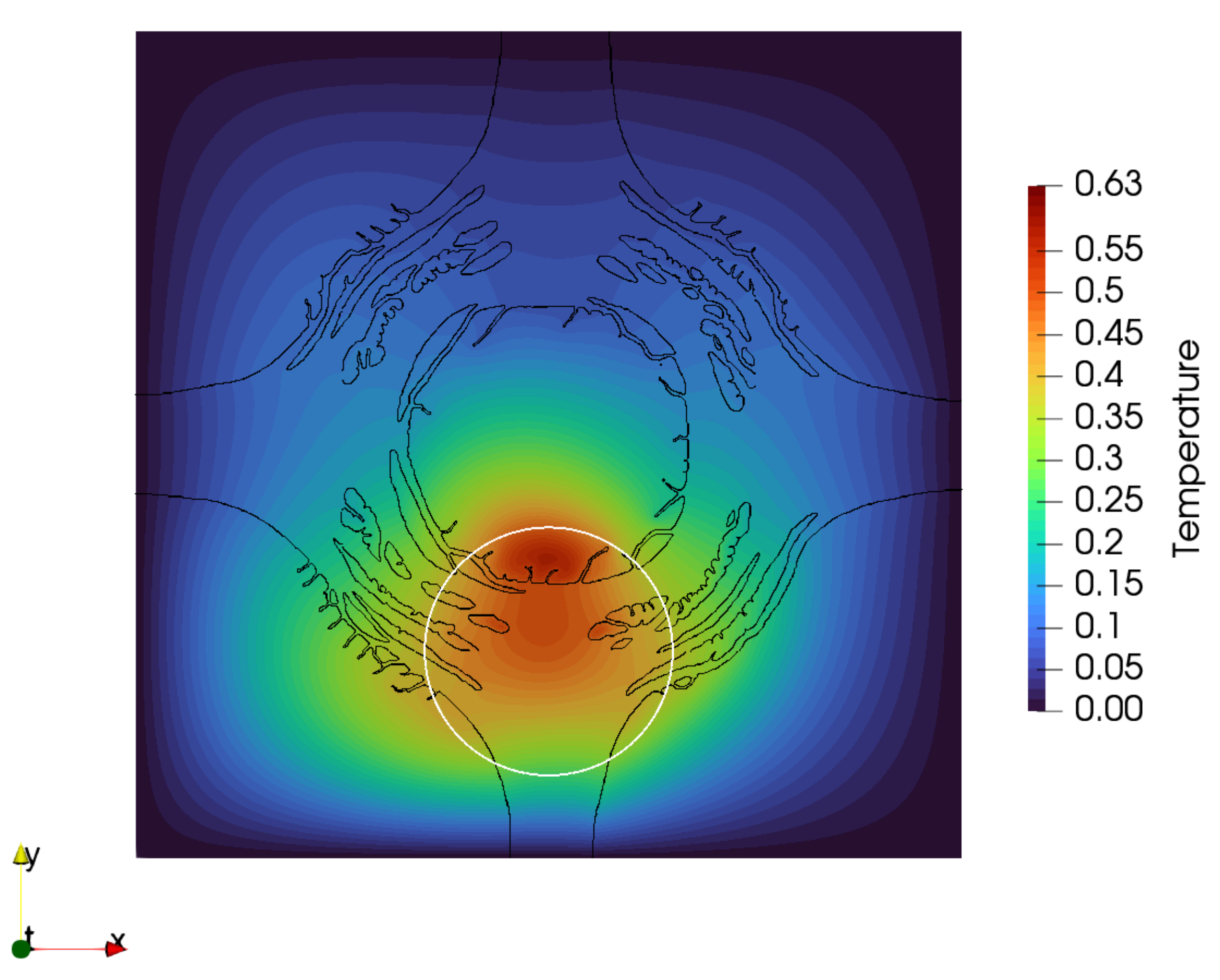}
    \label{fig:example2_constantDesign-d}
    }
    \caption{Optimised design for Example 2A: (a) two-dimensional spatial representation of design field with a bounding box showing the edges of the design domain; (b) (2+1)D space-time representation of design with temperature field and a bounding box showing the edges of the design domain; (c-d) temperature fields at two different times with black lines showing the contour of the design}
    \label{fig:example2_constantDesign}
\end{figure}
Figure \ref{fig:example2_constantDesign} shows the optimised design and temperature history with a time-constant design for Example 2A. It is seen that the optimised design has material placed underneath the circular movement pattern of the heat source, with four conductive pillars connecting to the outer cold walls when the distance is the shortest. Most of the conductive material is concentrated at the four regions closest to the walls and then material is saved in-between with a porous structure of conducting members of varying size.

\subsubsection{Time-varying design}  \label{sec:results_exp2_timevar}

To demonstrate the full capabilities of the proposed space-time approach, we will now allow the design field to fully vary in space-time for Examples 2(B-D). This will cause the coefficients of the governing equation to vary in both space and time, a significant test of the discretisation and the solver.
For space-time varying design fields, it is observed to be necessary to increase the regularisation in the time direction. When using the minimum filter radius in time, very fast changes in the time direction are seen. Therefore, for the following results we set $r_{t} = 0.3$ units of time, which does result in an increase in computational time for solving the filter equations. The volume fraction is decreased to $v_f = 0.1$ to give more clear and interesting results for the time-varying design case.

\begin{figure}
    \centering
    \subfloat[Space-time representation]{
    \includegraphics[width=0.55\linewidth]{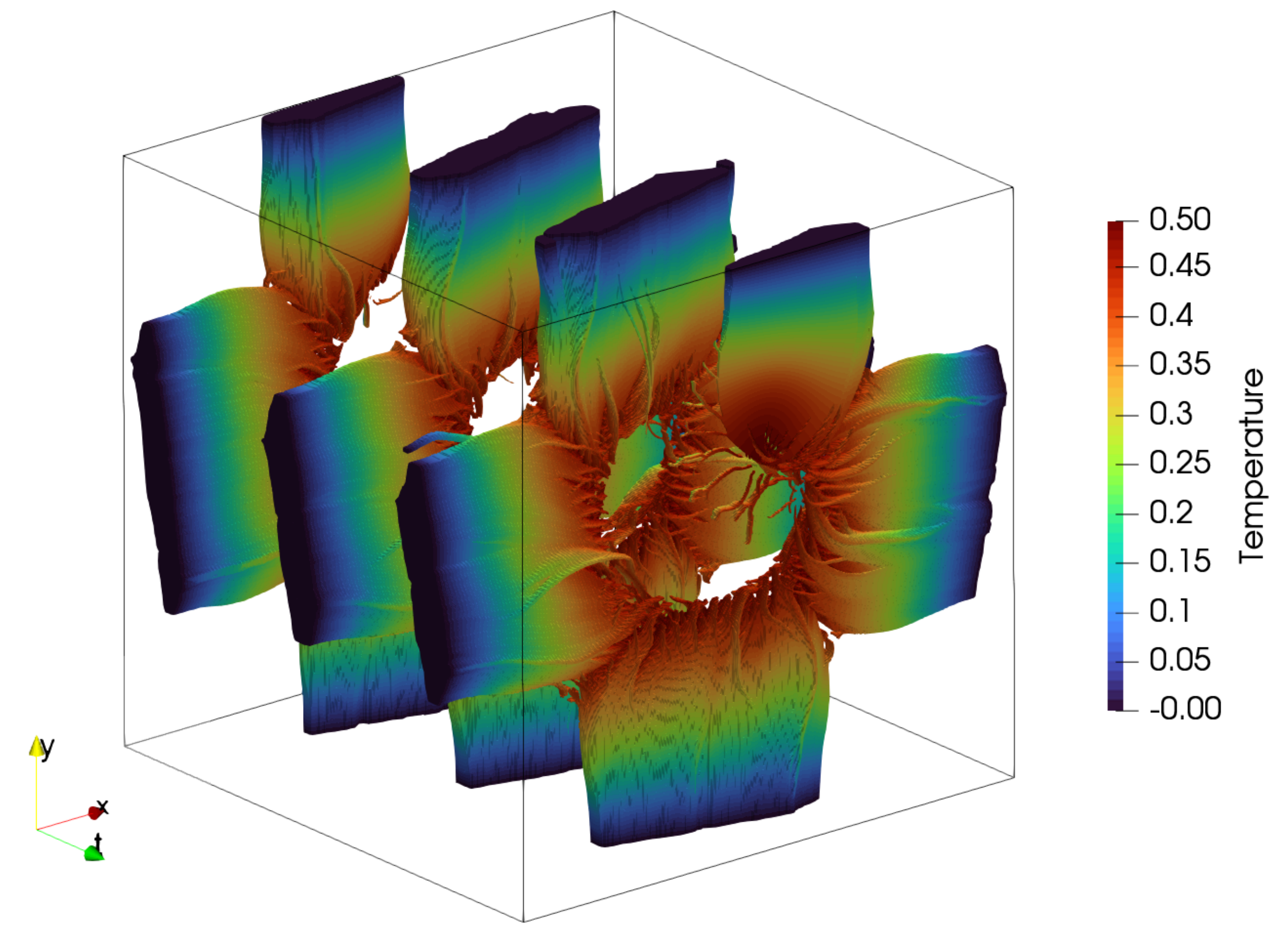}
    \label{fig:example2_varyingDesign_spacetimeDesign-a}
    }
    \hfill
    \subfloat[Zoomed-in view]{
    \includegraphics[width=0.4\linewidth]{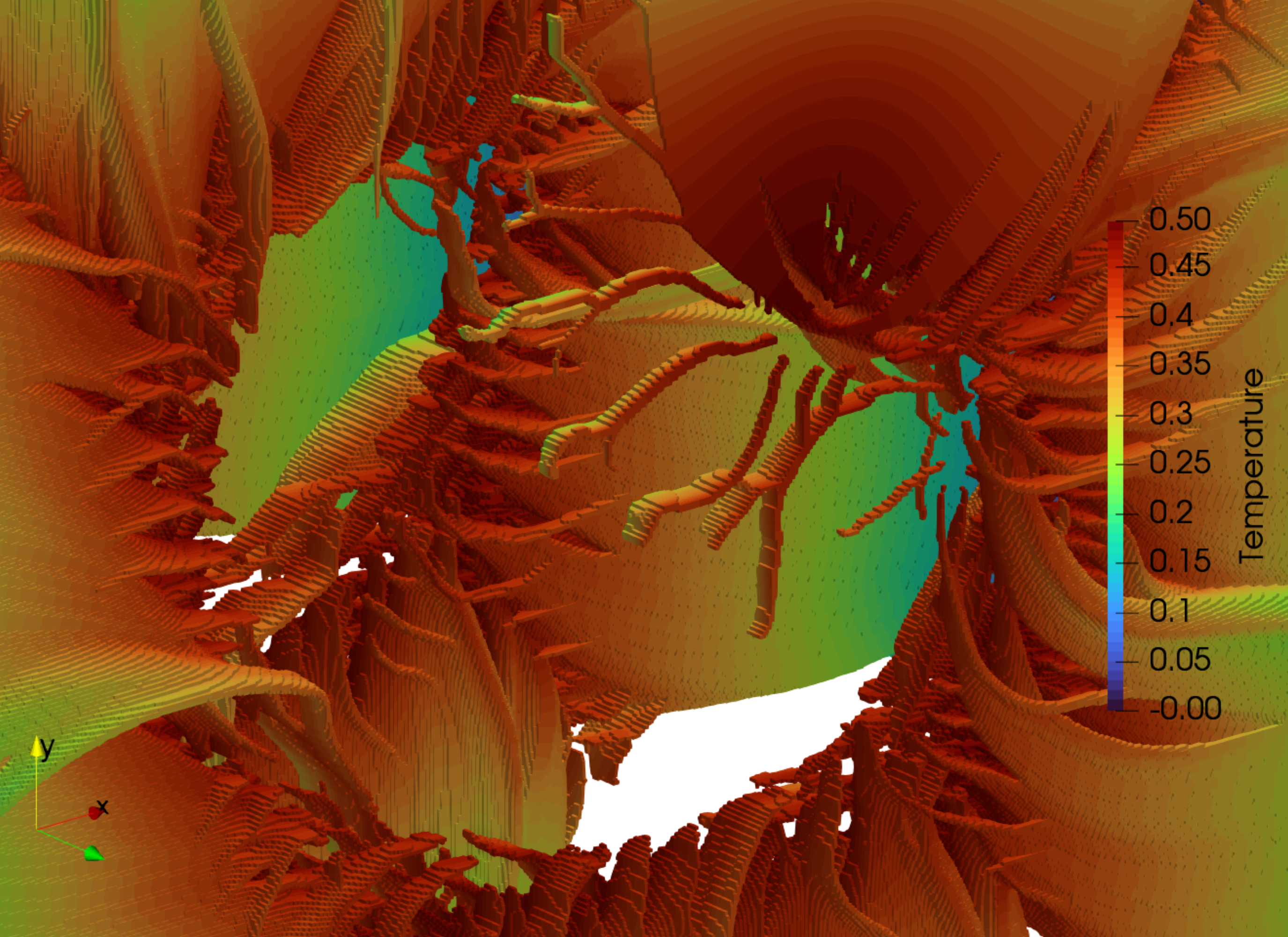}
    \label{fig:example2_varyingDesign_spacetimeDesign-b}
    }
    \caption{Optimised time-varying space-time design for Example 2B, with a black bounding box showing the design domain and coloured according to the temperature field: (a) overall view of entire space-time domain; (b) zoomed-in view showing fine scale details.}
    \label{fig:example2_varyingDesign_spacetimeDesign}
\end{figure}
Figure \ref{fig:example2_varyingDesign_spacetimeDesign} shows the optimised time-varying design for Example 2B. It can be seen from Figure \ref{fig:example2_varyingDesign_spacetimeDesign-a}, that the conductive material forms a spiral that follows the moving heat source and always connects to the nearest cold boundary. Despite the larger filter radius in time, small scale features can still be observed in Figure \ref{fig:example2_varyingDesign_spacetimeDesign-b} because the projection removes any strict length scale control, but they are not oscillatory as observed with a minimal filter radius.
\begin{figure}
    \centering
    \subfloat[$t = 2.75$]{
    \includegraphics[height=0.325\linewidth]{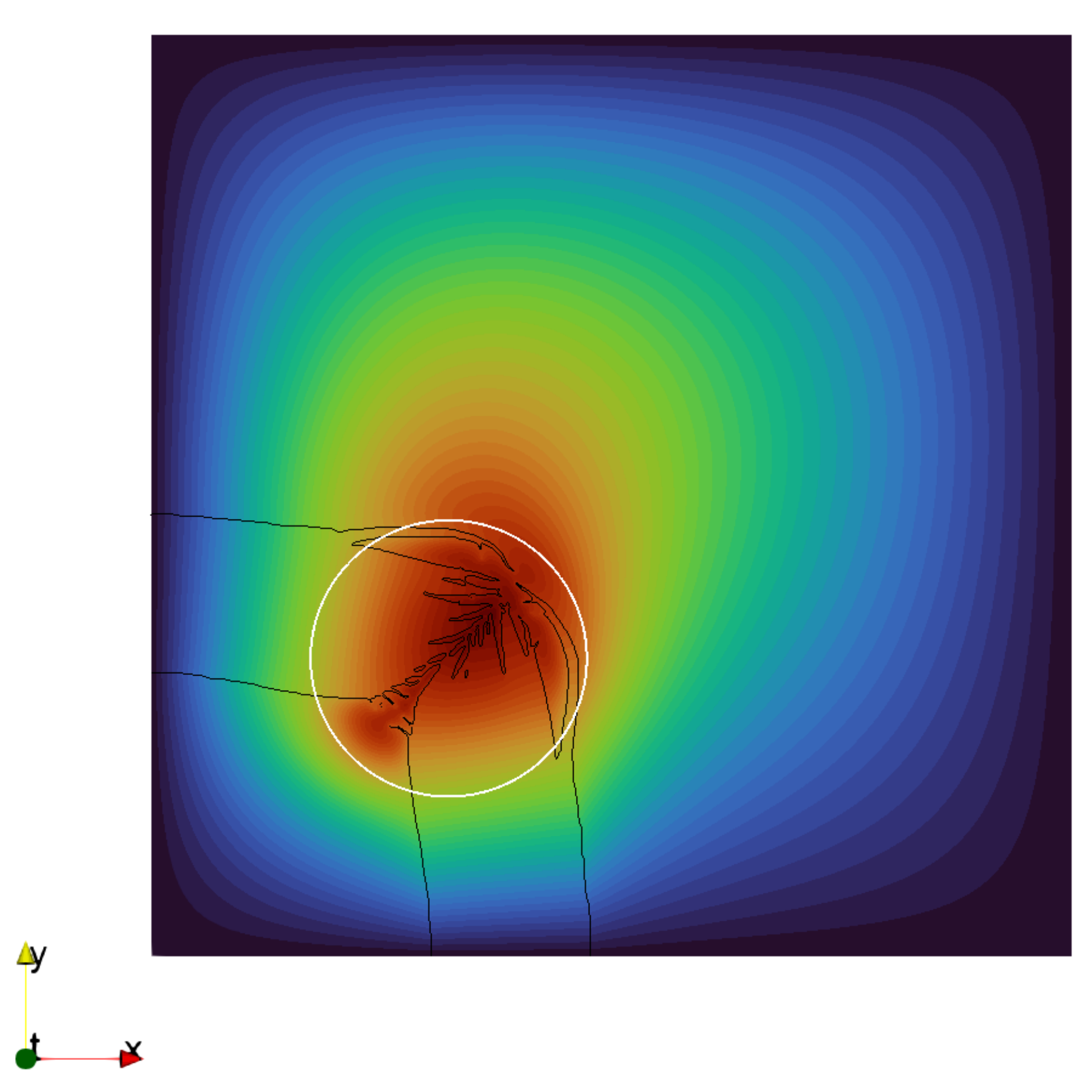}
    \label{fig:example2_varyingDesign_temp-a}
    }
    \subfloat[$t = 2.82$]{
    \includegraphics[height=0.325\linewidth]{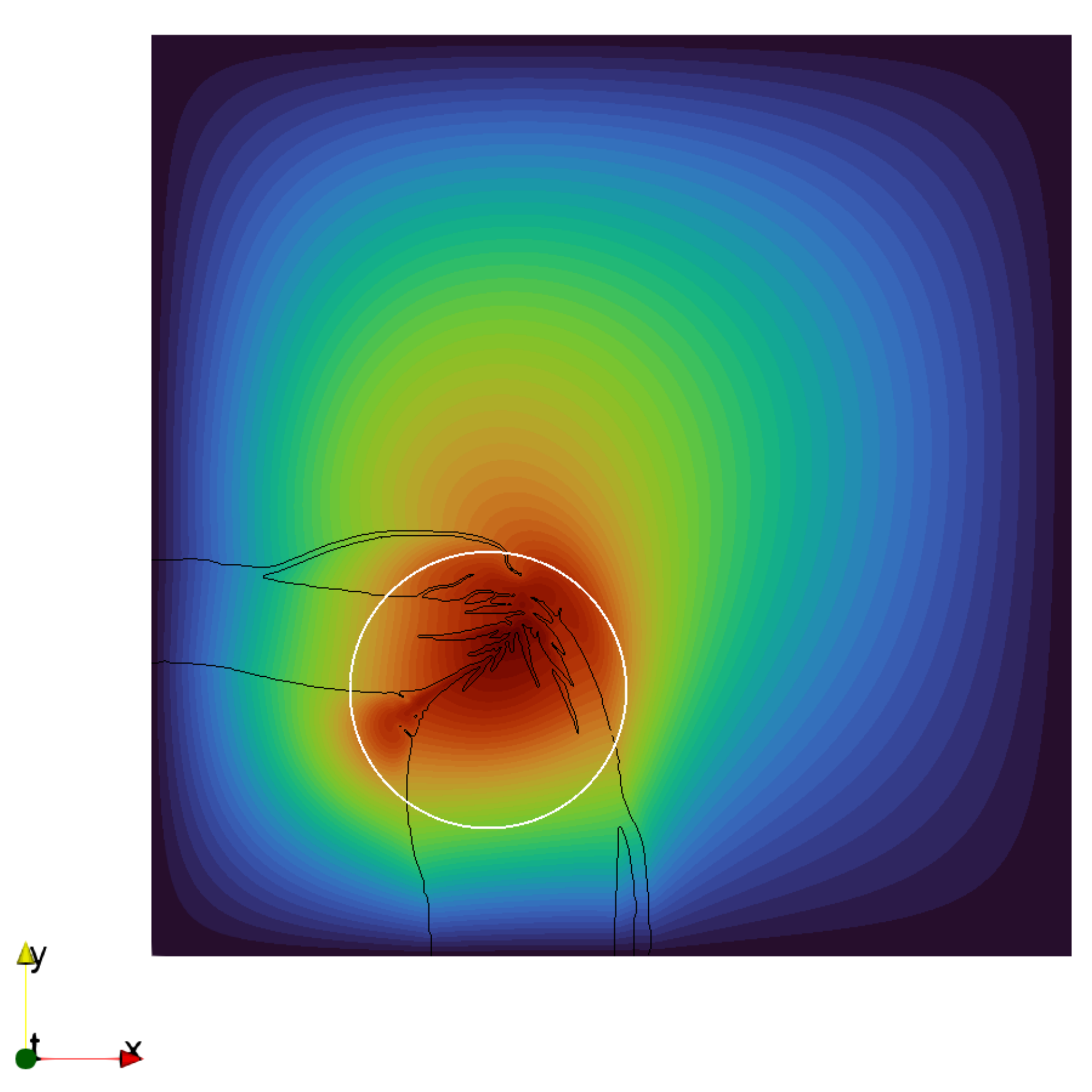}
    \label{fig:example2_varyingDesign_temp-b}
    }
    \subfloat[$t = 3.0$]{
    \includegraphics[height=0.325\linewidth]{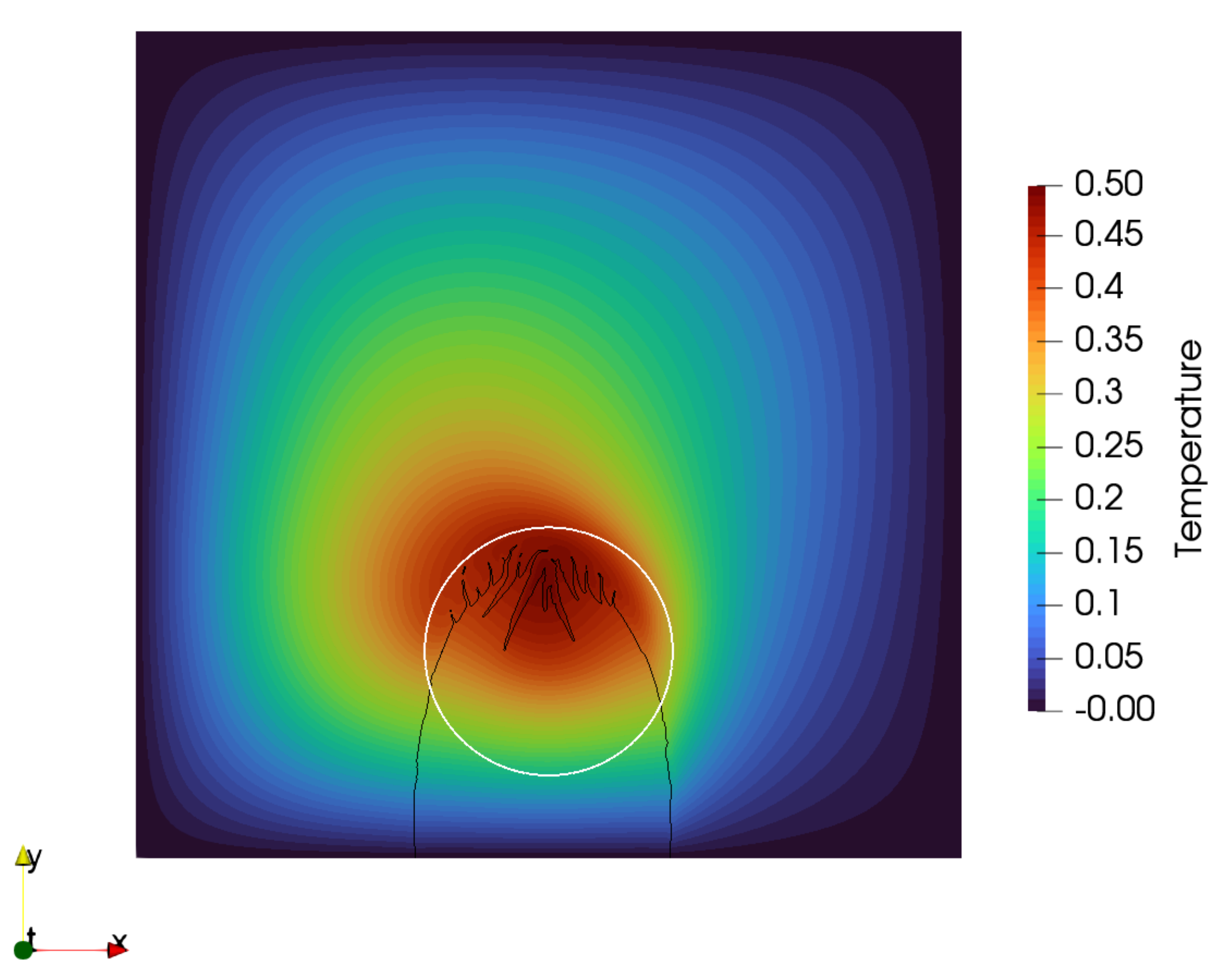}
    \label{fig:example2_varyingDesign_temp-c}
    }
    \caption{Time slices of the space-time domain to illustrate the movement of the conductive material for Example 2A. The colours show the temperature field, the black lines show the contours of the design, and the white circle shows the approximate location of the heat source.}
    \label{fig:example2_varyingDesign_temp}
\end{figure}
Figure \ref{fig:example2_varyingDesign_temp} further illustrates that the conductive material follows the moving heat source through time slices of the space-time domain. At the first time slice (Figure \ref{fig:example2_varyingDesign_temp-a}), the distance is the same to the left and bottom cold boundaries, so conductive material is placed to connect the heat source to both boundaries. As the heat source moves around the circle and comes closer to the bottom boundary, the material shifts (Figure \ref{fig:example2_varyingDesign_temp-b}) and finally concentrates to form a single connection to the bottom boundary (Figure \ref{fig:example2_varyingDesign_temp-c}).

To further show that the proposed framework can also handle other material parameters, the volumetric capacity is once again increased to $\mathcal{C}_\textrm{con} = 100 $ for Example 2C. It is not necessary to increase $p_\mathcal{C}$ as for Example 1C and it remains set to 2.
\begin{figure}
    \centering
    \subfloat[Space-time representation]{
    \includegraphics[width=0.55\linewidth]{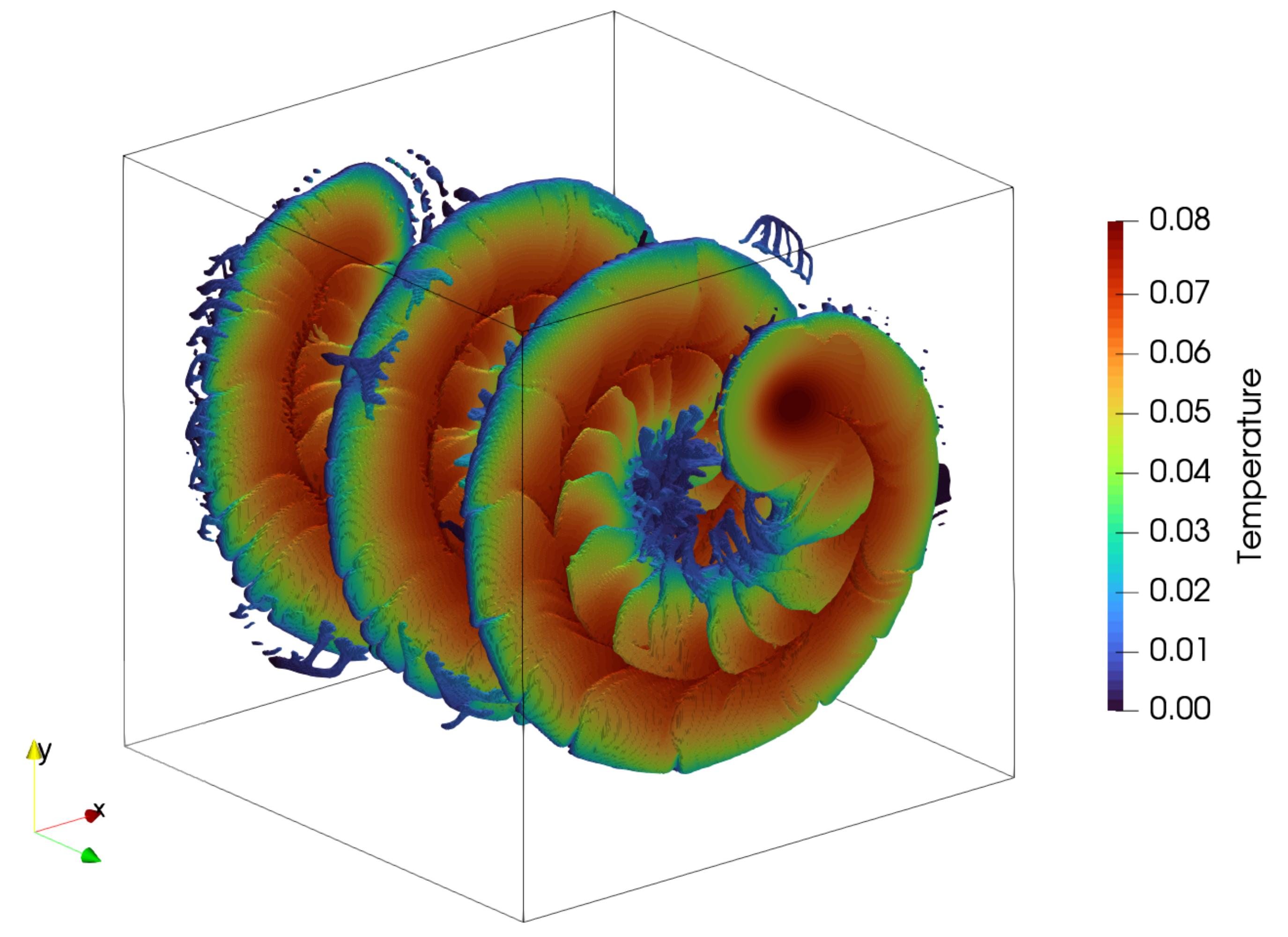}
    \label{fig:example2_highCap_spacetimeDesign-a}
    }
    \hfill
    \subfloat[Zoomed-in view]{
    \includegraphics[width=0.4\linewidth]{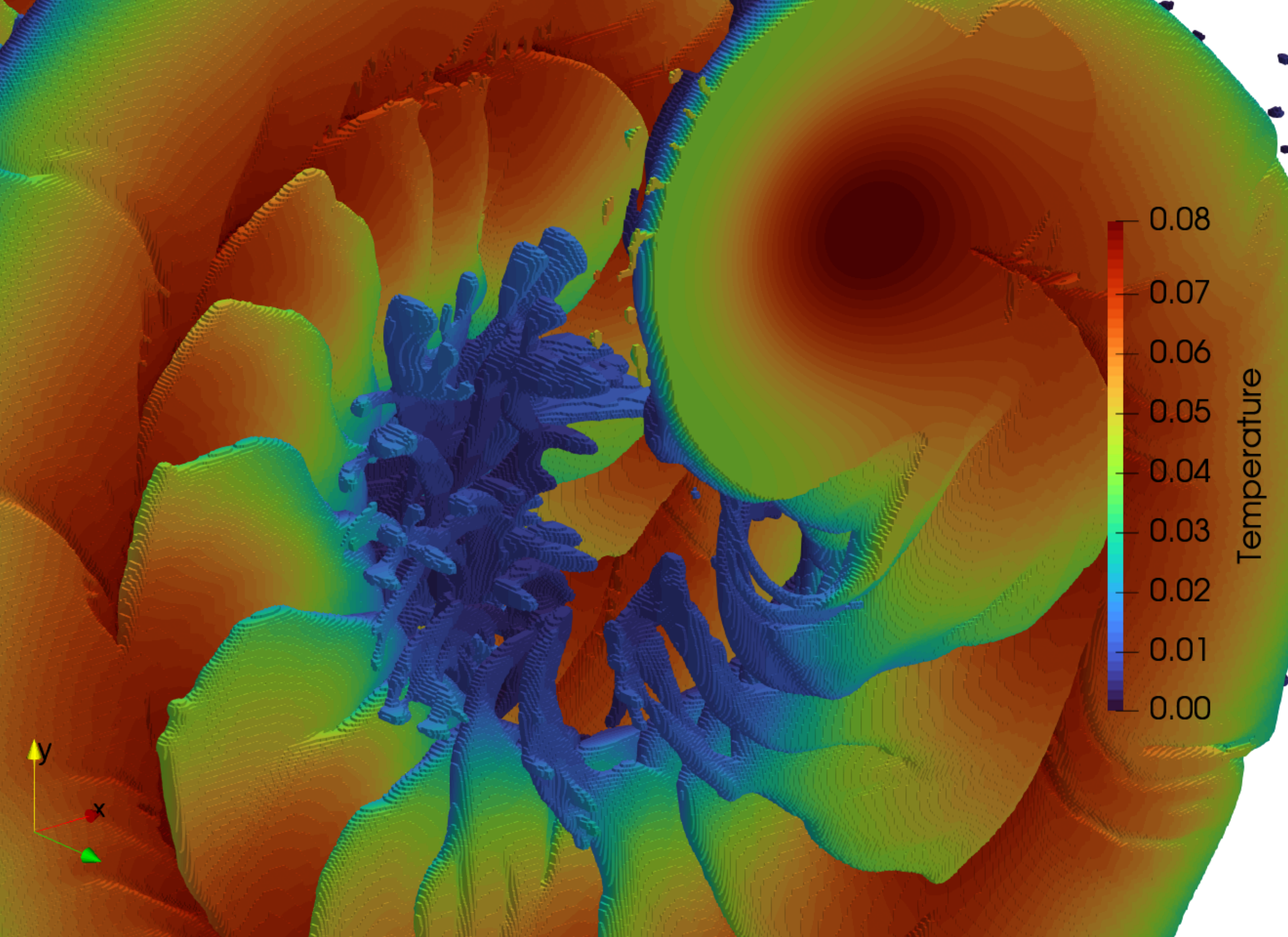}
    \label{fig:example2_highCap_spacetimeDesign-b}
    }
    \caption{Optimised time-varying space-time design for Example 2C, with a black bounding box showing the design domain and coloured according to the temperature field: (a) overall view of entire space-time domain; (b) zoomed in view showing fine scale details.}
    \label{fig:example2_highCap_spacetimeDesign}
\end{figure}
Figure \ref{fig:example2_highCap_spacetimeDesign} shows the optimised time-varying design for the higher capacity case. It can be seen that the capacitive material forms a spiral that follows the moving heat source, but it never connects to the cold boundaries - in contrast to Figure \ref{fig:example1_higherCap}. 
\begin{figure}
    \centering
    \subfloat[$t = 2.5$]{
    \includegraphics[height=0.325\linewidth]{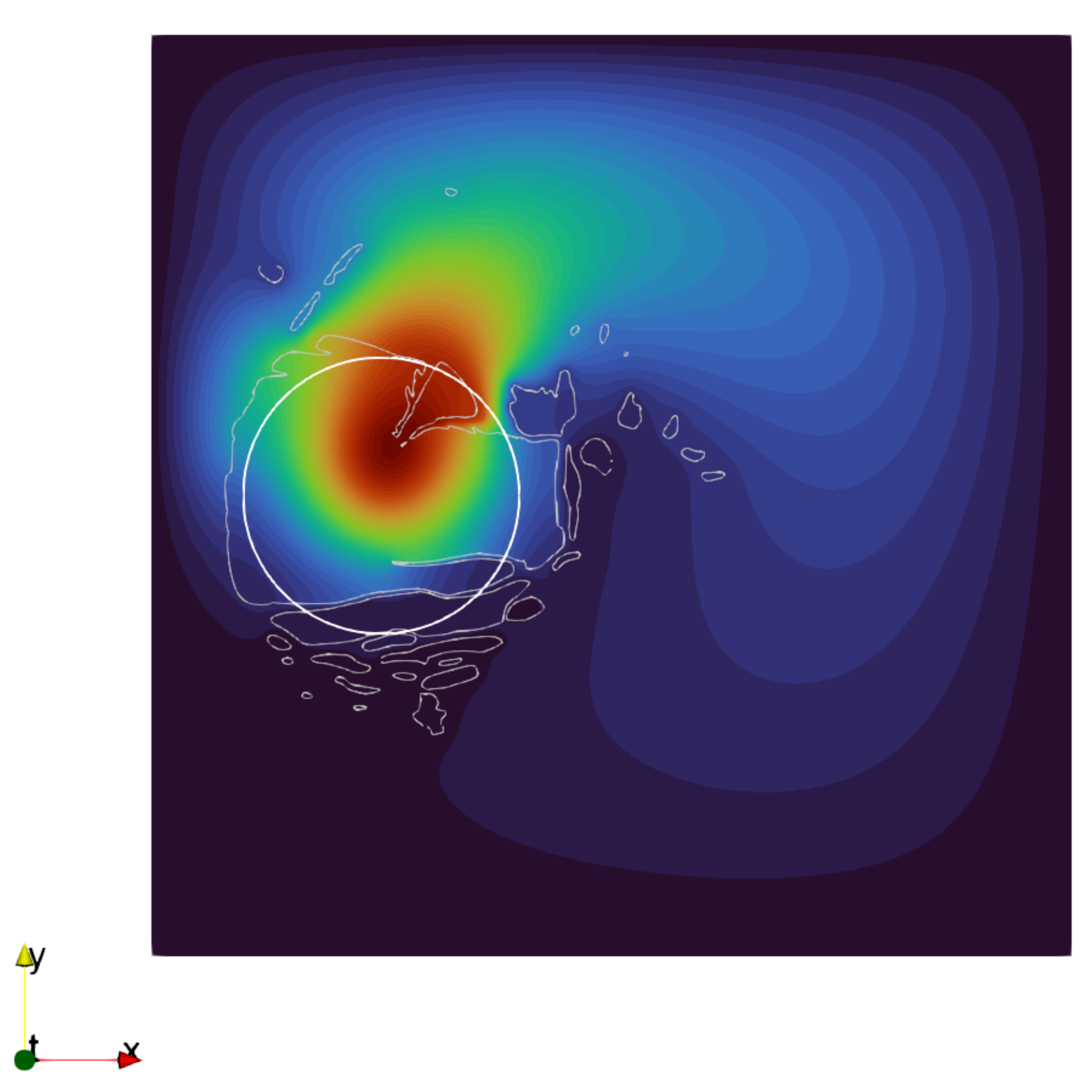}
    \label{fig:example2_highCap_temp-a}
    }
    \subfloat[$t = 2.75$]{
    \includegraphics[height=0.325\linewidth]{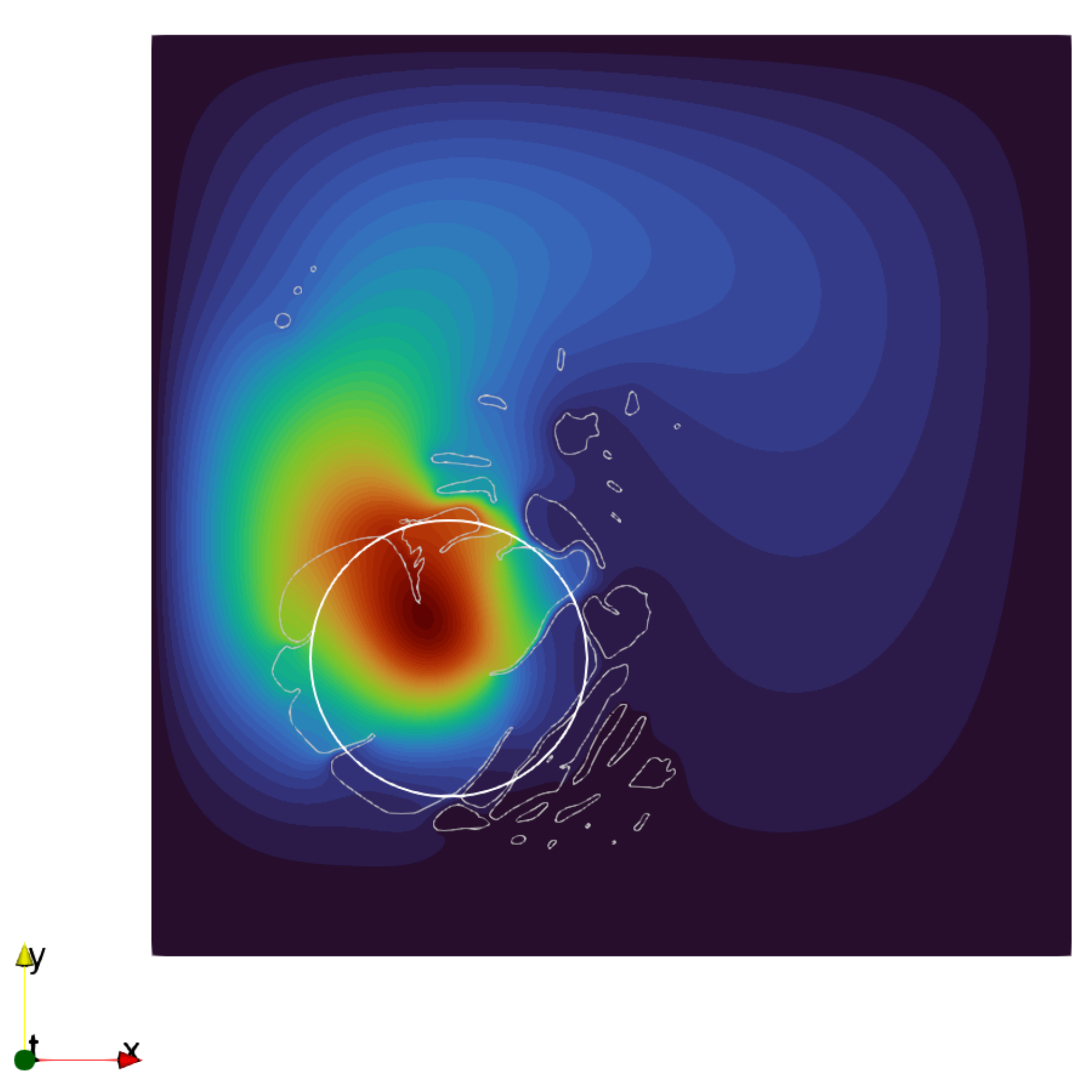}
    \label{fig:example2_highCap_temp-b}
    }
    \subfloat[$t = 3.0$]{
    \includegraphics[height=0.325\linewidth]{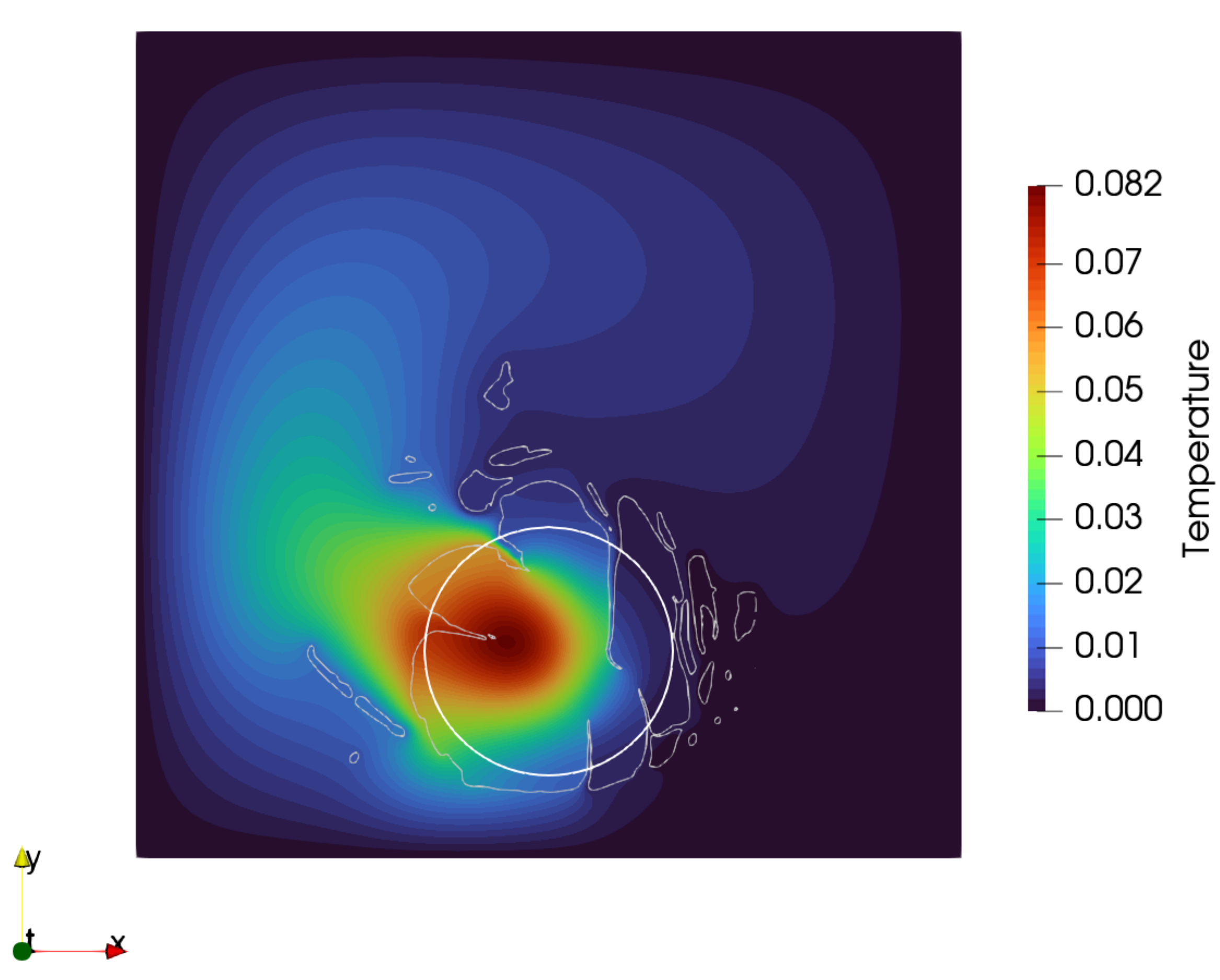}
    \label{fig:example2_highCap_temp-c}
    }
    \caption{Time slices of the space-time domain to illustrate the movement of the conductive material for Example 2C with a high volumetric capacity. The colours show the temperature field, the gray lines show the contours of the design, and the white circle shows the approximate location of the heat source.}
    \label{fig:example2_highCap_temp}
\end{figure}
Figure \ref{fig:example2_highCap_temp} shows three time slices, where it can be seen that the instantaneous design never connects to the cold boundaries and qualitatively resembles some form of weather formation moving around.
It appears that the temperature of the moving capacitative domain remains more or less constant over time, which is counter-intuitive as the temperature should be increasing since it cannot get rid of the energy, when not connecting to the cold boundaries. This was observed in Example 1D (Figure \ref{fig:example1_higherCap_tempHist}), where the temperature of the capacitive islands increases over time. The lack of increasing temperature over time is most likely due to energy conservation not being ensured for design fields changing over time, which would require to take the time-derivative of the volumetric capacity into account - which it is currently not. It should be highlighted that this is only an issue for time-varying design fields.

\begin{table}
    \centering
    \begin{tabular}{ccc}
        Level & Reference (2B) & High cap. (2C) \\ \hline
        0 & $640\times 640\times 1280$ & $640\times 640\times 1280$ \\
        1 & $320\times 320\times 1280$ & $320\times 320\times 1280$ \\
        2 & $160\times 160\times 1280$ & $160\times 160\times 1280$ \\
        3 & $80\times 80\times 1280$ & $80\times 80\times 1280$ \\
        4 & $40\times 40\times 1280$ & $40\times 40\times 1280$ \\
        5 & $20\times 20\times 1280$ & $40\times 40\times 640$ \\
        6 & $10\times 10\times 1280$ & $20\times 20\times 640$ \\
        7 & $10\times 10\times 640$ & $20\times 20\times 320$ \\
    \end{tabular}
    \caption{Multigrid hierarchies for the reference (Example 2B, Figure \ref{fig:example2_varyingDesign_spacetimeDesign}) and high capacity (Example 2C, Figure \ref{fig:example2_highCap_spacetimeDesign}) cases.}
    \label{tab:example2_multigrid_hierarchies}
\end{table}

\begin{table}
    \centering
    \begin{tabular}{ccccc}
         & \multicolumn{2}{c}{Avg. linear it.} &  & \\
        Problem & State & Adjoint & Avg. time per it. [s] & Total time [h:mm:s] \\ \hline
        Reference (2B) & 5.26 & 6.23 & 22.2 & 0:37:01 \\
        High capacity (2C) & 8.21 & 12.90 & 54.4 & 1:30:36 
    \end{tabular}
    \caption{Solver performance metrics for the two different material cases of Example 2(B,C).}
    \label{tab:example2_multigrid_performance}
\end{table}
Table \ref{tab:example2_multigrid_performance} shows chosen performance metrics for the two cases with the mesh hierarchies listed previously in Table \ref{tab:example2_multigrid_hierarchies}. It can be seen that the computational performance varies significantly between the two cases. The high capacity case requires $1.8\times$ the total linear iterations (state plus adjoint), where especially the adjoint problem adds to this increase by doubling. The time per iteration, and thus total time, increases by $2.4\times$, which is mostly due to the increase in linear iterations but additional factors must contribute to the last increase. This could, for example, be the size of the problem or the number of GMRES iterations on the coarsest grid.
\begin{figure}
    \centering
    \includegraphics[height=0.4\linewidth]{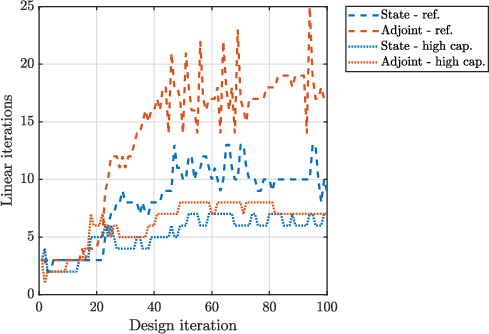}
    \caption{History of the number of state and adjoint linear iterations for the two different material cases of Example 2(B,C) over the optimisation iterations.}
    \label{fig:example2_comparison_linits}
\end{figure}
Figure \ref{fig:example2_comparison_linits} shows how the number of linear iterations changes over the optimisation process. It is clear that for the high capacity case, especially the adjoint solver is hard pressed. 
It is well known that multigrid using simple linear basis functions results in contrast-dependent behaviour and it is common to observe that the number of linear iterations increase over the design process \citep{Amir2014}, since the design field starts homogeneous and slowly becomes more heterogeneous and discrete with high contrast in the coefficients. This can be circumvented using a spectral coarse basis \citep{Lazarov2014,Alexandersen2015} and may be relevant to space-time problems with varying features in time such as the present problems.

\subsubsection{Very high resolution}  \label{sec:results_exp2_highres}

The space-time mesh is now further refined to $1280\times1280\times2560$, which yields a total of $4,\!194,\!304,\!000$ elements and $4,\!202,\!501,\!121$ DOFs. The design field is allowed to vary in space-time, thus, giving 4.19 billion design variables.

First, for Example 2D, the problem is solved with only the minimal filter radius. Using 200 nodes (25600 cores), the computational time is 41 minutes and 54 seconds (2514 seconds) excluding data export. Additional tests indicates that time-stepping on 4 nodes is the best case and will take approximately 6 hours, 53 minutes and 40 seconds ($24,\!820$ seconds). This yields a $9.1\times$ speed-up at a $5.5\times$ relative cost. Using 500 nodes (64000 cores) reduces the computational time to just 17 minutes and 15 seconds (1035 seconds), equivalent to $24.0\times$ speed-up at $5.2\times$ the relative cost. It should be noted that the computational time for the time-stepping approach is estimated using just 10 time steps for 10 design iterations and scaled accordingly, exactly as in Section \ref{sec:result_scaling}, which yields a highly favourable time estimate and equivalently conservative speed-up estimate, given that more nodes are needed to store 2560 time steps in practise.

Secondly, for Example 2E, the problem is solved with a filter radius in the time-direction of $r_{t} = 0.0166$ units of time. This increases the computational time on 500 nodes (64000 cores) to 32 minutes, which is mostly due to an increased number of linear iterations to solve the filter equation due to increased anisotropy.
\begin{figure}
    \centering
    \subfloat[Space-time representation]{
    \includegraphics[width=0.9\linewidth]{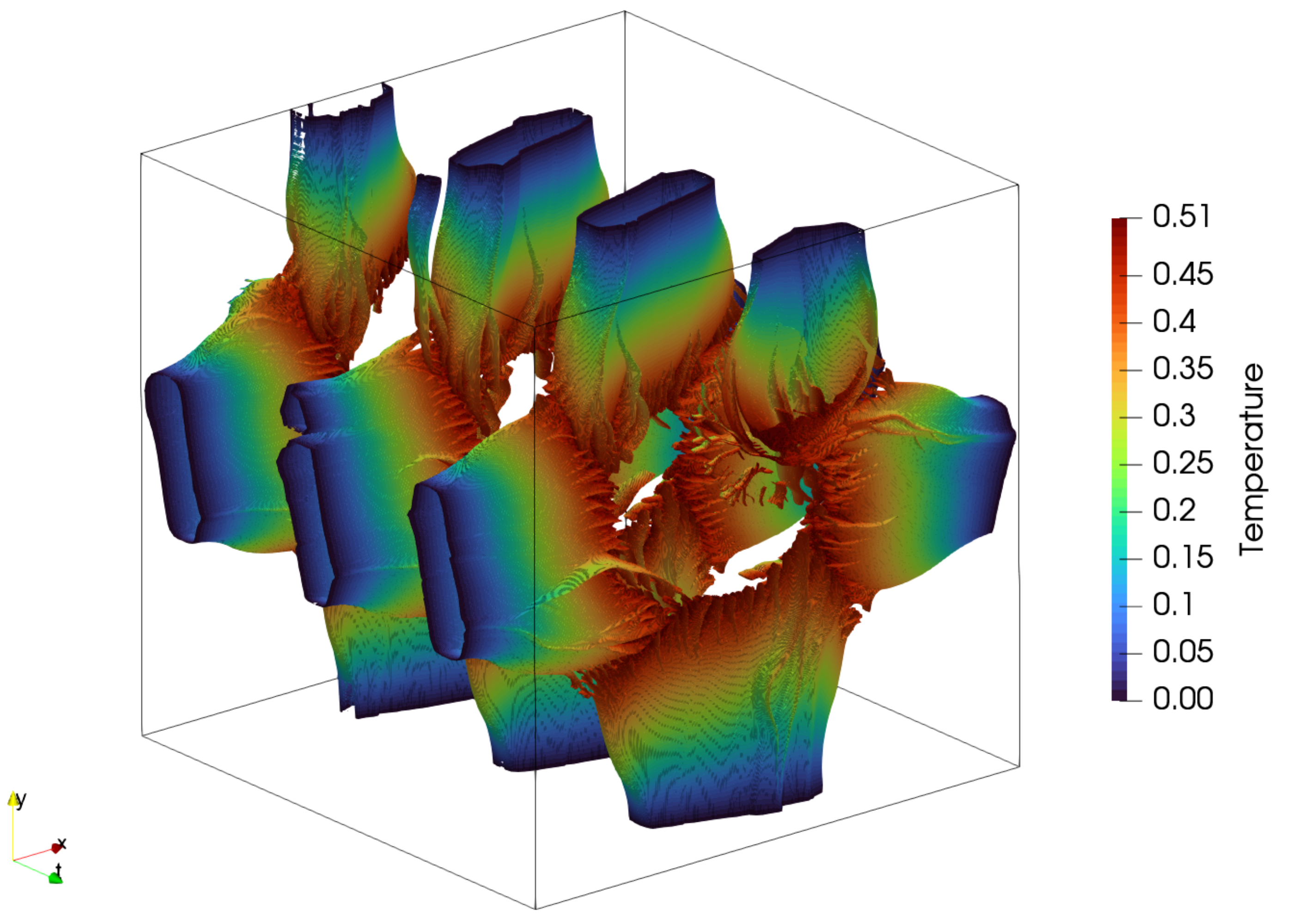}
    \label{fig:example2_highRes_spacetimeDesign-a}
    }
    \\
    \subfloat[Zoomed-in view]{
    \includegraphics[width=0.75\linewidth]{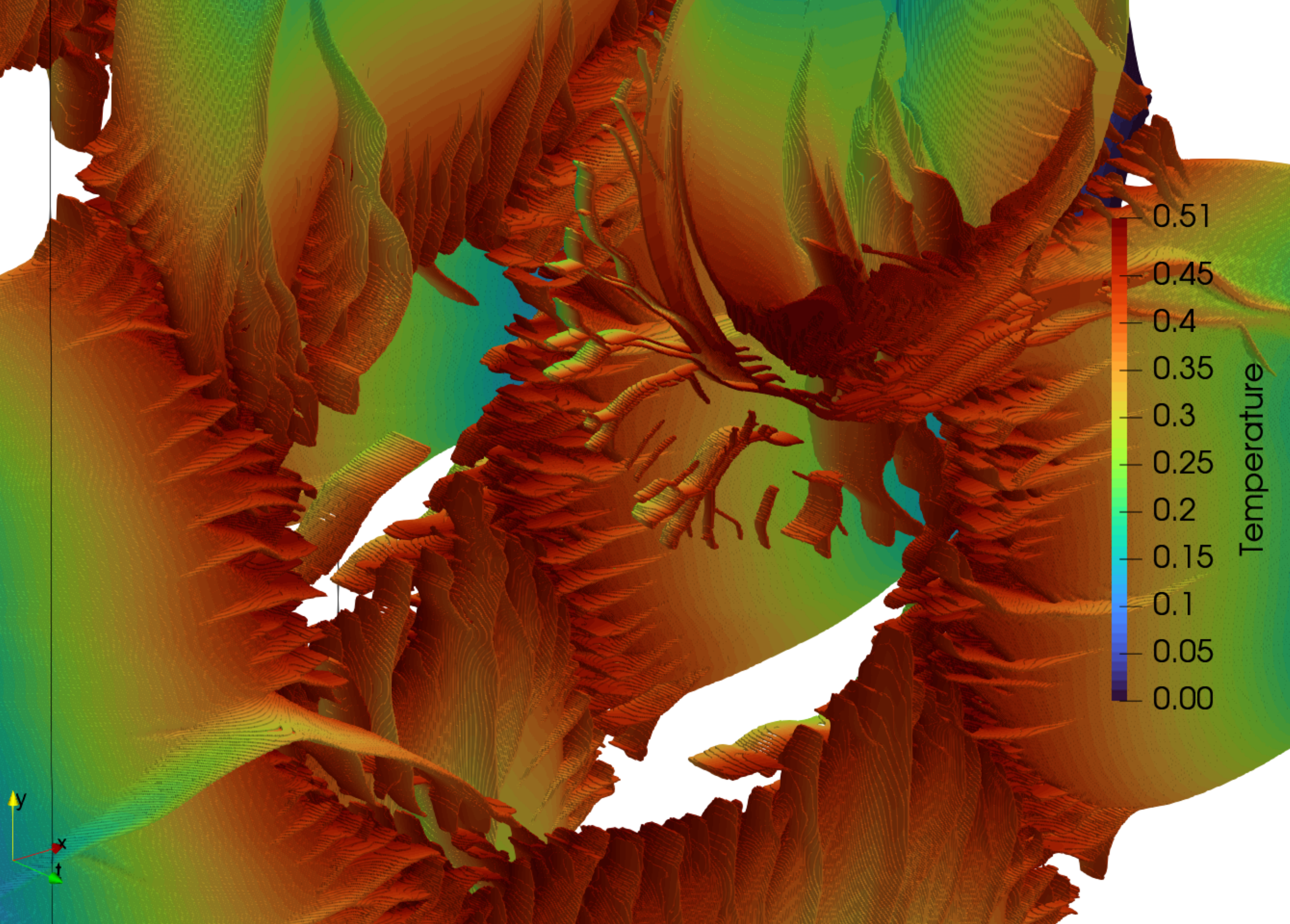}
    \label{fig:example2_highRes_spacetimeDesign-b}
    }
    \caption{Optimised time-varying space-time design for Example 2E, with a black bounding box showing the design domain and coloured according to the temperature field: (a) view of entire space-time domain; (b) zoomed in view showing fine scale details. It should be noted that only a ``shell'' is shown with the elements at the surface, due to visualisation issues if all elements are included.}
    \label{fig:example2_highRes_spacetimeDesign}
\end{figure}
Figure \ref{fig:example2_highRes_spacetimeDesign} shows the optimised space-time design for Example 2E. It bears an overall resemblance with Example 2B in Figure \ref{fig:example2_varyingDesign_spacetimeDesign}, but with even finer and thinner details. 
It should be noted that only a ``shell'' is shown with the elements at the surface, due to visualisation issues if all elements are included. Generally, visualisation becomes a real issue for high resolution problems due to the shear amount of data, as also noted by \citet{Aage2017}.

\begin{figure}
    \centering
    \subfloat[Whole time slice]{
    \includegraphics[width=0.75\linewidth]{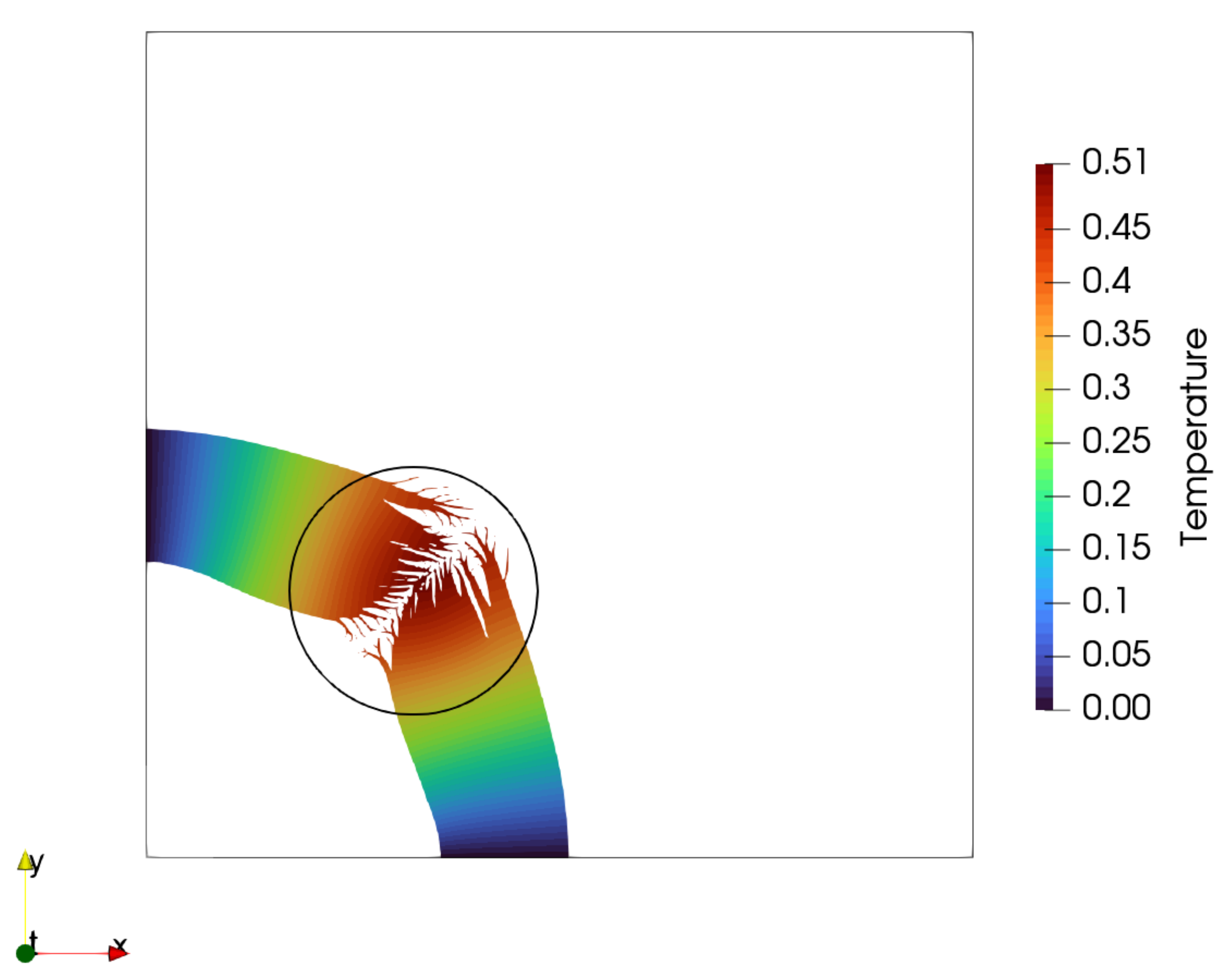}
    \label{fig:example2_highRes_temp-a}
    }
    \hfill
    \subfloat[Zoomed-in view]{
    \includegraphics[width=0.4\linewidth]{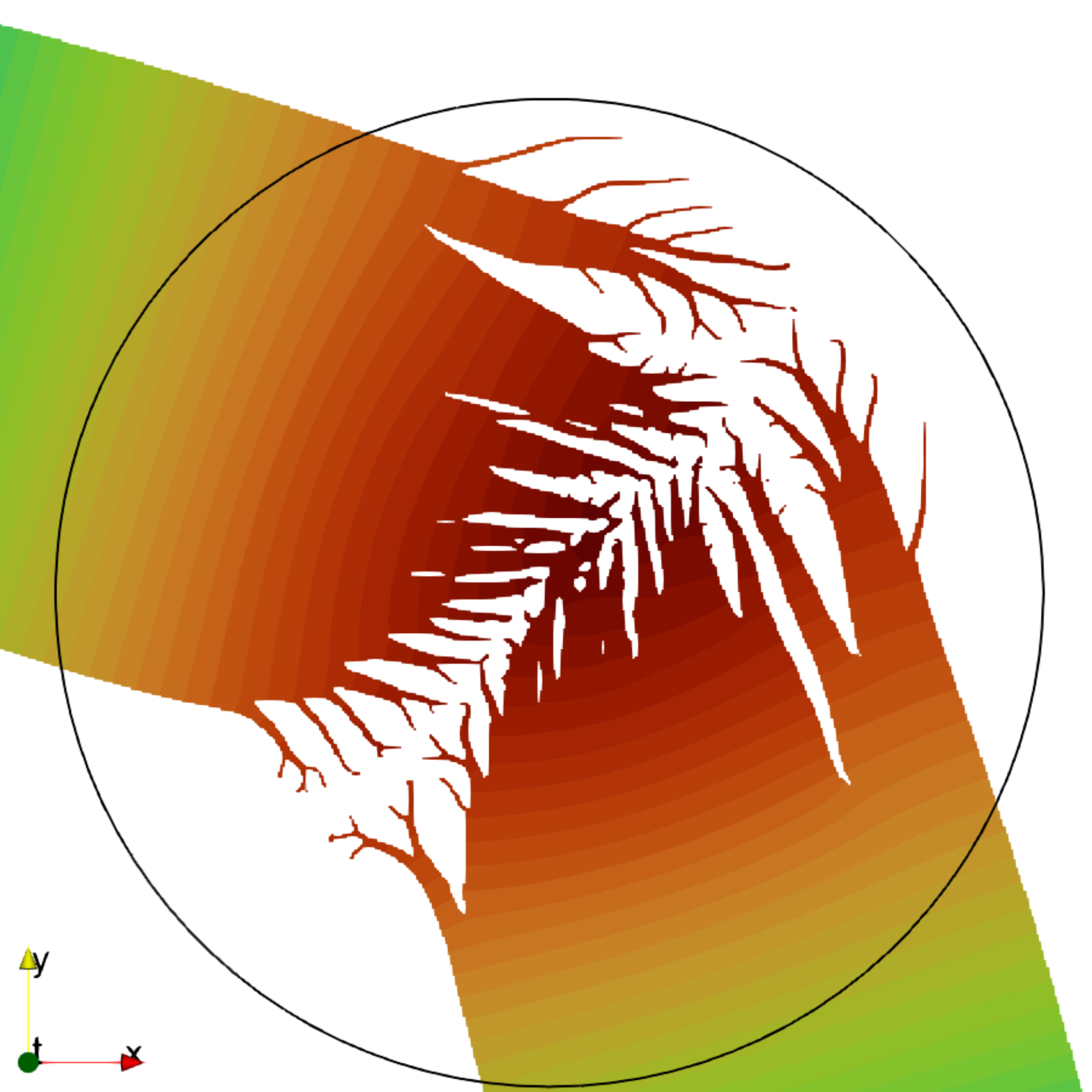}
    \label{fig:example2_highRes_temp-b}
    }
    \hfill
    \subfloat[Finite elements]{
    \includegraphics[width=0.4\linewidth]{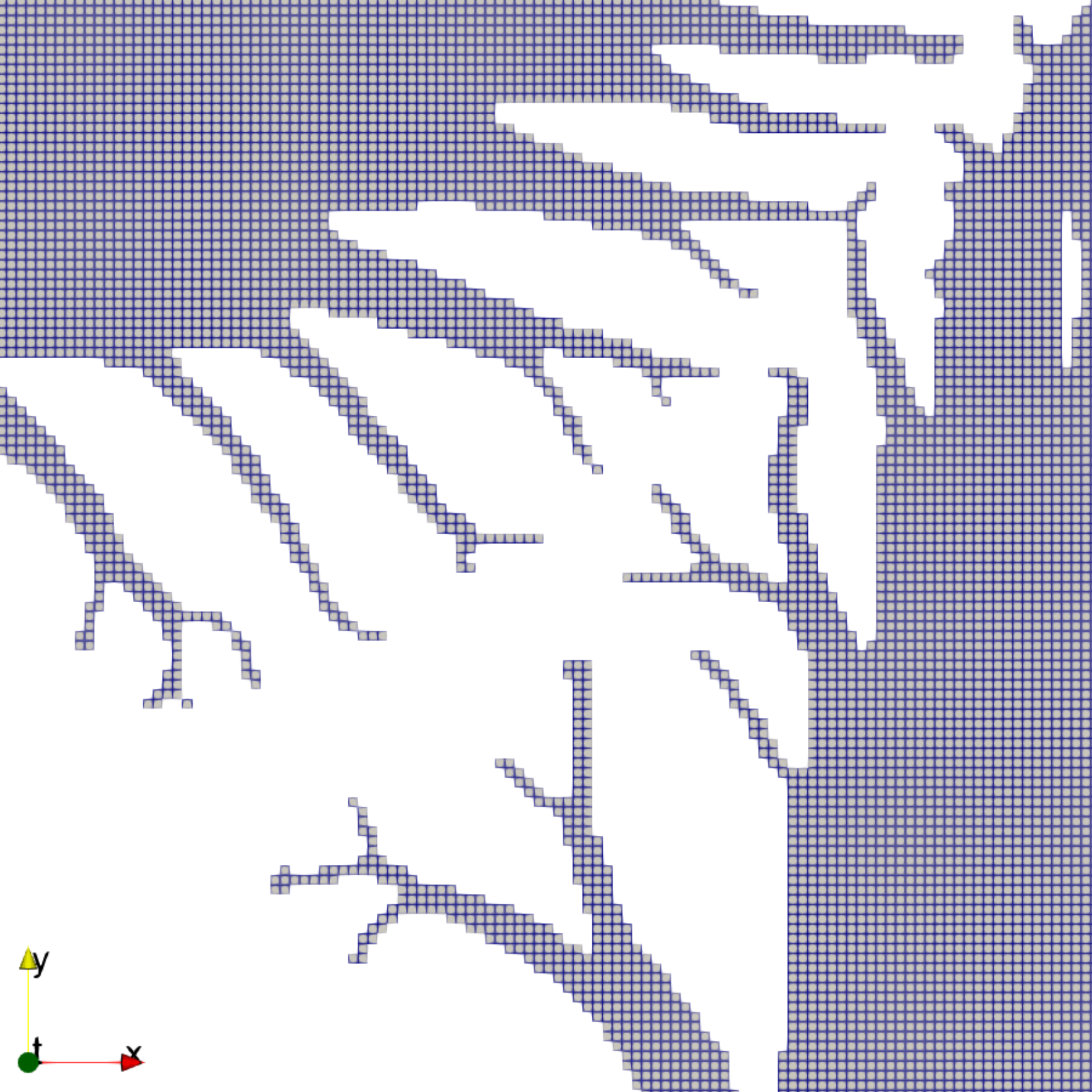}
    \label{fig:example2_highRes_temp-c}
    }
    \caption{Time slice at $t=2.75$ of the thresholded space-time design to show the detail level for Example 2D: (a) the whole time slice showing only the design; (b) zoomed-in view; (c) further zoomed-in view showing the finite elements. For (a-b) the colours show the temperature field in the design and the black circle shows the approximate location of the heat source.}
    \label{fig:example2_highRes_temp}
\end{figure}
Figure \ref{fig:example2_highRes_temp} shows a time slice of the thresholded space-time design. As before, the design resembles that of Example 2B shown in Figure \ref{fig:example2_varyingDesign_temp-a}, but with finer features. Figure \ref{fig:example2_highRes_temp-b} shows a zoomed-in view of the fine-scale features of the instantaneous design under the heat source. Figure \ref{fig:example2_highRes_temp-c} gives a further zoomed-in view of the fine-scale details, showing the size of the finite elements. It can clearly be seen that conductive features at a size of 1-2 elements exists (due to the loss of strict length scale enforcement due to projection), which for this case is at a size of 7.8-15.6 thousandths of the domain width.
Using Equation \ref{eq:diffusionTimescale}, the timescale at this lengthscale is around $8\times 10^{-5}$, which is actually quite a bit smaller than the element size in the time direction, $\Delta t = 2.3\times 10^{-3}$. However, most of the conduction does not happen across the branches, but along them and the element size is probably sufficient in practise.

\section{Discussion and conclusions} \label{sec:discussion}

The presented results of the proposed space-time approach show great promise to significantly reduce the time-to-solution for topology optimisation of time-dependent thermal conduction problems. Given that many physical problems can be described using a transient diffusion equation similar to thermal conduction, it is expected that the methodology can be extended to these relatively easily, e.g. mass diffusion, electrical charge diffusion, and porous media flow.

From the scaling studies in Section \ref{sec:result_scaling}, we observe that the traditional time-stepping approach is faster for a relatively low amount of cores (below 1000), but that it scales very poorly beyond a single node due to the two-dimensional problem being rather small. This is the main reason behind parallelising also in time, since the spatial problem alone is simply too small to efficiently use domain decomposition and parallel programming.

When switching to the proposed parallel space-time approach, we observe a $50\times$ speed-up on 400 nodes compared to using 8 nodes, the minimum currently required to store everything in memory. The extreme memory requirements are partly due to the very large systems of equations needed to be stored in memory, as well as the storage of the forward and adjoint solutions simultaneously to compute sensitivities. However, the current code is not optimised for reduced memory consumption and several auxiliary elemental and nodal fields are currently stored for development purposes, that may not be needed in a final production deployment of the code.

Comparing the two methods in Section \ref{sec:result_comparison}, we observe that when using the proposed space-time approach, we can achieve significant speed-ups relative to the reference time-stepping method. But this speed-up does come at an increased computational cost, specifically an increase in the relative number of core-hours used, a common ``currency'' on supercomputers and cloud computing platforms. For instance, we observed a $25.2\times$ speed-up at $4.8\times$ the cost and a $52.1\times$ speed-up at $7.7\times$ the cost.
It is worth mentioning that the computational cost in core-hours can be translated to energy consumption of running the job. Specific estimates are available for each job on LUMI, but we will restrict ourselves to highlighting that a $4.8\times$ increase in computational cost is roughly equal to a $4.8\times$ increase in energy consumption.

If we imagine the method to be used in an industrial context, low cost is key. However, ``time is money'' is also a common adage. While the increased computational cost can be directly translated to monetary cost on most systems, we still argue that the additional cost is well worth the significant decrease in time-to-solution to increase development speed. In Section \ref{sec:results_exp1_finermesh}, it is estimated that a time-stepping solution would take around 15 hours, while at $2.4\times$ the cost the time-to-solution can be reduced to just 50 minutes. Although it can be argued that 15 hours is probably tolerable in an industrial context, imagine extrapolating this to more complicated problems where an optimisation may ordinarily take multiple days. Bear in mind though, that the scalability and performance are currently only observed for the particular problems treated in this article and any changes to the problems or physics may change these conclusions.

When increasing the resolution further to 4.2 billion DOFs, we observed that the computational time was as low as 17 minutes on 500 nodes providing an approximate $24.0\times$ speed-up at $5.2\times$ the cost. This shows that the proposed space-time methodology and parallel implementation clearly allows us to solve very large (2+1)D problems in an incredibly short time - potentially leading to significant reductions in development time when matured for industrial use-cases over the coming years.

One could perhaps argue that the problems treated herein are over-resolved in time and that this can be part of providing favourable scaling \citep{Gotschel2021}. For Example 1, the period/wavelength of the oscillating heat source is $\approx 0.13$ and the element size in the time direction is $\frac{1}{1280} \approx 7.8 \times 10^{-4}$, giving $\sim161$ elements per period - which could be argued as being over-resolved. However, if one looks at the Fourier number at the element level (non-dimensional ``time step'' relative to the diffusive timescale), $Fo_{e} = \frac{k\Delta t}{\mathcal{C}{\Delta x}^2}$, the mesh may be too coarse. Although implicit time schemes, as treated herein, are unconditionally stable, the accuracy is still dependent on the time step size. It can be recommended that the element Fourier number should be less than 0.1 \citep{Myers1978}, implying an element size in the time direction of no more than $8.14\times10^{-8}$ in our case - meaning that the problem is significantly under-resolved. Further complexity is added to this discussion by the varying material properties in space (and potentially space-time), as well as the non-standard time scheme produced by the continuous Galerkin finite element method.

From the literature survey, it was observed that surprisingly few references discuss the computational time when treating topology optimisation of time-dependent problems - despite long computational times being an inherent issue. In our opinion, all future papers on topology optimisation of time-dependent systems must list the computational time at a bare minimum, preferably also discuss the computational cost involved and how it may be reduced. Of course many things influence the actual computational time, such as chosen methods, specific implementation, code optimisation, and so forth, so a direct comparisons across methods may be difficult. However, listing the computational time will still give a realistic expectation for readers, users and industry.

\subsection{Limitations}

The current framework is limited to rectangular domains and regular meshes. This allows for simple and efficient implementation of the geometric multigrid approach using semi-coarsening. However, obviously this limits the problems that the current framework can treat without modifications. One strength of STFE is the ability to locally refine in space-time, essentially allowing different ``time-step'' sizes in different parts of the domain based on the local dynamics of the solution. This ability is currently lost out on.

When treating time-dependent design fields varying in space-time, we believe that the current implementation has some issues with energy conservation in the time direction. Given that the heat capacity can change in time, discontinuously for a point in space from one time slab to the next, an additional term accounting for the time gradient of the capacity should be included.
This is mostly important if extending the method to simulate and optimise manufacturing processes like additive manufacturing \citep{Wang2020}.
The current framework and the presented results are limited to transient heat conduction. Linear diffusion equations are famously simple to solve and, thus, the presented speed-ups may not carry over to more complicated problems, like non-linear problems, or other physics, like fluid flow or dynamic mechanics.

\subsection{Future work}

Future work will be to thoroughly investigate and understand the impact of the many multigrid and domain decomposition parameters (number of levels, mesh size on coarsest grid, partitioning and load-balancing across all levels, etc.). To improve the current methodology and implementation, it should be investigated whether there exists better smoothers for the multigrid levels, as well as a better preconditioner for the solver on the coarsest grid. For large problems and large numbers of cores (beyond 512 nodes), we cannot rely on the coarsest grid being small enough for a Jacobi-preconditoned GMRES to converge satisfactorily. Lastly, since memory usage is a significant limitation of the current implementation, less memory-hungry methods such as matrix-free methods can be explored.

\section*{Acknowledgements}
The first author wishes to thank Professor Stefan Vandewalle for his support and input, as well as the rest of the NUMA group at KU Leuven for their hospitality during two visits in 2024 while working on the presented work.

The work was funded by Independent Research Fund Denmark (DFF) through a Sapere Aude Research Leader grant (3123-00020B) for the COMFORT project (COmputational Morphogenesis FOR Time-dependent problems).
We acknowledge the Danish e-Infrastructure Consortium (DeiC), Denmark, for awarding this project access to the LUMI supercomputer, owned by the EuroHPC Joint Undertaking, hosted by CSC (Finland) and the LUMI consortium through: DeiC-SDU-N5-2023036 and DeiC-SDU-N5-2025156.
Initial testing took place on the Sophia cluster through the DeiC resource allocations: DeiC-SDU-L-5 and DeiC-SDU-L5-9.

\section*{CRediT authorship contribution statement}
\textbf{Joe Alexandersen:} Conceptualisation, data curation, formal analysis, funding acquisition, investigation, methodology, project administration, resources, software, supervision, visualisation, writing - original draft, writing - review and editing. 
\textbf{Magnus Appel:} Methodology, software, writing - original draft, writing - review and editing. 

\section*{Declaration of competing interests}
We declare that we have no known competing financial interests or personal relationships that could have influenced the work reported in this paper. Any opinions are solely those of the authors and not the funding bodies.

\section*{Data statement}
We believe that the methodology is sufficiently described in this paper for reproducibility. Readers are always welcome to contact the authors for further elaboration and discussion. The code will be made publicly available at a later stage, when it is fully mature.

\bibliographystyle{elsarticle-num-names} 
\bibliography{main}

\begin{thebibliography}{102}
\expandafter\ifx\csname natexlab\endcsname\relax\def\natexlab#1{#1}\fi
\providecommand{\url}[1]{\texttt{#1}}
\providecommand{\href}[2]{#2}
\providecommand{\path}[1]{#1}
\providecommand{\DOIprefix}{doi:}
\providecommand{\ArXivprefix}{arXiv:}
\providecommand{\URLprefix}{URL: }
\providecommand{\Pubmedprefix}{pmid:}
\providecommand{\doi}[1]{\href{http://dx.doi.org/#1}{\path{#1}}}
\providecommand{\Pubmed}[1]{\href{pmid:#1}{\path{#1}}}
\providecommand{\bibinfo}[2]{#2}
\ifx\xfnm\relax \def\xfnm[#1]{\unskip,\space#1}\fi
\bibitem[{Bendsøe and Kikuchi(1988)}]{Bendsoee1988}
\bibinfo{author}{M.~P. Bendsøe}, \bibinfo{author}{N.~Kikuchi},
\newblock \bibinfo{title}{Generating optimal topologies in structural design using a homogenization method},
\newblock \bibinfo{journal}{Computer Methods in Applied Mechanics and Engineering} \bibinfo{volume}{71} (\bibinfo{year}{1988}) \bibinfo{pages}{197--224}. \DOIprefix\doi{https://doi.org/10.1016/0045-7825(88)90086-2}.
\bibitem[{Bendsøe and Sigmund(2004)}]{Bendsoee2004}
\bibinfo{author}{M.~P. Bendsøe}, \bibinfo{author}{O.~Sigmund}, \bibinfo{title}{Topology Optimization: Theory, Methods, and Applications}, \bibinfo{edition}{2} ed., \bibinfo{publisher}{Springer-Verlag Berlin Heidelberg}, \bibinfo{year}{2004}. \DOIprefix\doi{10.1007/978-3-662-05086-6}.
\bibitem[{Deaton and Grandhi(2014)}]{Deaton2014}
\bibinfo{author}{J.~D. Deaton}, \bibinfo{author}{R.~V. Grandhi},
\newblock \bibinfo{title}{A survey of structural and multidisciplinary continuum topology optimization: post 2000},
\newblock \bibinfo{journal}{Structural and Multidisciplinary Optimization} \bibinfo{volume}{49} (\bibinfo{year}{2014}) \bibinfo{pages}{1--38}. \DOIprefix\doi{10.1007/s00158-013-0956-z}.
\bibitem[{Amdahl(1967)}]{Amdahl1967}
\bibinfo{author}{G.~M. Amdahl},
\newblock \bibinfo{title}{Validity of the single processor approach to achieving large scale computing capabilities},
\newblock in: \bibinfo{booktitle}{Proceedings of the April 18-20, 1967, Spring Joint Computer Conference}, AFIPS '67 (Spring), \bibinfo{publisher}{Association for Computing Machinery}, \bibinfo{year}{1967}, p. \bibinfo{pages}{483–485}. \DOIprefix\doi{10.1145/1465482.1465560}.
\bibitem[{Mukherjee et~al.(2021)Mukherjee, Lu, Raghavan, Breitkopf, Dutta, Xiao, and Zhang}]{Mukherjee2021}
\bibinfo{author}{S.~Mukherjee}, \bibinfo{author}{D.~Lu}, \bibinfo{author}{B.~Raghavan}, \bibinfo{author}{P.~Breitkopf}, \bibinfo{author}{S.~Dutta}, \bibinfo{author}{M.~Xiao}, \bibinfo{author}{W.~Zhang},
\newblock \bibinfo{title}{Accelerating large-scale topology optimization: state-of-the-art and challenges},
\newblock \bibinfo{journal}{Archives of Computational Methods in Engineering}  (\bibinfo{year}{2021}) \bibinfo{pages}{1--23}. \DOIprefix\doi{10.1007/s11831-021-09544-3}.
\bibitem[{Borrvall and Petersson(2001)}]{Borrvall2001}
\bibinfo{author}{T.~Borrvall}, \bibinfo{author}{J.~Petersson},
\newblock \bibinfo{title}{Large-scale topology optimization in 3d using parallel computing},
\newblock \bibinfo{journal}{Computer Methods in Applied Mechanics and Engineering} \bibinfo{volume}{190} (\bibinfo{year}{2001}) \bibinfo{pages}{6201--6229}. \DOIprefix\doi{10.1016/S0045-7825(01)00216-X}.
\bibitem[{Evgrafov et~al.(2008)Evgrafov, Rupp, Maute, and Dunn}]{Evgrafov2008}
\bibinfo{author}{A.~Evgrafov}, \bibinfo{author}{C.~J. Rupp}, \bibinfo{author}{K.~Maute}, \bibinfo{author}{M.~L. Dunn},
\newblock \bibinfo{title}{Large-scale parallel topology optimization using a dual-primal substructuring solver},
\newblock \bibinfo{journal}{Structural and Multidisciplinary Optimization} \bibinfo{volume}{36} (\bibinfo{year}{2008}) \bibinfo{pages}{329--345}. \DOIprefix\doi{10.1007/s00158-007-0190-7}.
\bibitem[{Aage and Lazarov(2013)}]{Aage2013}
\bibinfo{author}{N.~Aage}, \bibinfo{author}{B.~S. Lazarov},
\newblock \bibinfo{title}{Parallel framework for topology optimization using the method of moving asymptotes},
\newblock \bibinfo{journal}{Structural and Multidisciplinary Optimization} \bibinfo{volume}{47} (\bibinfo{year}{2013}) \bibinfo{pages}{493--505}. \DOIprefix\doi{10.1007/s00158-012-0869-2}.
\bibitem[{Amir et~al.(2014)Amir, Aage, and Lazarov}]{Amir2014}
\bibinfo{author}{O.~Amir}, \bibinfo{author}{N.~Aage}, \bibinfo{author}{B.~S. Lazarov},
\newblock \bibinfo{title}{On {multigrid-CG} for efficient topology optimization},
\newblock \bibinfo{journal}{Structural and Multidisciplinary Optimization} \bibinfo{volume}{49} (\bibinfo{year}{2014}) \bibinfo{pages}{815--829}. \DOIprefix\doi{10.1007/s00158-013-1015-5}.
\bibitem[{Aage et~al.(2015)Aage, Andreassen, and Lazarov}]{Aage2015}
\bibinfo{author}{N.~Aage}, \bibinfo{author}{E.~Andreassen}, \bibinfo{author}{B.~S. Lazarov},
\newblock \bibinfo{title}{Topology optimization using {PETSc}: An easy-to-use, fully parallel, open source topology optimization framework},
\newblock \bibinfo{journal}{Structural and Multidisciplinary Optimization} \bibinfo{volume}{51} (\bibinfo{year}{2015}) \bibinfo{pages}{565--572}. \DOIprefix\doi{10.1007/s00158-014-1157-0}.
\bibitem[{Balay et~al.(2025)Balay, Abhyankar, Adams, Benson, Brown, Brune, Buschelman, Constantinescu, Dalcin, Dener, Eijkhout, Faibussowitsch, Gropp, Hapla, Isaac, Jolivet, Karpeev, Kaushik, Knepley, Kong, Kruger, May, McInnes, Mills, Mitchell, Munson, Roman, Rupp, Sanan, Sarich, Smith, Suh, Zampini, Zhang, Zhang, and Zhang}]{petsc-user-ref}
\bibinfo{author}{S.~Balay}, \bibinfo{author}{S.~Abhyankar}, \bibinfo{author}{M.~F. Adams}, \bibinfo{author}{S.~Benson}, \bibinfo{author}{J.~Brown}, \bibinfo{author}{P.~Brune}, \bibinfo{author}{K.~Buschelman}, \bibinfo{author}{E.~Constantinescu}, \bibinfo{author}{L.~Dalcin}, \bibinfo{author}{A.~Dener}, \bibinfo{author}{V.~Eijkhout}, \bibinfo{author}{J.~Faibussowitsch}, \bibinfo{author}{W.~D. Gropp}, \bibinfo{author}{V.~Hapla}, \bibinfo{author}{T.~Isaac}, \bibinfo{author}{P.~Jolivet}, \bibinfo{author}{D.~Karpeev}, \bibinfo{author}{D.~Kaushik}, \bibinfo{author}{M.~G. Knepley}, \bibinfo{author}{F.~Kong}, \bibinfo{author}{S.~Kruger}, \bibinfo{author}{D.~A. May}, \bibinfo{author}{L.~C. McInnes}, \bibinfo{author}{R.~T. Mills}, \bibinfo{author}{L.~Mitchell}, \bibinfo{author}{T.~Munson}, \bibinfo{author}{J.~E. Roman}, \bibinfo{author}{K.~Rupp}, \bibinfo{author}{P.~Sanan}, \bibinfo{author}{J.~Sarich}, \bibinfo{author}{B.~F. Smith}, \bibinfo{author}{H.~Suh}, \bibinfo{author}{S.~Zampini}, \bibinfo{author}{H.~Zhang},
  \bibinfo{author}{H.~Zhang}, \bibinfo{author}{J.~Zhang}, \bibinfo{title}{{PETSc/TAO} Users Manual}, \bibinfo{type}{Technical Report} \bibinfo{number}{ANL-21/39 - Revision 3.23}, Argonne National Laboratory, \bibinfo{year}{2025}. \DOIprefix\doi{10.2172/2476320}.
\bibitem[{Aage et~al.(2017)Aage, Andreassen, Lazarov, and Sigmund}]{Aage2017}
\bibinfo{author}{N.~Aage}, \bibinfo{author}{E.~Andreassen}, \bibinfo{author}{B.~S. Lazarov}, \bibinfo{author}{O.~Sigmund},
\newblock \bibinfo{title}{Giga-voxel computational morphogenesis for structural design},
\newblock \bibinfo{journal}{Nature} \bibinfo{volume}{550} (\bibinfo{year}{2017}) \bibinfo{pages}{84--86}. \DOIprefix\doi{10.1038/nature23911}.
\bibitem[{Baandrup et~al.(2020)Baandrup, Sigmund, Polk, and Aage}]{Baandrup2020}
\bibinfo{author}{M.~Baandrup}, \bibinfo{author}{O.~Sigmund}, \bibinfo{author}{H.~Polk}, \bibinfo{author}{N.~Aage},
\newblock \bibinfo{title}{Closing the gap towards super-long suspension bridges using computational morphogenesis},
\newblock \bibinfo{journal}{Nature Communications} \bibinfo{volume}{11} (\bibinfo{year}{2020}) \bibinfo{pages}{2735}. \DOIprefix\doi{10.1038/s41467-020-16599-6}.
\bibitem[{Alexandersen et~al.(2016)Alexandersen, Sigmund, and Aage}]{Alexandersen2016}
\bibinfo{author}{J.~Alexandersen}, \bibinfo{author}{O.~Sigmund}, \bibinfo{author}{N.~Aage},
\newblock \bibinfo{title}{Large scale three-dimensional topology optimisation of heat sinks cooled by natural convection},
\newblock \bibinfo{journal}{International Journal of Heat and Mass Transfer} \bibinfo{volume}{100} (\bibinfo{year}{2016}) \bibinfo{pages}{876--891}. \DOIprefix\doi{10.1016/j.ijheatmasstransfer.2016.05.013}.
\bibitem[{Høghøj et~al.(2020)Høghøj, Nørhave, Alexandersen, Sigmund, and Andreasen}]{Hoeghoej2020}
\bibinfo{author}{L.~C. Høghøj}, \bibinfo{author}{D.~R. Nørhave}, \bibinfo{author}{J.~Alexandersen}, \bibinfo{author}{O.~Sigmund}, \bibinfo{author}{C.~S. Andreasen},
\newblock \bibinfo{title}{Topology optimization of two fluid heat exchangers},
\newblock \bibinfo{journal}{International Journal of Heat and Mass Transfer} \bibinfo{volume}{163} (\bibinfo{year}{2020}) \bibinfo{pages}{120543}. \DOIprefix\doi{https://doi.org/10.1016/j.ijheatmasstransfer.2020.120543}.
\bibitem[{Rogi{\'e} and Andreasen(2022)}]{Rogie2022}
\bibinfo{author}{B.~Rogi{\'e}}, \bibinfo{author}{C.~S. Andreasen},
\newblock \bibinfo{title}{Design complexity tradeoffs in topology optimization of forced convection laminar flow heat sinks},
\newblock \bibinfo{journal}{Structural and Multidisciplinary Optimization} \bibinfo{volume}{66} (\bibinfo{year}{2022}) \bibinfo{pages}{6}. \DOIprefix\doi{10.1007/s00158-022-03449-w}.
\bibitem[{Zhou et~al.(2025)Zhou, Ye, Liu, Fu, and Chung}]{Zhou2025}
\bibinfo{author}{Y.~Zhou}, \bibinfo{author}{C.~Ye}, \bibinfo{author}{Y.~Liu}, \bibinfo{author}{S.~Fu}, \bibinfo{author}{E.~T. Chung}, \bibinfo{title}{A robust solver for large-scale heat transfer topology optimization}, \bibinfo{year}{2025}. \URLprefix \url{https://arxiv.org/abs/2410.06850}. \href{http://arxiv.org/abs/2410.06850}{{\tt arXiv:2410.06850}}.
\bibitem[{Herrero-P{\'e}rez and Pic{\'o}-Vicente(2023)}]{Herrero-Perez2023}
\bibinfo{author}{D.~Herrero-P{\'e}rez}, \bibinfo{author}{S.~G. Pic{\'o}-Vicente},
\newblock \bibinfo{title}{A parallel geometric multigrid method for adaptive topology optimization},
\newblock \bibinfo{journal}{Structural and Multidisciplinary Optimization} \bibinfo{volume}{66} (\bibinfo{year}{2023}) \bibinfo{pages}{225}. \DOIprefix\doi{10.1007/s00158-023-03675-w}.
\bibitem[{Luo et~al.(2024)Luo, Yang, and Wang}]{Luo2024}
\bibinfo{author}{S.~Luo}, \bibinfo{author}{F.~Yang}, \bibinfo{author}{Y.~Wang},
\newblock \bibinfo{title}{An efficient isogeometric topology optimization based on the adaptive damped geometric multigrid method},
\newblock \bibinfo{journal}{Advances in Engineering Software} \bibinfo{volume}{196} (\bibinfo{year}{2024}) \bibinfo{pages}{103712}. \DOIprefix\doi{10.1016/j.advengsoft.2024.103712}.
\bibitem[{Min et~al.(1999)Min, Kikuchi, Park, Kim, and Chang}]{Min1999}
\bibinfo{author}{S.~Min}, \bibinfo{author}{N.~Kikuchi}, \bibinfo{author}{Y.~C. Park}, \bibinfo{author}{S.~Kim}, \bibinfo{author}{S.~Chang},
\newblock \bibinfo{title}{Optimal topology design of structures under dynamic loads},
\newblock \bibinfo{journal}{Structural optimization} \bibinfo{volume}{17} (\bibinfo{year}{1999}) \bibinfo{pages}{208--218}. \DOIprefix\doi{10.1007/BF01195945}.
\bibitem[{Li et~al.(2001)Li, Steven, and Xie}]{Li2001}
\bibinfo{author}{Q.~Li}, \bibinfo{author}{G.~Steven}, \bibinfo{author}{Y.~M. Xie},
\newblock \bibinfo{title}{Thermoelastic topology optimization for problems with varying temperature fields},
\newblock \bibinfo{journal}{Journal of Thermal Stresses} \bibinfo{volume}{24} (\bibinfo{year}{2001}) \bibinfo{pages}{347--366}. \DOIprefix\doi{10.1080/01495730151078153}.
\bibitem[{Zhuang et~al.(2013)Zhuang, Xiong, and Ding}]{Zhuang2013}
\bibinfo{author}{C.~Zhuang}, \bibinfo{author}{Z.~Xiong}, \bibinfo{author}{H.~Ding},
\newblock \bibinfo{title}{Topology optimization of the transient heat conduction problem on a triangular mesh},
\newblock \bibinfo{journal}{Numerical Heat Transfer, Part B: Fundamentals} \bibinfo{volume}{64} (\bibinfo{year}{2013}) \bibinfo{pages}{239--262}. \DOIprefix\doi{10.1080/10407790.2013.791785}.
\bibitem[{Zhuang and Xiong(2015)}]{Zhuang2015}
\bibinfo{author}{C.~Zhuang}, \bibinfo{author}{Z.~Xiong},
\newblock \bibinfo{title}{Temperature-constrained topology optimization of transient heat conduction problems},
\newblock \bibinfo{journal}{Numerical Heat Transfer, Part B: Fundamentals} \bibinfo{volume}{68} (\bibinfo{year}{2015}) \bibinfo{pages}{366--385}. \DOIprefix\doi{10.1080/10407790.2015.1033306}.
\bibitem[{Wu et~al.(2019)Wu, Zhang, and Liu}]{Wu2019}
\bibinfo{author}{S.~Wu}, \bibinfo{author}{Y.~Zhang}, \bibinfo{author}{S.~Liu},
\newblock \bibinfo{title}{Topology optimization for minimizing the maximum temperature of transient heat conduction structure},
\newblock \bibinfo{journal}{Structural and Multidisciplinary Optimization} \bibinfo{volume}{60} (\bibinfo{year}{2019}) \bibinfo{pages}{69--82}. \DOIprefix\doi{10.1007/s00158-019-02196-9}.
\bibitem[{Zeng et~al.(2020)Zeng, Wang, Yang, and Alexandersen}]{Zeng2020}
\bibinfo{author}{T.~Zeng}, \bibinfo{author}{H.~Wang}, \bibinfo{author}{M.~Yang}, \bibinfo{author}{J.~Alexandersen},
\newblock \bibinfo{title}{Topology optimization of heat sinks for instantaneous chip cooling using a transient pseudo-3d thermofluid model},
\newblock \bibinfo{journal}{International Journal of Heat and Mass Transfer} \bibinfo{volume}{154} (\bibinfo{year}{2020}) \bibinfo{pages}{119681}. \DOIprefix\doi{10.1016/j.ijheatmasstransfer.2020.119681}.
\bibitem[{Srinivas and Ananthasuresh(2006)}]{Srinivas2006}
\bibinfo{author}{V.~Srinivas}, \bibinfo{author}{G.~Ananthasuresh},
\newblock \bibinfo{title}{Analysis and topology optimization of heat sinks with a phase- change material on {COMSOL} multiphysics platform},
\newblock in: \bibinfo{booktitle}{Proceedings of the {COMSOL} Users Conference 2006 Bangalore}, \bibinfo{year}{2006}. \URLprefix \url{https://www.researchgate.net/publication/267679913}.
\bibitem[{Pizzolato et~al.(2017)Pizzolato, Sharma, Maute, Sciacovelli, and Verda}]{Pizzolato2017}
\bibinfo{author}{A.~Pizzolato}, \bibinfo{author}{A.~Sharma}, \bibinfo{author}{K.~Maute}, \bibinfo{author}{A.~Sciacovelli}, \bibinfo{author}{V.~Verda},
\newblock \bibinfo{title}{Topology optimization for heat transfer enhancement in latent heat thermal energy storage},
\newblock \bibinfo{journal}{International Journal of Heat and Mass Transfer} \bibinfo{volume}{113} (\bibinfo{year}{2017}) \bibinfo{pages}{875--888}. \DOIprefix\doi{10.1016/j.ijheatmasstransfer.2017.05.098}.
\bibitem[{Lundgaard et~al.(2019)Lundgaard, Engelbrecht, and Sigmund}]{Lundgaard2019}
\bibinfo{author}{C.~Lundgaard}, \bibinfo{author}{K.~Engelbrecht}, \bibinfo{author}{O.~Sigmund},
\newblock \bibinfo{title}{A density-based topology optimization methodology for thermal energy storage systems},
\newblock \bibinfo{journal}{Structural and Multidisciplinary Optimization} \bibinfo{volume}{60} (\bibinfo{year}{2019}) \bibinfo{pages}{2189--2204}. \DOIprefix\doi{10.1007/s00158-019-02375-8}.
\bibitem[{Yao et~al.(2021)Yao, Zhao, Zhao, Wang, and Li}]{Yao2021}
\bibinfo{author}{Q.~Yao}, \bibinfo{author}{C.~Zhao}, \bibinfo{author}{Y.~Zhao}, \bibinfo{author}{H.~Wang}, \bibinfo{author}{W.~Li},
\newblock \bibinfo{title}{Topology optimization for heat transfer enhancement in latent heat storage},
\newblock \bibinfo{journal}{International Journal of Thermal Sciences} \bibinfo{volume}{159} (\bibinfo{year}{2021}) \bibinfo{pages}{106578}. \DOIprefix\doi{10.1016/j.ijthermalsci.2020.106578}.
\bibitem[{Christensen and Alexandersen(2023)}]{Christensen2023}
\bibinfo{author}{M.~B.~M. Christensen}, \bibinfo{author}{J.~Alexandersen}, \bibinfo{title}{{Topology optimisation of heat sinks embedded with phase-change material for minimising temperature oscillations}}, \bibinfo{year}{2023}. \URLprefix \url{https://hal.science/hal-04185641}, \bibinfo{note}{working paper or preprint}.
\bibitem[{Li et~al.(2004)Li, Saitou, and Kikuchi}]{Li2004}
\bibinfo{author}{Y.~Li}, \bibinfo{author}{K.~Saitou}, \bibinfo{author}{N.~Kikuchi},
\newblock \bibinfo{title}{Topology optimization of thermally actuated compliant mechanisms considering time-transient effect},
\newblock \bibinfo{journal}{Finite Elements in Analysis and Design} \bibinfo{volume}{40} (\bibinfo{year}{2004}) \bibinfo{pages}{1317--1331}. \DOIprefix\doi{10.1016/j.finel.2003.05.002}.
\bibitem[{Guo et~al.(2024)Guo, Cheng, Wang, Lai, and Chen}]{Guo2024}
\bibinfo{author}{Y.~Guo}, \bibinfo{author}{S.~Cheng}, \bibinfo{author}{Y.~Wang}, \bibinfo{author}{X.~Lai}, \bibinfo{author}{L.~Chen},
\newblock \bibinfo{title}{Topology optimization for transient thermoelastic structures under time-dependent loads},
\newblock \bibinfo{journal}{Engineering with Computers} \bibinfo{volume}{40} (\bibinfo{year}{2024}) \bibinfo{pages}{1677--1693}. \DOIprefix\doi{10.1007/s00366-023-01878-9}.
\bibitem[{Michaleris et~al.(1994)Michaleris, Tortorelli, and Vidal}]{Michaleris1994}
\bibinfo{author}{P.~Michaleris}, \bibinfo{author}{D.~A. Tortorelli}, \bibinfo{author}{C.~A. Vidal},
\newblock \bibinfo{title}{Tangent operators and design sensitivity formulations for transient non-linear coupled problems with applications to elastoplasticity},
\newblock \bibinfo{journal}{International Journal for Numerical Methods in Engineering} \bibinfo{volume}{37} (\bibinfo{year}{1994}) \bibinfo{pages}{2471--2499}. \DOIprefix\doi{10.1002/nme.1620371408}.
\bibitem[{Dahl et~al.(2008)Dahl, Jensen, and Sigmund}]{Dahl2008}
\bibinfo{author}{J.~Dahl}, \bibinfo{author}{J.~S. Jensen}, \bibinfo{author}{O.~Sigmund},
\newblock \bibinfo{title}{Topology optimization for transient wave propagation problems in one dimension},
\newblock \bibinfo{journal}{Structural and Multidisciplinary Optimization} \bibinfo{volume}{36} (\bibinfo{year}{2008}) \bibinfo{pages}{585--595}. \DOIprefix\doi{10.1007/s00158-007-0192-5}.
\bibitem[{Elesin et~al.(2014)Elesin, Lazarov, Jensen, and Sigmund}]{Elesin2014}
\bibinfo{author}{Y.~Elesin}, \bibinfo{author}{B.~S. Lazarov}, \bibinfo{author}{J.~S. Jensen}, \bibinfo{author}{O.~Sigmund},
\newblock \bibinfo{title}{Time domain topology optimization of {3D} nanophotonic devices},
\newblock \bibinfo{journal}{Photonics and Nanostructures - Fundamentals and Applications} \bibinfo{volume}{12} (\bibinfo{year}{2014}) \bibinfo{pages}{23--33}. \DOIprefix\doi{https://doi.org/10.1016/j.photonics.2013.07.008}.
\bibitem[{Coffin and Maute(2016)}]{Coffin2016}
\bibinfo{author}{P.~Coffin}, \bibinfo{author}{K.~Maute},
\newblock \bibinfo{title}{A level-set method for steady-state and transient natural convection problems},
\newblock \bibinfo{journal}{Structural and Multidisciplinary Optimization} \bibinfo{volume}{53} (\bibinfo{year}{2016}) \bibinfo{pages}{1047--1067}. \DOIprefix\doi{10.1007/s00158-015-1377-y}.
\bibitem[{Kristiansen and Aage(2022)}]{Kristiansen2022}
\bibinfo{author}{H.~Kristiansen}, \bibinfo{author}{N.~Aage},
\newblock \bibinfo{title}{An open-source framework for large-scale transient topology optimization using petsc},
\newblock \bibinfo{journal}{Structural and Multidisciplinary Optimization} \bibinfo{volume}{65} (\bibinfo{year}{2022}) \bibinfo{pages}{295}. \DOIprefix\doi{10.1007/s00158-022-03312-y}.
\bibitem[{Makhija and Beran(2020)}]{Makhija2020}
\bibinfo{author}{D.~S. Makhija}, \bibinfo{author}{P.~S. Beran},
\newblock \bibinfo{title}{Concurrent shape and topology optimization for unsteady conjugate heat transfer},
\newblock \bibinfo{journal}{Structural and Multidisciplinary Optimization} \bibinfo{volume}{62} (\bibinfo{year}{2020}) \bibinfo{pages}{1275--1297}. \DOIprefix\doi{10.1007/s00158-020-02554-y}.
\bibitem[{Griewank(1992)}]{Griewank1992}
\bibinfo{author}{A.~Griewank},
\newblock \bibinfo{title}{Achieving logarithmic growth of temporal and spatial complexity in reverse automatic differentiation} \bibinfo{volume}{1} (\bibinfo{year}{1992}) \bibinfo{pages}{35--54}.
\bibitem[{Griewank and Walther(2000)}]{Griewank2000}
\bibinfo{author}{A.~Griewank}, \bibinfo{author}{A.~Walther},
\newblock \bibinfo{title}{Algorithm 799: revolve: an implementation of checkpointing for the reverse or adjoint mode of computational differentiation},
\newblock \bibinfo{journal}{ACM Trans. Math. Softw.} \bibinfo{volume}{26} (\bibinfo{year}{2000}) \bibinfo{pages}{19–45}. \DOIprefix\doi{10.1145/347837.347846}.
\bibitem[{Margetis et~al.(2023)Margetis, Papoutsis-Kiachagias, and Giannakoglou}]{Margetis2023}
\bibinfo{author}{A.-S.~I. Margetis}, \bibinfo{author}{E.~M. Papoutsis-Kiachagias}, \bibinfo{author}{K.~C. Giannakoglou},
\newblock \bibinfo{title}{Reducing memory requirements of unsteady adjoint by synergistically using check-pointing and compression},
\newblock \bibinfo{journal}{International Journal for Numerical Methods in Fluids} \bibinfo{volume}{95} (\bibinfo{year}{2023}) \bibinfo{pages}{23--43}. \DOIprefix\doi{https://doi.org/10.1002/fld.5136}.
\bibitem[{Yamaleev et~al.(2010)Yamaleev, Diskin, and Nielsen}]{Yamaleev2010}
\bibinfo{author}{N.~K. Yamaleev}, \bibinfo{author}{B.~Diskin}, \bibinfo{author}{E.~J. Nielsen},
\newblock \bibinfo{title}{Local-in-time adjoint-based method for design optimization of unsteady flows},
\newblock \bibinfo{journal}{Journal of Computational Physics} \bibinfo{volume}{229} (\bibinfo{year}{2010}) \bibinfo{pages}{5394--5407}. \DOIprefix\doi{10.1016/j.jcp.2010.03.045}.
\bibitem[{Chen et~al.(2017)Chen, Yaji, Yamada, Izui, and Nishiwaki}]{Chen2017}
\bibinfo{author}{C.~Chen}, \bibinfo{author}{K.~Yaji}, \bibinfo{author}{T.~Yamada}, \bibinfo{author}{K.~Izui}, \bibinfo{author}{S.~Nishiwaki},
\newblock \bibinfo{title}{Local-in-time adjoint-based topology optimization of unsteady fluid flows using the lattice boltzmann method},
\newblock \bibinfo{journal}{Mechanical Engineering Journal} \bibinfo{volume}{4} (\bibinfo{year}{2017}) \bibinfo{pages}{17--00120--17--00120}. \DOIprefix\doi{10.1299/mej.17-00120}.
\bibitem[{Yaji et~al.(2018)Yaji, Ogino, Chen, and Fujita}]{Yaji2018}
\bibinfo{author}{K.~Yaji}, \bibinfo{author}{M.~Ogino}, \bibinfo{author}{C.~Chen}, \bibinfo{author}{K.~Fujita},
\newblock \bibinfo{title}{Large-scale topology optimization incorporating local-in-time adjoint-based method for unsteady thermal-fluid problem},
\newblock \bibinfo{journal}{Structural and Multidisciplinary Optimization} \bibinfo{volume}{58} (\bibinfo{year}{2018}) \bibinfo{pages}{817--822}. \DOIprefix\doi{10.1007/s00158-018-1922-6}.
\bibitem[{Theulings et~al.(2024)Theulings, Maas, Noël, van Keulen, and Langelaar}]{Theulings2024}
\bibinfo{author}{M.~J.~B. Theulings}, \bibinfo{author}{R.~Maas}, \bibinfo{author}{L.~Noël}, \bibinfo{author}{F.~van Keulen}, \bibinfo{author}{M.~Langelaar},
\newblock \bibinfo{title}{Reducing time and memory requirements in topology optimization of transient problems},
\newblock \bibinfo{journal}{International Journal for Numerical Methods in Engineering} \bibinfo{volume}{125} (\bibinfo{year}{2024}) \bibinfo{pages}{e7461}. \DOIprefix\doi{10.1002/nme.7461}.
\bibitem[{Hyun and Wang(2019)}]{Hyun2019}
\bibinfo{author}{J.~Hyun}, \bibinfo{author}{S.~Wang},
\newblock \bibinfo{title}{Systematically engineered thermal metastructure for rapid heat dissipation/diffusion by considering the thermal eigenvalue},
\newblock \bibinfo{journal}{Applied Thermal Engineering} \bibinfo{volume}{157} (\bibinfo{year}{2019}) \bibinfo{pages}{113487}. \DOIprefix\doi{https://doi.org/10.1016/j.applthermaleng.2019.03.058}.
\bibitem[{Hyun and Kim(2021)}]{Hyun2021}
\bibinfo{author}{J.~Hyun}, \bibinfo{author}{H.~A. Kim},
\newblock \bibinfo{title}{Level-set topology optimization for effective control of transient conductive heat response using eigenvalue},
\newblock \bibinfo{journal}{International Journal of Heat and Mass Transfer} \bibinfo{volume}{176} (\bibinfo{year}{2021}) \bibinfo{pages}{121374}. \DOIprefix\doi{10.1016/j.ijheatmasstransfer.2021.121374}.
\bibitem[{Onodera and Yamada(2025)}]{Onodera2025}
\bibinfo{author}{S.~Onodera}, \bibinfo{author}{T.~Yamada},
\newblock \bibinfo{title}{A topology optimization method for managing transient thermal and vibration effects with eigenvalues and steady-state constraints},
\newblock \bibinfo{journal}{International Journal of Heat and Mass Transfer} \bibinfo{volume}{246} (\bibinfo{year}{2025}) \bibinfo{pages}{127083}. \DOIprefix\doi{10.1016/j.ijheatmasstransfer.2025.127083}.
\bibitem[{Isiklar et~al.(2024)Isiklar, Christiansen, and Sigmund}]{Isiklar2024}
\bibinfo{author}{G.~Isiklar}, \bibinfo{author}{R.~E. Christiansen}, \bibinfo{author}{O.~Sigmund}, \bibinfo{title}{Topology optimization of thermal initial value problems exploiting efficient harmonic analysis}, \bibinfo{year}{2024}. \URLprefix \url{https://www.researchsquare.com/article/rs-3984636/v1}, \bibinfo{note}{preprint}.
\bibitem[{Yan et~al.(2024)Yan, Liu, and Yan}]{Yan2024}
\bibinfo{author}{K.~Yan}, \bibinfo{author}{D.~Liu}, \bibinfo{author}{J.~Yan},
\newblock \bibinfo{title}{Topology optimization method for transient heat conduction using the lyapunov equation},
\newblock \bibinfo{journal}{International Journal of Heat and Mass Transfer} \bibinfo{volume}{231} (\bibinfo{year}{2024}) \bibinfo{pages}{125815}. \DOIprefix\doi{10.1016/j.ijheatmasstransfer.2024.125815}.
\bibitem[{Hooijkamp and Keulen(2018)}]{Hooijkamp2018}
\bibinfo{author}{E.~C. Hooijkamp}, \bibinfo{author}{F.~v. Keulen},
\newblock \bibinfo{title}{Topology optimization for linear thermo-mechanical transient problems: Modal reduction and adjoint sensitivities},
\newblock \bibinfo{journal}{International Journal for Numerical Methods in Engineering} \bibinfo{volume}{113} (\bibinfo{year}{2018}) \bibinfo{pages}{1230--1257}. \DOIprefix\doi{10.1002/nme.5635}.
\bibitem[{van~der Kolk et~al.(2018)van~der Kolk, Hooijkamp, Langelaar, and van Keulen}]{Kolk2018}
\bibinfo{author}{M.~van~der Kolk}, \bibinfo{author}{E.~C. Hooijkamp}, \bibinfo{author}{M.~Langelaar}, \bibinfo{author}{F.~van Keulen},
\newblock \bibinfo{title}{Using exact particular solutions and modal reduction in topology optimization of transient thermo-mechanical problems},
\newblock in: \bibinfo{editor}{A.~Schumacher}, \bibinfo{editor}{T.~Vietor}, \bibinfo{editor}{S.~Fiebig}, \bibinfo{editor}{K.-U. Bletzinger}, \bibinfo{editor}{K.~Maute} (Eds.), \bibinfo{booktitle}{Advances in Structural and Multidisciplinary Optimization}, \bibinfo{year}{2018}, pp. \bibinfo{pages}{1027--1041}.
\bibitem[{Li et~al.(2024)Li, Yin, Jiang, Zhang, and Wang}]{Li2024}
\bibinfo{author}{S.~Li}, \bibinfo{author}{J.~Yin}, \bibinfo{author}{X.~Jiang}, \bibinfo{author}{Y.~Zhang}, \bibinfo{author}{H.~Wang},
\newblock \bibinfo{title}{A novel reduced basis method for adjoint sensitivity analysis of dynamic topology optimization},
\newblock \bibinfo{journal}{Engineering Analysis with Boundary Elements} \bibinfo{volume}{162} (\bibinfo{year}{2024}) \bibinfo{pages}{403--419}. \DOIprefix\doi{10.1016/j.enganabound.2024.03.001}.
\bibitem[{Li et~al.(2025)Li, Yin, Zhang, and Wang}]{Li2025}
\bibinfo{author}{S.~Li}, \bibinfo{author}{J.~Yin}, \bibinfo{author}{Y.~Zhang}, \bibinfo{author}{H.~Wang},
\newblock \bibinfo{title}{An online reduced-order method for dynamic sensitivity analysis},
\newblock \bibinfo{journal}{Engineering Analysis with Boundary Elements} \bibinfo{volume}{175} (\bibinfo{year}{2025}) \bibinfo{pages}{106198}. \DOIprefix\doi{10.1016/j.enganabound.2025.106198}.
\bibitem[{Gander(2015)}]{Gander2015}
\bibinfo{author}{M.~J. Gander},
\newblock \bibinfo{title}{50 years of time parallel time integration},
\newblock in: \bibinfo{editor}{T.~Carraro}, \bibinfo{editor}{M.~Geiger}, \bibinfo{editor}{S.~K{\"o}rkel}, \bibinfo{editor}{R.~Rannacher} (Eds.), \bibinfo{booktitle}{Multiple Shooting and Time Domain Decomposition Methods}, \bibinfo{publisher}{Springer International Publishing}, \bibinfo{address}{Cham}, \bibinfo{year}{2015}, pp. \bibinfo{pages}{69--113}.
\bibitem[{Ong and Schroder(2020)}]{Ong2020}
\bibinfo{author}{B.~W. Ong}, \bibinfo{author}{J.~B. Schroder},
\newblock \bibinfo{title}{Applications of time parallelization},
\newblock \bibinfo{journal}{Computing and Visualization in Science} \bibinfo{volume}{23} (\bibinfo{year}{2020}) \bibinfo{pages}{11}. \DOIprefix\doi{10.1007/s00791-020-00331-4}.
\bibitem[{Lions et~al.(2001)Lions, Maday, and Turinici}]{Lions2001}
\bibinfo{author}{J.-L. Lions}, \bibinfo{author}{Y.~Maday}, \bibinfo{author}{G.~Turinici},
\newblock \bibinfo{title}{R\'{e}solution d'edp par un sch\'{e}ma en temps parar\'{e}el},
\newblock \bibinfo{journal}{Comptes Rendus de l'Acad\'{e}mie des Sciences - Series I - Mathematics} \bibinfo{volume}{332} (\bibinfo{year}{2001}) \bibinfo{pages}{661--668}. \URLprefix \url{https://www.sciencedirect.com/science/article/pii/S0764444200017936}. \DOIprefix\doi{https://doi.org/10.1016/S0764-4442(00)01793-6}.
\bibitem[{Friedhoff et~al.(2012)Friedhoff, Falgout, Kolev, MacLachlan, and Schroder}]{Friedhoff2012}
\bibinfo{author}{S.~Friedhoff}, \bibinfo{author}{R.~D. Falgout}, \bibinfo{author}{T.~Kolev}, \bibinfo{author}{S.~MacLachlan}, \bibinfo{author}{J.~B. Schroder}, \bibinfo{title}{A multigrid-in-time algorithm for solving evolution equations in parallel}, \bibinfo{type}{Technical Report}, Lawrence Livermore National Lab.(LLNL), Livermore, CA (United States), \bibinfo{year}{2012}.
\bibitem[{Womble(1990)}]{Womble1990}
\bibinfo{author}{D.~E. Womble},
\newblock \bibinfo{title}{A time-stepping algorithm for parallel computers},
\newblock \bibinfo{journal}{SIAM Journal on Scientific and Statistical Computing} \bibinfo{volume}{11} (\bibinfo{year}{1990}) \bibinfo{pages}{824--837}. \URLprefix \url{https://doi.org/10.1137/0911049}. \DOIprefix\doi{10.1137/0911049}.
\bibitem[{Emmett and Minion(2012)}]{Emmett2012}
\bibinfo{author}{M.~Emmett}, \bibinfo{author}{M.~Minion},
\newblock \bibinfo{title}{Toward an efficient parallel in time method for partial differential equations},
\newblock \bibinfo{journal}{Communications in Applied Mathematics and Computational Science} \bibinfo{volume}{7} (\bibinfo{year}{2012}) \bibinfo{pages}{105--132}.
\bibitem[{Hahne et~al.(2023)Hahne, Polenz, Kulchytska-Ruchka, Friedhoff, Ulbrich, and Sch{\"o}ps}]{Hahne2023}
\bibinfo{author}{J.~Hahne}, \bibinfo{author}{B.~Polenz}, \bibinfo{author}{I.~Kulchytska-Ruchka}, \bibinfo{author}{S.~Friedhoff}, \bibinfo{author}{S.~Ulbrich}, \bibinfo{author}{S.~Sch{\"o}ps},
\newblock \bibinfo{title}{Parallel-in-time optimization of induction motors},
\newblock \bibinfo{journal}{Journal of Mathematics in Industry} \bibinfo{volume}{13} (\bibinfo{year}{2023}) \bibinfo{pages}{6}.
\bibitem[{Du et~al.(2013)Du, Sarkis, Schaerer, and Szyld}]{Du2013}
\bibinfo{author}{X.~Du}, \bibinfo{author}{M.~Sarkis}, \bibinfo{author}{C.~E. Schaerer}, \bibinfo{author}{D.~B. Szyld},
\newblock \bibinfo{title}{Inexact and truncated parareal-in-time krylov subspace methods for parabolic optimal control problems},
\newblock \bibinfo{journal}{Electronic Transactions on Numerical Analysis} \bibinfo{volume}{40} (\bibinfo{year}{2013}) \bibinfo{pages}{36--57}.
\bibitem[{G{\"o}tschel and Minion(2018)}]{Minion2018}
\bibinfo{author}{S.~G{\"o}tschel}, \bibinfo{author}{M.~L. Minion},
\newblock \bibinfo{title}{Parallel-in-time for parabolic optimal control problems using pfasst},
\newblock in: \bibinfo{editor}{P.~E. Bj{\o}rstad}, \bibinfo{editor}{S.~C. Brenner}, \bibinfo{editor}{L.~Halpern}, \bibinfo{editor}{H.~H. Kim}, \bibinfo{editor}{R.~Kornhuber}, \bibinfo{editor}{T.~Rahman}, \bibinfo{editor}{O.~B. Widlund} (Eds.), \bibinfo{booktitle}{Domain Decomposition Methods in Science and Engineering XXIV}, \bibinfo{publisher}{Springer International Publishing}, \bibinfo{address}{Cham}, \bibinfo{year}{2018}, pp. \bibinfo{pages}{363--371}.
\bibitem[{S.~G\"unther and Schroder(2019)}]{Gunther2019}
\bibinfo{author}{N.~R.~G. S.~G\"unther}, \bibinfo{author}{J.~B. Schroder},
\newblock \bibinfo{title}{A non-intrusive parallel-in-time approach for simultaneous optimization with unsteady pdes},
\newblock \bibinfo{journal}{Optimization Methods and Software} \bibinfo{volume}{34} (\bibinfo{year}{2019}) \bibinfo{pages}{1306--1321}. \DOIprefix\doi{10.1080/10556788.2018.1504050}.
\bibitem[{Appel and Alexandersen(2024)}]{Appel2024}
\bibinfo{author}{M.~Appel}, \bibinfo{author}{J.~Alexandersen},
\newblock \bibinfo{title}{One-shot {Parareal} approach fro topology optimisation of transient heat flow},
\newblock \bibinfo{journal}{Preprint available on arXiv:2411.19030}  (\bibinfo{year}{2024}).
\bibitem[{Hackbusch(1983)}]{Hackbusch1983}
\bibinfo{author}{W.~Hackbusch},
\newblock \bibinfo{title}{Parabolic multi-grid methods},
\newblock in: \bibinfo{editor}{R.~Glowinski}, \bibinfo{editor}{J.-L. Lions} (Eds.), \bibinfo{booktitle}{Computing methods in applied sciences and engineering VI}, \bibinfo{publisher}{Elsevier science publishers B.V.}, \bibinfo{year}{1983}, pp. \bibinfo{pages}{189--197}.
\bibitem[{Falgout et~al.(2017)Falgout, Friedhoff, Kolev, MacLachlan, Schroder, and Vandewalle}]{Falgout2017}
\bibinfo{author}{R.~D. Falgout}, \bibinfo{author}{S.~Friedhoff}, \bibinfo{author}{T.~V. Kolev}, \bibinfo{author}{S.~P. MacLachlan}, \bibinfo{author}{J.~B. Schroder}, \bibinfo{author}{S.~Vandewalle},
\newblock \bibinfo{title}{Multigrid methods with space--time concurrency},
\newblock \bibinfo{journal}{Computing and Visualization in Science} \bibinfo{volume}{18} (\bibinfo{year}{2017}) \bibinfo{pages}{123--143}. \DOIprefix\doi{10.1007/s00791-017-0283-9}.
\bibitem[{Horton and Vandewalle(1995)}]{Horton1995}
\bibinfo{author}{G.~Horton}, \bibinfo{author}{S.~Vandewalle},
\newblock \bibinfo{title}{A space-time multigrid method for parabolic partial differential equations},
\newblock \bibinfo{journal}{SIAM Journal on Scientific Computing} \bibinfo{volume}{16} (\bibinfo{year}{1995}) \bibinfo{pages}{848--864}.
\bibitem[{Appel and Alexandersen(2025)}]{Appel2025}
\bibinfo{author}{M.~Appel}, \bibinfo{author}{J.~Alexandersen},
\newblock \bibinfo{title}{Space-time multigrid methods suitable for topology optimisation of transient heat conduction},
\newblock \bibinfo{journal}{Preprint available on arXiv:2505.10168}  (\bibinfo{year}{2025}).
\bibitem[{Alexandersen(2023)}]{Alexandersen2023}
\bibinfo{author}{J.~Alexandersen}, \bibinfo{title}{Towards fast topology optimisation of transient problems: parallel space-time multigrid}, \bibinfo{year}{2023}. \URLprefix \url{https://easychair.org/smart-program/CM2023/2023-04-18.html\#talk:214721}. \DOIprefix\doi{10.13140/RG.2.2.12680.33283}, \bibinfo{note}{21st Copper Mountain Conference on Multigrid Methods}.
\bibitem[{Alexandersen and Appel(2025)}]{Alexandersen2025}
\bibinfo{author}{J.~Alexandersen}, \bibinfo{author}{M.~Appel}, \bibinfo{title}{Towards fast topology optimisation of transient heat conduction using parallel space-time methods}, \bibinfo{year}{2025}. \DOIprefix\doi{10.13140/RG.2.2.26102.10563}, \bibinfo{note}{16th World Congress on Structural and Multidisciplinary Optimization}.
\bibitem[{Sigmund and Maute(2013)}]{Sigmund2013}
\bibinfo{author}{O.~Sigmund}, \bibinfo{author}{K.~Maute},
\newblock \bibinfo{title}{Topology optimization approaches},
\newblock \bibinfo{journal}{Structural and Multidisciplinary Optimization} \bibinfo{volume}{48} (\bibinfo{year}{2013}) \bibinfo{pages}{1031--1055}. \DOIprefix\doi{10.1007/s00158-013-0978-6}.
\bibitem[{Jensen(2009)}]{Jensen2009}
\bibinfo{author}{J.~S. Jensen},
\newblock \bibinfo{title}{Space–time topology optimization for one-dimensional wave propagation},
\newblock \bibinfo{journal}{Computer Methods in Applied Mechanics and Engineering} \bibinfo{volume}{198} (\bibinfo{year}{2009}) \bibinfo{pages}{705--715}. \DOIprefix\doi{10.1016/j.cma.2008.10.008}.
\bibitem[{Jensen(2010)}]{Jensen2010}
\bibinfo{author}{J.~Jensen},
\newblock \bibinfo{title}{Optimization of space-time material layout for 1d wave propagation with varying mass and stiffness parameters},
\newblock volume~\bibinfo{volume}{39}, \bibinfo{publisher}{Sciendo}, \bibinfo{year}{2010}, pp. \bibinfo{pages}{599--614}. \bibinfo{note}{Workshop On Optimization with PDE Constraints ; Conference date: 01-01-2008 Through 01-01-2008}.
\bibitem[{Wang et~al.(2020)Wang, Munro, Wang, van Keulen, and Wu}]{Wang2020}
\bibinfo{author}{W.~Wang}, \bibinfo{author}{D.~Munro}, \bibinfo{author}{C.~C.~L. Wang}, \bibinfo{author}{F.~van Keulen}, \bibinfo{author}{J.~Wu},
\newblock \bibinfo{title}{Space-time topology optimization for additive manufacturing},
\newblock \bibinfo{journal}{Structural and Multidisciplinary Optimization} \bibinfo{volume}{61} (\bibinfo{year}{2020}) \bibinfo{pages}{1--18}. \DOIprefix\doi{10.1007/s00158-019-02420-6}.
\bibitem[{Hulme(1972)}]{Hulme1972}
\bibinfo{author}{B.~L. Hulme},
\newblock \bibinfo{title}{One-step piecewise polynomial galerkin methods for initial value problems},
\newblock \bibinfo{journal}{Mathematics of Computation} \bibinfo{volume}{26} (\bibinfo{year}{1972}) \bibinfo{pages}{415--426}. \DOIprefix\doi{10.2307/2005168}.
\bibitem[{Winther(1981)}]{Winther1981}
\bibinfo{author}{R.~Winther},
\newblock \bibinfo{title}{A stable finite element method for initial-boundary value problems for first-order hyperbolic systems},
\newblock \bibinfo{journal}{Mathematics of Computation - Math. Comput.} \bibinfo{volume}{36} (\bibinfo{year}{1981}) \bibinfo{pages}{65--65}. \DOIprefix\doi{10.1090/S0025-5718-1981-0595042-6}.
\bibitem[{Aziz and Monk(1989)}]{Aziz1989}
\bibinfo{author}{A.~K. Aziz}, \bibinfo{author}{P.~Monk},
\newblock \bibinfo{title}{Continuous finite elements in space and time for the heat equation},
\newblock \bibinfo{journal}{Mathematics of Computation} \bibinfo{volume}{52} (\bibinfo{year}{1989}) \bibinfo{pages}{255--274}.
\bibitem[{Hulbert and Hughes(1990)}]{Hulbert1990}
\bibinfo{author}{G.~M. Hulbert}, \bibinfo{author}{T.~J. Hughes},
\newblock \bibinfo{title}{Space-time finite element methods for second-order hyperbolic equations},
\newblock \bibinfo{journal}{Computer Methods in Applied Mechanics and Engineering} \bibinfo{volume}{84} (\bibinfo{year}{1990}) \bibinfo{pages}{327--348}. \DOIprefix\doi{10.1016/0045-7825(90)90082-W}.
\bibitem[{Steinbach and Yang(2019)}]{Steinbach2019}
\bibinfo{author}{O.~Steinbach}, \bibinfo{author}{H.~Yang}, \bibinfo{title}{7. Space-time finite element methods for parabolic evolution equations: discretization, a posteriori error estimation, adaptivity and solution}, \bibinfo{publisher}{De Gruyter}, \bibinfo{address}{Berlin, Boston}, \bibinfo{year}{2019}, pp. \bibinfo{pages}{207--248}. \DOIprefix\doi{10.1515/9783110548488-007}.
\bibitem[{Langer et~al.(2019)Langer, Neum{\"u}ller, and Schafelner}]{Langer2019}
\bibinfo{author}{U.~Langer}, \bibinfo{author}{M.~Neum{\"u}ller}, \bibinfo{author}{A.~Schafelner}, \bibinfo{title}{Space-Time Finite Element Methods for Parabolic Evolution Problems with Variable Coefficients}, \bibinfo{publisher}{Springer International Publishing}, \bibinfo{address}{Cham}, \bibinfo{year}{2019}, pp. \bibinfo{pages}{247--275}. \DOIprefix\doi{10.1007/978-3-030-14244-5\_13}.
\bibitem[{von Danwitz et~al.(2023)von Danwitz, Voulis, Hosters, and Behr}]{Danwitz2023}
\bibinfo{author}{M.~von Danwitz}, \bibinfo{author}{I.~Voulis}, \bibinfo{author}{N.~Hosters}, \bibinfo{author}{M.~Behr},
\newblock \bibinfo{title}{Time-continuous and time-discontinuous space-time finite elements for advection-diffusion problems},
\newblock \bibinfo{journal}{International Journal for Numerical Methods in Engineering} \bibinfo{volume}{124} (\bibinfo{year}{2023}) \bibinfo{pages}{3117--3144}. \DOIprefix\doi{10.1002/nme.7241}.
\bibitem[{Brooks and Hughes(1982)}]{Brooks1982}
\bibinfo{author}{A.~N. Brooks}, \bibinfo{author}{T.~J. Hughes},
\newblock \bibinfo{title}{Streamline upwind/petrov-galerkin formulations for convection dominated flows with particular emphasis on the incompressible navier-stokes equations},
\newblock \bibinfo{journal}{Computer Methods in Applied Mechanics and Engineering} \bibinfo{volume}{32} (\bibinfo{year}{1982}) \bibinfo{pages}{199--259}. \DOIprefix\doi{10.1016/0045-7825(82)90071-8}.
\bibitem[{Strang and Fix(1973)}]{Strang1973}
\bibinfo{author}{G.~Strang}, \bibinfo{author}{G.~Fix}, \bibinfo{title}{An Analysis of the Finite Element Method}, \bibinfo{publisher}{Prentice-Hall}, \bibinfo{year}{1973}.
\bibitem[{Hughes(1987)}]{Hughes1987}
\bibinfo{author}{T.~J. Hughes}, \bibinfo{title}{The finite element method: linear static and dynamic finite element analysis}, \bibinfo{publisher}{Prentice-Hall}, \bibinfo{year}{1987}.
\bibitem[{Hansbo(2000)}]{Hansbo2000}
\bibinfo{author}{P.~Hansbo},
\newblock \bibinfo{title}{A crank–nicolson type space–time finite element method for computing on moving meshes},
\newblock \bibinfo{journal}{Journal of Computational Physics} \bibinfo{volume}{159} (\bibinfo{year}{2000}) \bibinfo{pages}{274--289}. \DOIprefix\doi{10.1006/jcph.2000.6436}.
\bibitem[{Saad(1993)}]{Saad1993}
\bibinfo{author}{Y.~Saad},
\newblock \bibinfo{title}{A flexible inner-outer preconditioned gmres algorithm},
\newblock \bibinfo{journal}{SIAM Journal on Scientific Computing} \bibinfo{volume}{14} (\bibinfo{year}{1993}) \bibinfo{pages}{461--469}.
\bibitem[{Saad and Schultz(1986)}]{Saad1986}
\bibinfo{author}{Y.~Saad}, \bibinfo{author}{M.~H. Schultz},
\newblock \bibinfo{title}{Gmres: A generalized minimal residual algorithm for solving nonsymmetric linear systems},
\newblock \bibinfo{journal}{SIAM Journal on Scientific and Statistical Computing} \bibinfo{volume}{7} (\bibinfo{year}{1986}) \bibinfo{pages}{856--869}. \DOIprefix\doi{10.1137/0907058}.
\bibitem[{Gander and Neum\"{u}ller(2016)}]{Gander2016_stmg_br}
\bibinfo{author}{M.~J. Gander}, \bibinfo{author}{M.~Neum\"{u}ller},
\newblock \bibinfo{title}{Analysis of a new space-time parallel multigrid algorithm for parabolic problems},
\newblock \bibinfo{journal}{SIAM Journal on Scientific Computing} \bibinfo{volume}{38} (\bibinfo{year}{2016}) \bibinfo{pages}{A2173--A2208}. \URLprefix \url{https://doi.org/10.1137/15M1046605}. \DOIprefix\doi{10.1137/15M1046605}.
\bibitem[{Franco et~al.(2018)Franco, Gaspar, {Villela Pinto}, and Rodrigo}]{franco2018_stmg_with_CN}
\bibinfo{author}{S.~R. Franco}, \bibinfo{author}{F.~J. Gaspar}, \bibinfo{author}{M.~A. {Villela Pinto}}, \bibinfo{author}{C.~Rodrigo},
\newblock \bibinfo{title}{Multigrid method based on a space-time approach with standard coarsening for parabolic problems},
\newblock \bibinfo{journal}{Applied Mathematics and Computation} \bibinfo{volume}{317} (\bibinfo{year}{2018}) \bibinfo{pages}{25--34}. \URLprefix \url{https://www.sciencedirect.com/science/article/pii/S0096300317305945}. \DOIprefix\doi{https://doi.org/10.1016/j.amc.2017.08.043}.
\bibitem[{Lazarov and Sigmund(2011)}]{Lazarov2011}
\bibinfo{author}{B.~S. Lazarov}, \bibinfo{author}{O.~Sigmund},
\newblock \bibinfo{title}{Filters in topology optimization based on helmholtz-type differential equations},
\newblock \bibinfo{journal}{International Journal for Numerical Methods in Engineering} \bibinfo{volume}{86} (\bibinfo{year}{2011}) \bibinfo{pages}{765--781}. \DOIprefix\doi{10.1002/nme.3072}.
\bibitem[{Wang et~al.(2020)Wang, Zhou, Tian, and Wang}]{Wang2020b}
\bibinfo{author}{B.~Wang}, \bibinfo{author}{Y.~Zhou}, \bibinfo{author}{K.~Tian}, \bibinfo{author}{G.~Wang},
\newblock \bibinfo{title}{Novel implementation of extrusion constraint in topology optimization by helmholtz-type anisotropic filter},
\newblock \bibinfo{journal}{Structural and Multidisciplinary Optimization} \bibinfo{volume}{62} (\bibinfo{year}{2020}) \bibinfo{pages}{2091--2100}. \DOIprefix\doi{10.1007/s00158-020-02597-1}.
\bibitem[{Guest et~al.(2004)Guest, Pr{\'e}vost, and Belytschko}]{Guest2004}
\bibinfo{author}{J.~K. Guest}, \bibinfo{author}{J.~H. Pr{\'e}vost}, \bibinfo{author}{T.~Belytschko},
\newblock \bibinfo{title}{Achieving minimum length scale in topology optimization using nodal design variables and projection functions},
\newblock \bibinfo{journal}{International journal for numerical methods in engineering} \bibinfo{volume}{61} (\bibinfo{year}{2004}) \bibinfo{pages}{238--254}.
\bibitem[{Wang et~al.(2011)Wang, Lazarov, and Sigmund}]{Wang2011}
\bibinfo{author}{F.~Wang}, \bibinfo{author}{B.~S. Lazarov}, \bibinfo{author}{O.~Sigmund},
\newblock \bibinfo{title}{On projection methods, convergence and robust formulations in topology optimization},
\newblock \bibinfo{journal}{Structural and multidisciplinary optimization} \bibinfo{volume}{43} (\bibinfo{year}{2011}) \bibinfo{pages}{767--784}.
\bibitem[{Svanberg(1987)}]{Svanberg1987}
\bibinfo{author}{K.~Svanberg},
\newblock \bibinfo{title}{The method of moving asymptotes—a new method for structural optimization},
\newblock \bibinfo{journal}{International Journal for Numerical Methods in Engineering} \bibinfo{volume}{24} (\bibinfo{year}{1987}) \bibinfo{pages}{359--373}. \URLprefix \url{https://onlinelibrary.wiley.com/doi/abs/10.1002/nme.1620240207}. \DOIprefix\doi{https://doi.org/10.1002/nme.1620240207}.
\bibitem[{Guest et~al.(2011)Guest, Asadpoure, and Ha}]{Guest2011}
\bibinfo{author}{J.~K. Guest}, \bibinfo{author}{A.~Asadpoure}, \bibinfo{author}{S.-H. Ha},
\newblock \bibinfo{title}{Eliminating beta-continuation from heaviside projection and density filter algorithms},
\newblock \bibinfo{journal}{Structural and Multidisciplinary Optimization} \bibinfo{volume}{44} (\bibinfo{year}{2011}) \bibinfo{pages}{443--453}.
\bibitem[{LUM(2025)}]{LUMI_specs}
\bibinfo{title}{Lumi-c specifications}, \bibinfo{howpublished}{\url{https://docs.lumi-supercomputer.eu/hardware/lumic/}}, \bibinfo{year}{2025}. \bibinfo{note}{Accessed: 2025-07-15}.
\bibitem[{Yan et~al.(2018)Yan, Wang, and Sigmund}]{Yan2018}
\bibinfo{author}{S.~Yan}, \bibinfo{author}{F.~Wang}, \bibinfo{author}{O.~Sigmund},
\newblock \bibinfo{title}{On the non-optimality of tree structures for heat conduction},
\newblock \bibinfo{journal}{International Journal of Heat and Mass Transfer} \bibinfo{volume}{122} (\bibinfo{year}{2018}) \bibinfo{pages}{660--680}. \DOIprefix\doi{10.1016/j.ijheatmasstransfer.2018.01.114}.
\bibitem[{Lazarov(2014)}]{Lazarov2014}
\bibinfo{author}{B.~S. Lazarov},
\newblock \bibinfo{title}{Topology optimization using multiscale finite element method for high-contrast media},
\newblock in: \bibinfo{editor}{I.~Lirkov}, \bibinfo{editor}{S.~Margenov}, \bibinfo{editor}{J.~Wa{\'{s}}niewski} (Eds.), \bibinfo{booktitle}{Large-Scale Scientific Computing}, \bibinfo{publisher}{Springer Berlin Heidelberg}, \bibinfo{address}{Berlin, Heidelberg}, \bibinfo{year}{2014}, pp. \bibinfo{pages}{339--346}. \DOIprefix\doi{10.1007/978-3-662-43880-0\_38}.
\bibitem[{Alexandersen and Lazarov(2015)}]{Alexandersen2015}
\bibinfo{author}{J.~Alexandersen}, \bibinfo{author}{B.~S. Lazarov},
\newblock \bibinfo{title}{Topology optimisation of manufacturable microstructural details without length scale separation using a spectral coarse basis preconditioner},
\newblock \bibinfo{journal}{Computer Methods in Applied Mechanics and Engineering} \bibinfo{volume}{290} (\bibinfo{year}{2015}) \bibinfo{pages}{156--182}. \DOIprefix\doi{0.1016/j.cma.2015.02.028}.
\bibitem[{G{\"o}tschel et~al.(2021)G{\"o}tschel, Minion, Ruprecht, and Speck}]{Gotschel2021}
\bibinfo{author}{S.~G{\"o}tschel}, \bibinfo{author}{M.~Minion}, \bibinfo{author}{D.~Ruprecht}, \bibinfo{author}{R.~Speck},
\newblock \bibinfo{title}{Twelve ways to fool the masses when giving parallel-in-time results},
\newblock in: \bibinfo{editor}{B.~Ong}, \bibinfo{editor}{J.~Schroder}, \bibinfo{editor}{J.~Shipton}, \bibinfo{editor}{S.~Friedhoff} (Eds.), \bibinfo{booktitle}{Parallel-in-Time Integration Methods}, \bibinfo{publisher}{Springer International Publishing}, \bibinfo{address}{Cham}, \bibinfo{year}{2021}, pp. \bibinfo{pages}{81--94}.
\bibitem[{Myers(1978)}]{Myers1978}
\bibinfo{author}{G.~E. Myers},
\newblock \bibinfo{title}{The critical time step for finite-element solutions to two-dimensional heat-conduction transients},
\newblock \bibinfo{journal}{Journal of Heat Transfer} \bibinfo{volume}{100} (\bibinfo{year}{1978}) \bibinfo{pages}{120--127}. \DOIprefix\doi{10.1115/1.3450485}.

\end{thebibliography}

\end{document}